\documentclass[12pt]{article}

\usepackage[utf8]{inputenc}
\usepackage[T1]{fontenc}
\usepackage{lmodern}
\usepackage[margin=1in]{geometry}
\usepackage{setspace}
\usepackage{amsmath, amssymb, amsthm}
\usepackage{booktabs}
\usepackage{array}
\usepackage{float}
\usepackage{graphicx}
\usepackage{subcaption}
\usepackage{xcolor}
\usepackage{tikz}
\usepackage{hyperref}
\usepackage[authoryear]{natbib}
\hypersetup{
    colorlinks=true,
    linkcolor=blue,
    citecolor=blue,
    urlcolor=blue
}

\newtheorem{proposition}{Proposition}
\newtheorem{assumption}{Assumption}

\title{Who Saw It Coming? Historical Experience and the 2021 Inflation Forecast Failure\thanks{The author acknowledges financial support from the Chaire en macro\'{e}conomie et pr\'{e}visions ESG UQAM. I thank Hugo Couture for excellent research assistance.}}
\author{Dalibor Stevanovi\'{c} \\
UQAM, CIRANO}
\date{\today}

\begin{document}

\maketitle

\begin{abstract}
This paper studies the 2021 U.S.\ inflation forecasting failure. I show that the failure was primarily driven by sample composition rather than functional-form misspecification: estimation samples dominated by the Great Moderation underweight supply-shock regimes, and expectations anchored to that regime were slow to recognize the shift. Three historically informed adjustments, an intercept correction, a similarity re-estimation on 1970s data, and a kernel-weighted estimator, substantially close the forecast gap, and the gains extend to eight additional U.S.\ price indices. Household survey respondents over 60, whose lifetime includes the 1970s, reported higher inflation expectations from early 2021, consistent with experience-based learning; younger cohorts remained anchored to the prevailing regime. A controlled experiment with large language models conditioned on ``experienced'' and ``young'' professional personas confirms that experiential priors generate significant forecast differences under a common training leakage assumption. Across all three exercises, the source of the prior mattered more than the sophistication of the model.
\end{abstract}

\medskip


\medskip


\thispagestyle{empty}
\clearpage
\setcounter{page}{1}

\section{Introduction}

The 2021 U.S.\ inflation surge was not anticipated by the vast majority of forecasters. Econometric models, machine learning algorithms, professional surveys, and central bank projections all predicted moderate inflation for the year ahead, while the realized average reached almost 7 percent. The consequences of missing such a regime change extend beyond forecast accuracy: delayed recognition of a supply-driven surge postpones the monetary response and requires steeper corrective action later. Recognizing a regime change promptly is therefore a first-order policy problem, but it is precisely what standard forecasting tools are not designed to do, since adaptive methods detect departures from the prevailing regime only after observing enough data from the new one.

\begin{figure}[!b]
\centering
\includegraphics[width=\textwidth]{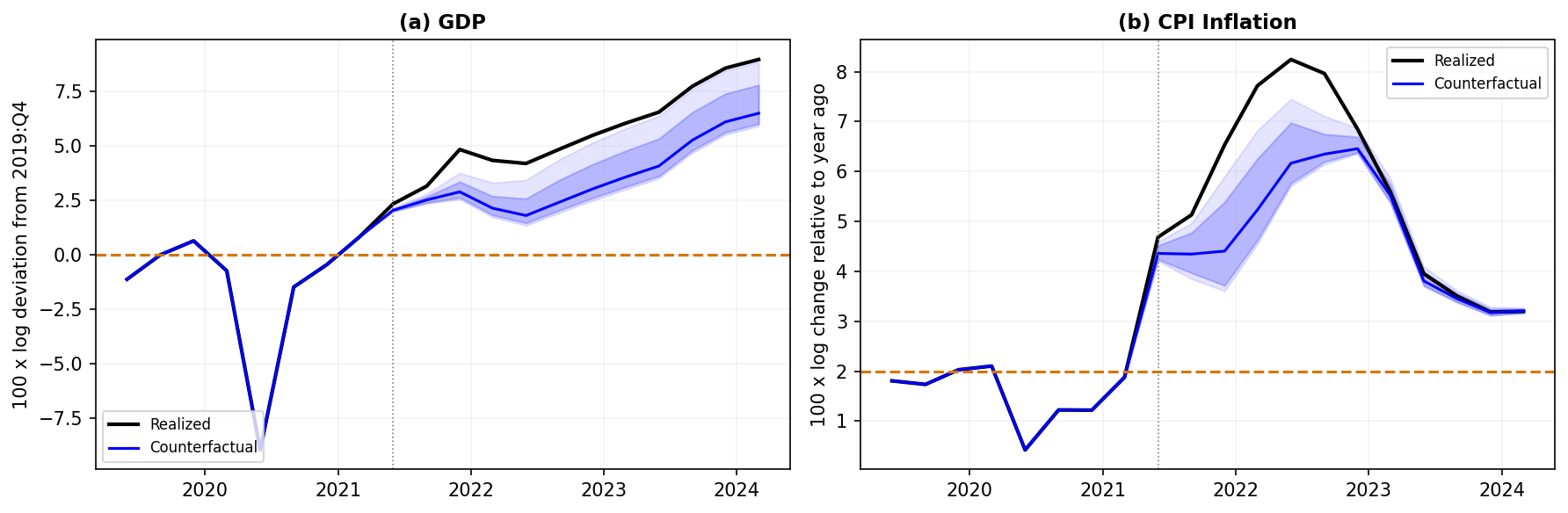}
\caption{Counterfactual under demand neutralization. Bivariate SVAR (GDP, CPI) estimated on 1997:Q1--2019:Q4 with sign restrictions. The counterfactual (blue) sets demand shocks to zero from 2021:Q2 to 2021:Q4. Shaded areas: 10--90\% and 25--75\% credible sets across 200 rotations.}
\label{fig:counterfactual}
\end{figure}

The cost of this delayed recognition was substantial. \citet{GiannonePrimiceri2024} show that demand forces were the primary driver of the post-COVID inflation surge. Figure~\ref{fig:counterfactual} illustrates this using a bivariate SVAR estimated on pre-pandemic data: setting demand shocks to zero from 2021:Q2 to 2021:Q4 reduces peak CPI inflation by approximately 2.5 percentage points. This counterfactual neutralizes only the demand component and does not account for the additional moderating effect that better-calibrated expectations would have had on the pass-through of supply shocks to prices and wages. \citet{GiannonePrimiceri2024} estimate that the Fed's accommodative stance contributed roughly 3 percentage points to peak inflation; \citet{CominJohnsonJones2023} and \citet{GagliardoneGertler2024} report comparable estimates. The American Rescue Plan of March 2021 was calibrated under the assumption that inflation would remain low. Younger households remained anchored to Great Moderation dynamics throughout 2021, while experienced cohorts adjusted earlier \citep{WeberEtAl2024}. At the firm level, \citet{YotzovBloomBunnMizenThwaites2024} estimate a 30\% pass-through from CPI changes to firm own-price expectations, a channel that is active only when inflation is elevated and media attention is high. Earlier recognition of the regime change would have moderated demand through monetary tightening, fiscal restraint, and faster expectation adjustment, while also reducing the pass-through of supply shocks to firm pricing decisions.

\begin{figure}[!h]
\centering
\includegraphics[width=\textwidth]{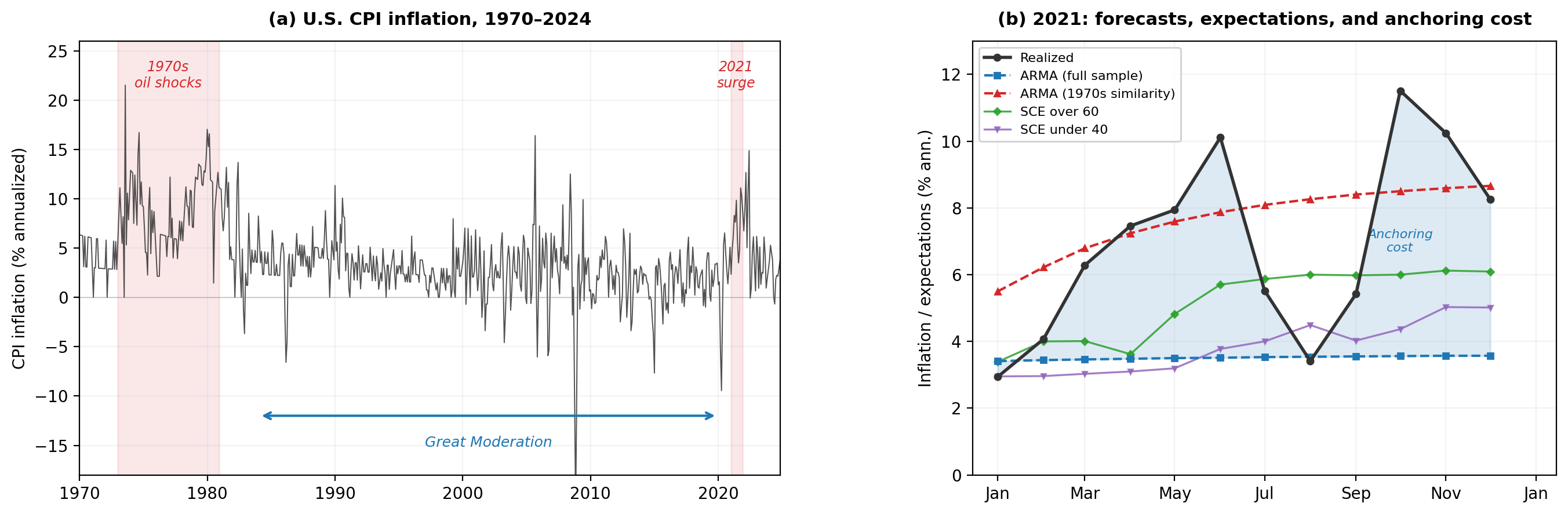}
\caption{Panel~(a): Annualized monthly CPI inflation, 1970--2024. The shaded areas mark the 1970s oil-shock episode (1973--1980) and the 2021 surge. Panel~(b): 2021 in detail. Black circles: realized inflation. Blue squares: ARMA(1,1) full-sample forecast. Red triangles: ARMA(1,1) estimated on 1973--1980 (similarity approach). Green diamonds: median one-year-ahead inflation expectation, SCE respondents over 60. Purple triangles: SCE respondents under 40. The blue shaded area represents the anchoring cost, the gap between the full-sample forecast and the realized path.}
\label{fig:intro_motivation}
\end{figure}

Figure~\ref{fig:intro_motivation} previews the main result. Panel~(a) shows U.S.\ CPI inflation from 1960 to 2024: the 1970s oil-shock episode and the 2021 surge, separated by nearly four decades of stable and predictable inflation during the Great Moderation. Panel~(b) zooms in on 2021. The ARMA(1,1) estimated on the full sample produces a flat forecast around 3.5 percent, missing the surge entirely; the shaded area between this forecast and the realized path is the cost of anchoring to Great Moderation dynamics. The same model re-estimated on 1970s data tracks the realized inflation far more closely. Household expectations mirror this pattern: SCE respondents over 60, whose lifetime includes the 1970s, reported expectations near 6 percent by mid-year, while respondents under 40 remained below 5 percent throughout.

I show that, for the class of models considered, forecast errors are primarily driven by sample composition rather than functional form misspecification. Estimation samples dominated by the Great Moderation underweight rare but economically relevant regimes, so that full-sample estimators converge to normal-regime parameters. The same mechanism operates on the expectations side: agents whose entire experience falls within the Great Moderation form priors that are well anchored to a low-inflation regime. Such anchoring is ordinarily a sign of credible monetary policy, but it becomes a source of forecast bias when the regime changes, because the experiential prior actively resists the inflationary signal rather than merely lacking information about it. Historically informed adjustments, based on past episodes with similar macroeconomic conditions, can generate substantial improvements in forecast accuracy when such regimes are identifiable in real time. These results identify the source of the failure and show that a simple, real-time implementable correction was sufficient to close most of the forecast gap.

The analysis proceeds in three parts. The first documents the forecasting failure across econometric models, machine learning methods, and institutional projections, and formalizes the source of the bias. Age-cohort data from the NY Fed Survey of Consumer Expectations show that respondents over 60, whose lifetime includes the 1970s, reported inflation expectations near 6 percent by mid-2021, while respondents under 40 remained below 5 percent throughout the year. FOMC statement sentiment, Wall Street Journal coverage, and Google Trends data all confirm that inflation attention remained low through the first months of 2021, consistent with anchored priors resisting the early inflationary signal.

The second shows that two simple adjustments to an ARMA benchmark, an intercept correction borrowing 1978--1979 forecast errors and a similarity re-estimation on 1973--1980 data, substantially close the forecast gap. The intercept correction yields 6.83 percent versus the realized 6.93. A kernel-weighted variant using oil price growth as a state variable confirms this result through a data-driven criterion. The 1970s analogy rests on real-time observable signals: by December 2020, supply-chain pressure had reached unprecedented levels, commodity prices were rising, and monetary policy was accommodative. These findings extend beyond headline CPI: the same framework applied to eight additional U.S.\ price indices confirms that the historically informed methods dominate unadjusted benchmarks, including for volatile producer price series where conventional methods miss by an order of magnitude.

The third part investigates the experience channel through a controlled experiment with two LLMs (Claude and ChatGPT), each conditioned on three professional economist personas (experienced, young, and neutral) at four data vintages. The object of interest is the persona gap $\Delta_t$, the difference in forecasts between the experienced and young personas. Under a common training leakage assumption, $\Delta_t$ is identified from the LLM outputs regardless of the training corpus. The experience-based learning model of \citet{MalmendierNagel2016} yields two testable implications: $\Delta_t > 0$ at early vintages and a narrowing of $\Delta_t$ as data accumulate. Both patterns are observed. The similarity and kernel-weighted methods and the MN experienced persona produce forecasts of the same functional form, iteration of a perceived AR(1), and differ only in how they weight historical observations, providing a formal link between the statistical and experiential approaches.

The paper makes three contributions. First, it identifies sample composition as the primary source of the 2021 forecast failure and shows that the bias can be corrected using economically motivated historical analogies. Second, it documents that the same experience channel operates across statistical methods, household expectations, and large language models: the source of the prior mattered more than the sophistication of the model. Third, it offers an interpretation of the forecast failure as the cost of expectations anchoring: full-sample estimation and experience-based priors formed entirely within the Great Moderation both assign negligible weight to high-inflation dynamics, producing the same downward bias when a supply-shock regime materializes.

\paragraph{Related literature.}
A large body of work has sought to decompose the 2021--2024 inflation surge into supply and demand components.\footnote{See \citet{diGiovanni2022} and \citet{Shapiro2022} on the supply side, and \citet{BenignoEggertsson2023} on the demand side.} An emerging consensus is that supply disruptions initiated the episode while demand pressures, amplified by fiscal stimulus and accommodative monetary policy, sustained it \citep{BlanchardBernanke2023, BallLeighMishra2022, GiannonePrimiceri2024}. The present paper does not attempt such a decomposition; it takes the mixed supply-demand nature of the episode as given and asks how a forecaster could have exploited historical analogies to improve real-time predictions.

A large literature studies forecasting under structural instability. Intercept corrections \citep{ClementsHendry2006}, window selection and observation weighting \citep{PesaranTimmermann2007, PTP2013, DendramisKapetaniosMarcellino2020}, and kernel-weighted estimation \citep{LPU2022} all address the problem of parameter shifts in the estimation sample. The present paper uses these tools for a different purpose: not to adapt to a break that has already occurred in the sample, but to borrow information from a past regime that resembles the one anticipated at the forecast origin. The distinction matters because the 2021 regime change occurs in the forecast period, not in the estimation window, so that standard break-detection methods are uninformative. In a related direction, \citet{GouletCouloumeGobelKlieber2024} develop tools to decompose ML forecasts into contributions from individual training observations. Their approach identifies which historical data points matter most for a given prediction; the present paper imposes this selection directly by targeting a specific historical episode.

The paper also relates to the growing literature on expectations and forecasting during the post-COVID episode. \citet{Reis2026} shows that a Phillips curve augmented with household expectations predicts the 2021--2024 inflation path well, and \citet{BriandMarcellinoStevanovic2025} show that media-based attention measures improve inflation forecasts during high-inflation episodes. Both exercises use a \emph{recursive} design in which the information set is updated each period. The present paper adopts a \emph{fixed-origin, multi-step} design closer to \citet{ForoniMarcellinoStevanovic2022}: the information set is frozen at December 2020 and forecasts cover horizons 1 through 12 without updating, a harder exercise, but arguably more relevant for real-time decision-making when the key question is the full trajectory over the coming year. On the expectations side, \citet{HajdiniKurmann2026} show that regime shifts under rational expectations generate predictable forecast errors whose sign depends on the persistence of the realized regime relative to agents' expectations. Their model accounts for the 1970s forecast errors through time-varying monetary policy credibility, but rejects for the post-2020 episode. The present paper takes a different approach: rather than testing whether forecast errors are consistent with rational expectations, it asks whether the forecasting failure can be traced to sample composition and whether historically informed estimation can close the gap.

Finally, the LLM experiment contributes to a nascent literature on LLM-based economic forecasting \citep{fariae2024ai, carriero2024macro, zarifhonarvar2025inflation, LundgaardHansenetal2025}. The persona-based design addresses the look-ahead bias concern raised by \citet{fariae2024ai} and \citet{LundgaardHansenetal2025} through a common training leakage assumption: because the training corpus, data, and instructions are identical across personas, any contamination from post-2021 outcomes shifts all persona outputs by the same amount and cancels in the persona gap. This complements the no-leakage and validation-based strategies formalized in \citet{LudwigMullainathanRambachan2025}.

The rest of the paper is organized as follows. Section~\ref{sec:failure} documents the forecasting failure across surveys, institutional projections, and household expectations. Section~\ref{sec:forecast_design} develops the historically informed approaches and presents the main empirical results. Section~\ref{sec:llm} reports the LLM experiment and its formal connection to experience-based learning. Section~\ref{sec:conclusion} concludes.

\section{The 2021 inflation forecast failure}
\label{sec:failure}

This section documents the breadth of the forecasting failure across professional surveys, institutional projections, and household expectations. 

\subsection{Surveys and official forecasts}
\label{sec:surveys}

Table~\ref{tab:other_forecasts} documents the forecasting failure across several survey-based and institutional forecasts. Because these sources differ in target definition (CPI inflation vs.\ unit costs), updating frequency (quarterly survey rounds vs.\ daily nowcasts), and horizon, the table is not a controlled comparison; its purpose is to illustrate the breadth of the miss across the forecasting landscape. The first column reports the realized annualized monthly CPI inflation, which averaged 6.93 percent in 2021; the remaining columns show that no forecasting source came close to this figure. The Survey of Professional Forecasters averages 3.89 percent, the Blue Chip consensus 3.37 percent, the CBO 2.65 percent, and the FOMC projections 3.83\%. Even the Cleveland Fed's inflation nowcast, which is among the most timely available, updated daily using high-frequency data on gasoline and oil prices alongside monthly CPI and PCE releases, averages only 4.49 percent.\footnote{The Cleveland Fed nowcasting model combines four components: core inflation extrapolated from recent trends, food price inflation, and gasoline price inflation derived from daily Brent crude and weekly retail gasoline prices, which are then integrated into a final CPI or PCE nowcast. See \url{https://www.clevelandfed.org/indicators-and-data/inflation-nowcasting}.} The FOMC, despite its comprehensive access to macroeconomic information and internal modeling resources, projected only 3.83\% on average, roughly half the realized outcome. Firms fared no better: the Atlanta Fed's Business Inflation Expectations survey averages only 2.82 percent for the year.\footnote{The Atlanta Fed survey asks firms about expected changes to unit costs over the next 12 months rather than expected inflation, making it not directly comparable to the household surveys discussed in Section~\ref{sec:experience}.} These nowcasts incorporate progressively more information as the year unfolds; by the fourth quarter, forecasters have observed most of the 2021 inflation path, yet the annual averages still fall well short of the realized outcome. The failure across this diverse set of forecasters suggests that the challenge in 2021 was not confined to any particular model or institution, but reflected a broader difficulty in recognizing that the economy had entered a regime poorly represented in recent experience.

\begin{table}[!t]
\centering
\caption{2021 CPI Inflation: Surveys and Institutions}
\label{tab:other_forecasts}
\setlength{\tabcolsep}{3.5pt}
\begin{tabular}{l|c|cccccc}
\hline
\hline
Date & Inflation & SPF & BlueChip & Clev.\ Fed  & CBO & FOMC & Atlanta \\
\hline
2021-01 & 2.94  &  & 2.0 (0.8) & 1.51 & & & 2.2 \\
2021-02 & 4.07  & 2.48 (1.39) & 2.3 (0.9) & 1.70 & 1.9 & & 2.2 \\
2021-03 & 6.28  &  & 2.4 (0.8) & 2.47 &  & 2.4 & 2.4 \\
2021-04 & 7.46  &  & 2.5 (0.8) & 3.46 & & & 2.5 \\
2021-05 & 7.94  & 3.38 (1.54) & 2.7 (0.7) & 4.55 & & & 2.8 \\
2021-06 & 10.11 &  & 3.3 (1.5) & 4.81 &  & 3.4 & 3.0 \\
2021-07 & 5.52  &  & 3.7 (1.2) & 5.23 & 3.4 & & 2.8 \\
2021-08 & 3.41  & 5.09 (2.27) & 4.0 (1.0) & 5.38 & & & 3.0 \\
2021-09 & 5.42  &  & 4.2 (1.0) & 5.40 &  & 4.2 & 3.1 \\
2021-10 & 11.51 &  & 4.3 (0.6) & 5.76 & & & 3.1 \\
2021-11 & 10.24 & 4.62 (1.25) & 4.4 (0.5) & 6.60 & & & 3.3 \\
2021-12 & 8.26  &  & 4.6 (0.2) & 6.94 &  & 5.3 & 3.4 \\
\hline
\textbf{2021 avg} & \bf{6.93} & \bf{3.89} & \bf{3.37} & \bf{4.49} & \bf{2.65} & \bf{3.83} & \bf{2.82} \\
\hline
\hline
\end{tabular}
\begin{flushleft}
\footnotesize
\textit{Notes:} The first column reports realized annualized monthly CPI inflation, $y_t = 1200 \times (\log \text{CPI}_t - \log \text{CPI}_{t-1})$, from the latest available vintage of the CPI data (downloaded from FRED on March 12, 2026). The remaining columns report nowcasts of current-year CPI inflation, except for FOMC which reports PCE inflation projections from the \emph{Summary of Economic Projections}, and Atlanta which reports median one-year-ahead expectations of changes to unit costs from the Federal Reserve Bank of Atlanta Business Inflation Expectations survey. Numbers in parentheses denote forecast dispersion: for the SPF (Survey of Professional Forecasters, median forecast), the interquartile range; the SPF is conducted during the second month of each quarter. For Blue Chip (consensus forecast), the difference between the average of the top~10 and bottom~10 forecasters.
The Cleveland Fed series is the Cleveland Federal Reserve nowcast.
FOMC projections are from the \emph{Summary of Economic Projections} (March, June, September, December 2021).
CBO projections are from \emph{An Update to the Budget and Economic Outlook} (February 2021, July 2021).
\end{flushleft}
\end{table}

Table~\ref{tab:other_forecasts} also reveals an important temporal pattern. Most surveys and institutions began revising their projections upward only during the second half of 2021, well after the inflation surge was already under way, consistent with gradual learning as new data arrived. At the same time, forecaster disagreement rose sharply: the SPF interquartile range widened from 1.39 in 2021Q1 to 2.27 in 2021Q3, its highest level in the sample, before narrowing to 1.25 in 2021Q4 as the consensus consolidated around elevated inflation. The Blue Chip dispersion follows a similar pattern, peaking at 1.5 in June 2021 when the first large CPI surprises arrived, then compressing steadily to 0.2 by December. The simultaneous rise in forecast levels and forecast disagreement is characteristic of signal extraction in a noisier environment: as the inflationary signal strengthened, forecasters updated at different speeds depending on how much weight they placed on the new information relative to their prior beliefs.

\begin{figure}[!t]
\centering
\includegraphics[width=0.95\textwidth]{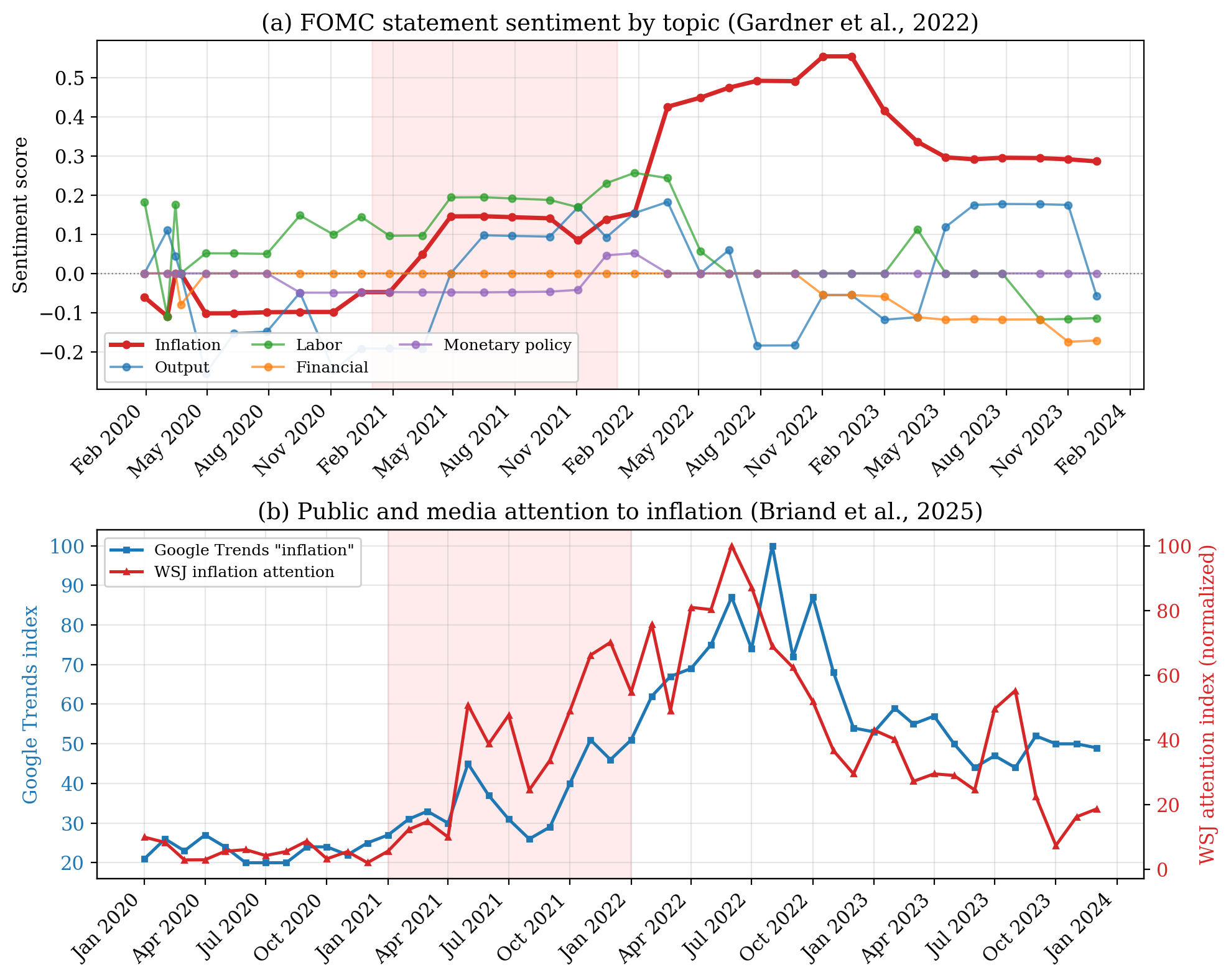}
\caption{Inflation attention and central bank sentiment, 2020--2023. Panel~(a) reports the FOMC statement sentiment index of \citet{Gardner2021}, decomposed by topic (inflation, output, labor, financial conditions, monetary policy). Panel~(b) shows the Google Trends index for ``inflation'' (left axis) and the WSJ inflation attention measure of \citet{BriandMarcellinoStevanovic2025} (right axis). The shaded area marks the year 2021.}
\label{fig:attention}
\end{figure}

Figure~\ref{fig:attention} provides corroborating evidence. Panel~(a) shows that the FOMC statement sentiment index of \citet{Gardner2021} remained dominated by labor market concerns through the first months of 2021. Inflation sentiment begins to rise around May but does not dominate until early 2022. Panel~(b) tells a consistent story: the Google Trends index for ``inflation'' and the Wall Street Journal attention measure of \citet{BriandMarcellinoStevanovic2025} both remained flat through the first four months of 2021, then rose sharply after the May CPI release. In both cases, inflation attention lagged the inflationary signal by several months.

The forecasting failure therefore did not stem from a lack of information: the relevant signals were available in real time. What was missing was the capacity to recognize that the economy had entered a different regime before it materialized in the data. If the information existed but was not exploited, the natural question becomes: who \emph{did} recognize the regime change?

\subsection{Household expectations and the experience channel}
\label{sec:experience}

\begin{figure}[!t]
\centering
\includegraphics[width=0.85\textwidth]{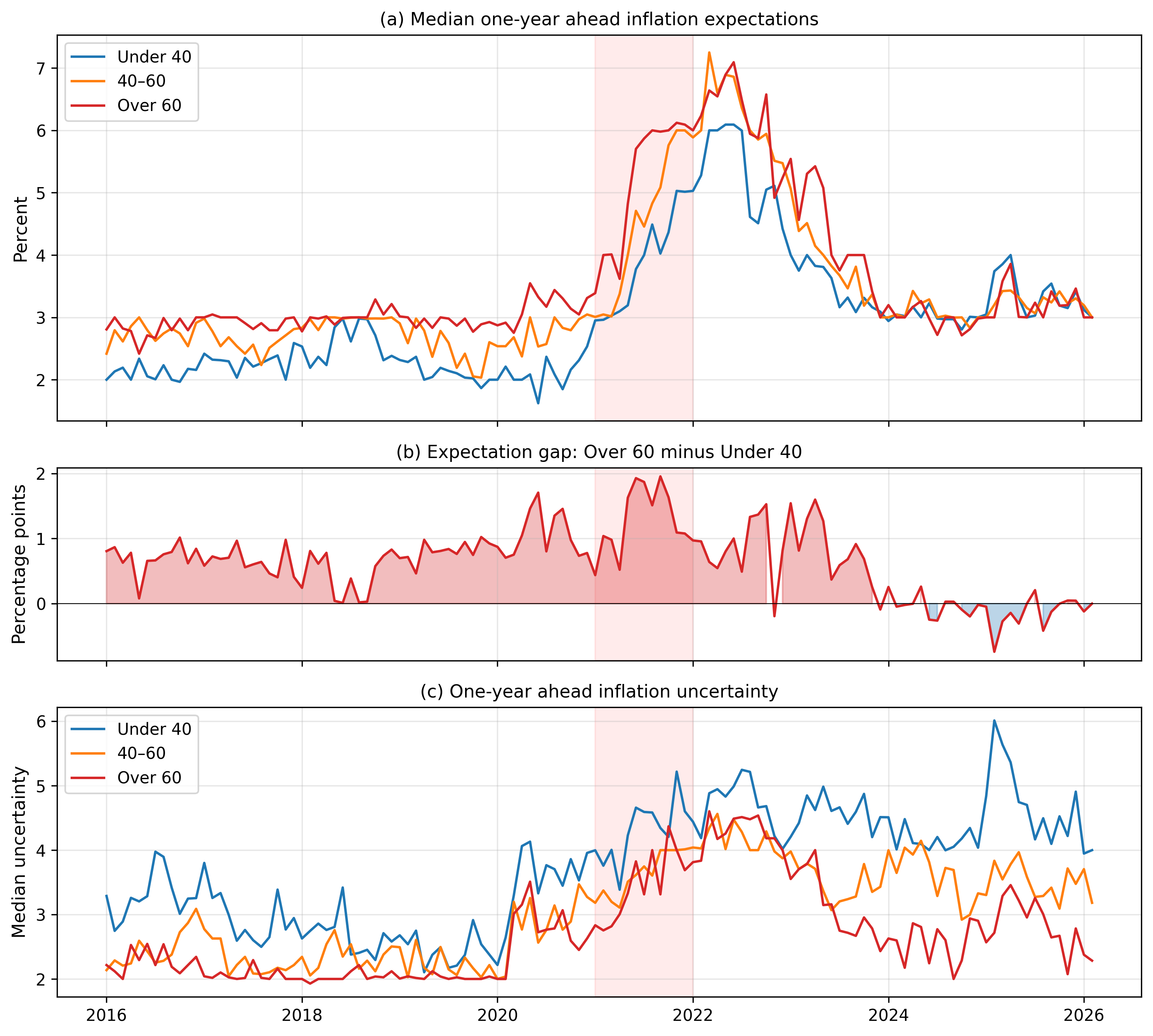}
\caption{Household inflation expectations and the experience channel (2016--2026), from the NY Fed Survey of Consumer Expectations. Panel~(a): Median one-year ahead inflation expectations by age group. Panel~(b): Expectation gap between respondents over 60 and under 40. Panel~(c): Median one-year ahead inflation uncertainty by age group. The shaded area marks 2021. Source: Federal Reserve Bank of New York.}
\label{fig:sce_age_gap}
\end{figure}

Figure~\ref{fig:sce_age_gap} decomposes one-year-ahead inflation expectations from the NY Fed's Survey of Consumer Expectations by age cohort. Panels~(a) and~(b) show that respondents over 60, the cohort whose lifetime experience includes the high-inflation years of the 1970s and early 1980s, consistently reported higher expectations than respondents under 40. The expectation gap, which had fluctuated around 0.5 percentage points before 2020, widened sharply in early 2021 as the over-60 cohort revised upward faster than younger respondents, peaking at nearly 2 percentage points by mid-year. The pre-existing level difference is itself a prediction of the experience-based learning model: agents whose estimation sample includes the 1970s carry a permanently higher perceived mean, which positions them closer to the realized outcome when a supply shock materializes. In February 2021, before any large CPI surprise had materialized, the over-60 median was already 4.0 percent compared to 3.0 for the under-40 group. The gap then gradually narrows through 2022--2023 as all age groups converge toward elevated expectations, and by 2024 it has essentially vanished, consistent with the experience-based learning framework of \citet{MalmendierNagel2016}: having now lived through their own supply-driven inflation episode, younger cohorts acquired the experiential prior that older respondents already possessed from the 1970s.

The experience channel has a natural counterpart: expectations anchoring. The under-40 cohort's low expectations in early 2021 did not reflect a mere absence of information about high-inflation regimes. Rather, these expectations were the rational output of an experiential prior built entirely on the Great Moderation, a period in which inflation targeting was credible and deviations from the target were small and transitory. In the \citet{MalmendierNagel2016} framework, this cohort's lifetime-weighted estimate of the inflation process concentrates on low-volatility, low-mean parameters. The resulting anchoring is well calibrated in normal times, but fragile to regime change: when a supply shock materializes, the prior assigns high probability to a transitory deviation and low probability to a persistent shift, producing forecasts consistent with the ``transitory inflation'' view that prevailed among professional forecasters and policymakers through mid-2021. \citet{Reis2022anchor} documents that long-run inflation expectations remained anchored throughout the episode, a finding consistent with the present argument. If this interpretation is correct, then the same anchoring that preserved long-run credibility may have contributed to the delayed recognition of the regime change at shorter horizons. \citet{HajdiniKurmann2026} formalize a closely related mechanism: under rational expectations with Markov-switching regimes, agents who assign high persistence to a low-inflation regime systematically under-predict inflation when a high-inflation regime materializes. Their framework shows that this under-prediction is not a departure from rationality but a direct consequence of the perceived regime transition probabilities. The experience-based learning model adds a layer to this result: the transition probabilities themselves are shaped by lifetime experience, so that agents whose sample contains only the Great Moderation perceive the low-inflation regime as near-absorbing.

Panel~(c) of Figure~\ref{fig:sce_age_gap} reports inflation uncertainty by age group. This measure is constructed from each respondent's individual density forecast: the SCE asks respondents to assign probabilities to predefined intervals of future inflation, and a parametric density is fitted to these probabilities. The reported uncertainty is the spread of this individual density, capturing how confident each respondent is in their own forecast, rather than the cross-sectional disagreement among respondents.\footnote{See \citet{Armantier2017} for a detailed description of the SCE methodology and the construction of individual density forecasts.} Throughout the sample, respondents over 60 report lower uncertainty than younger cohorts, typically 1 to 2 percentage points below the under-40 group. This pattern persists during the 2021 surge: the over-60 cohort maintains the lowest uncertainty even as it reports the highest point expectations. The combination of higher expectations and tighter individual density forecasts is consistent with more precise signal extraction by cohorts whose experiential prior encompasses a high-inflation regime.

This pattern is consistent with the experience-based learning framework of \citet{MalmendierNagel2016} and \citet{PedemonteTomaVerdugo2025}: agents who have lived through high inflation overweight that experience in forming expectations. The experience channel documented here for U.S.\ age cohorts has a cross-country counterpart. \citet{WeberEtAl2024} show through randomized control trials that households in high-inflation countries are already attentive to inflation and do not respond to information treatments, while those in low-inflation countries do, confirming that experience shapes attention. The genesis of the present paper illustrates the point. 

Having experienced the Yugoslav hyperinflation firsthand, and having just applied historical analogies to forecast the COVID recession in \citet{ForoniMarcellinoStevanovic2022}, I used the same approach in May 2021 with the 1970s as the reference episode, producing forecasts far above any available at the time. Whether this reflects a well-calibrated prior or coincidence is what the rest of the paper investigates. \citet{SalleGorodnichenkoCoibion2025} provide direct evidence on the causal channel: individuals who recall having lived through past inflationary episodes report higher inflation expectations and lower uncertainty. Even a simulated experience of historical inflation dynamics in a laboratory setting creates a ``pseudo-lifetime experience'' that durably shifts participants' beliefs, confirming that the link between memory and expectations is causal rather than merely correlational.

Together, these findings suggest that if lived experience improves household inflation expectations, a forecaster should be able to exploit the same logic systematically, borrowing information from the most relevant historical precedent rather than treating all past observations equally.


\section{Historically informed forecasting}
\label{sec:forecast_design}

This section develops the historically informed forecasting approaches and presents the main empirical results. The baseline predictive model is an ARMA(1,1) for annualized monthly CPI inflation:
\begin{equation}\label{eq:arma}
    y_t = \alpha + \rho y_{t-1} + \varepsilon_t + \theta \varepsilon_{t-1},
\end{equation}
where
$y_t = 1200(\log(CPI_t)-\log(CPI_{t-1}))$.

The choice of an ARMA(1,1) specification is supported by several strands of the literature. \citet{Lutkepohl1987} shows that marginalization of a finite-order VAR generally yields an ARMA process for each individual series, and \citet{DufourStevanovic2013} extend this result to factor models, showing that the implied univariate process for any observable, including inflation, is an ARMA rather than a pure AR. Empirically, \citet{NgPerron2001} document the presence of a large moving-average root in U.S. inflation, while \citet{StockWatson2007} show that the MA component has grown in importance since 1984. Finally, \citet{ForoniMarcelliniStevanovic2019} confirm that including an MA term significantly improves inflation forecasts in a mixed-frequency setting. In a comprehensive out-of-sample forecasting exercise covering a large number of models and a very long evaluation period, \citet{KotchoniLerouxStevanovic2019} find that the ARMA(1,1) is among the best-performing specifications for U.S.\ inflation forecasting.

The key premise is that the ARMA benchmark can be improved by drawing on \emph{historical analogues}: past episodes whose macroeconomic characteristics resemble those of the current environment. Rather than treating the entire postwar sample symmetrically, the historically informed approaches developed below selectively borrow information from periods of supply-driven inflation, most notably the 1970s oil-shock era, to adjust the baseline forecasts. 

\subsection{Forecast bias under rare regimes}
\label{sec:framework}

To see why full-sample estimation produces biased forecasts when a rare regime occurs, consider the ARMA(1,1) within each regime $S \in \{N, C\}$:
\begin{equation}\label{eq:arma_regime}
    \pi_t = \alpha_S + \rho_S\, \pi_{t-1} + \varepsilon_t + \theta_S\, \varepsilon_{t-1},
\end{equation}
where regime~$N$ denotes the stable, low-inflation dynamics of the Great Moderation and regime~$C$ denotes a supply-disruption episode with $\alpha_C > \alpha_N$. When the sample contains $T_N \gg T_C$ observations, the full-sample MLE is dominated by regime-$N$ parameters:
\begin{equation}\label{eq:plim}
    \hat{\alpha}^{FS} \xrightarrow{p} \frac{T_N}{T}\, \alpha_N + \frac{T_C}{T}\, \alpha_C \approx \alpha_N \quad \text{when } T_C / T \to 0,
\end{equation}
and the $h$-step-ahead forecast at a regime-$C$ origin is biased downward:
\begin{equation}\label{eq:forecast_bias}
    \text{Bias}_h = E[\pi_{T+h}|S_{T+1:T+h} = C] - \widehat{\pi}_{T+h|T}^{FS} > 0, \quad \forall\, h \geq 1.
\end{equation}
The same structure applies to expectations formation. An agent whose lifetime falls entirely within regime~$N$ computes a lifetime-weighted mean that converges to $\alpha_N$ by the same averaging logic as~\eqref{eq:plim}. The analogy suggests that anchored expectations and full-sample estimation share a common structure: both assign negligible weight to regime-$C$ parameters when regime-$C$ observations are rare in the relevant sample, whether that sample is a statistical estimation window or a lifetime of experience.

An important feature of the 2021 problem is that the regime change occurs \emph{in the forecast period}, not in the estimation sample: as of December~2020, the most recent observations still belong to regime~$N$. Standard break tests are uninformative: they detect breaks within the observed sample, not breaks that have not yet produced data. Adaptive methods face the same limitation: Markov-switching models assign negligible probability to regime~$C$ when recent data are from regime~$N$, and TVP models that update through the anti-inflationary 2020 observations are pulled in the wrong direction. More generally, any model that weights historical observations uniformly inherits this bias. For the model classes examined in this paper, the issue is not functional-form misspecification but \emph{sample composition}.

The historically informed approaches address this bias by reweighting the sample toward regime-$C$ observations. The intercept correction shifts the forecast by the bias observed during a reference episode; the similarity approach re-estimates all parameters on a regime-$C$ subsample; the kernel-weighted estimator uses a continuous weighting scheme that nests the full-sample estimator and the similarity approach as limiting cases. All three require the forecaster to identify, on economic grounds, that the current environment resembles a past supply-disruption regime.

\subsection{Why the 1970s?}
\label{sec:motivation}

\begin{figure}[!t]
\centering
\includegraphics[width=0.85\textwidth]{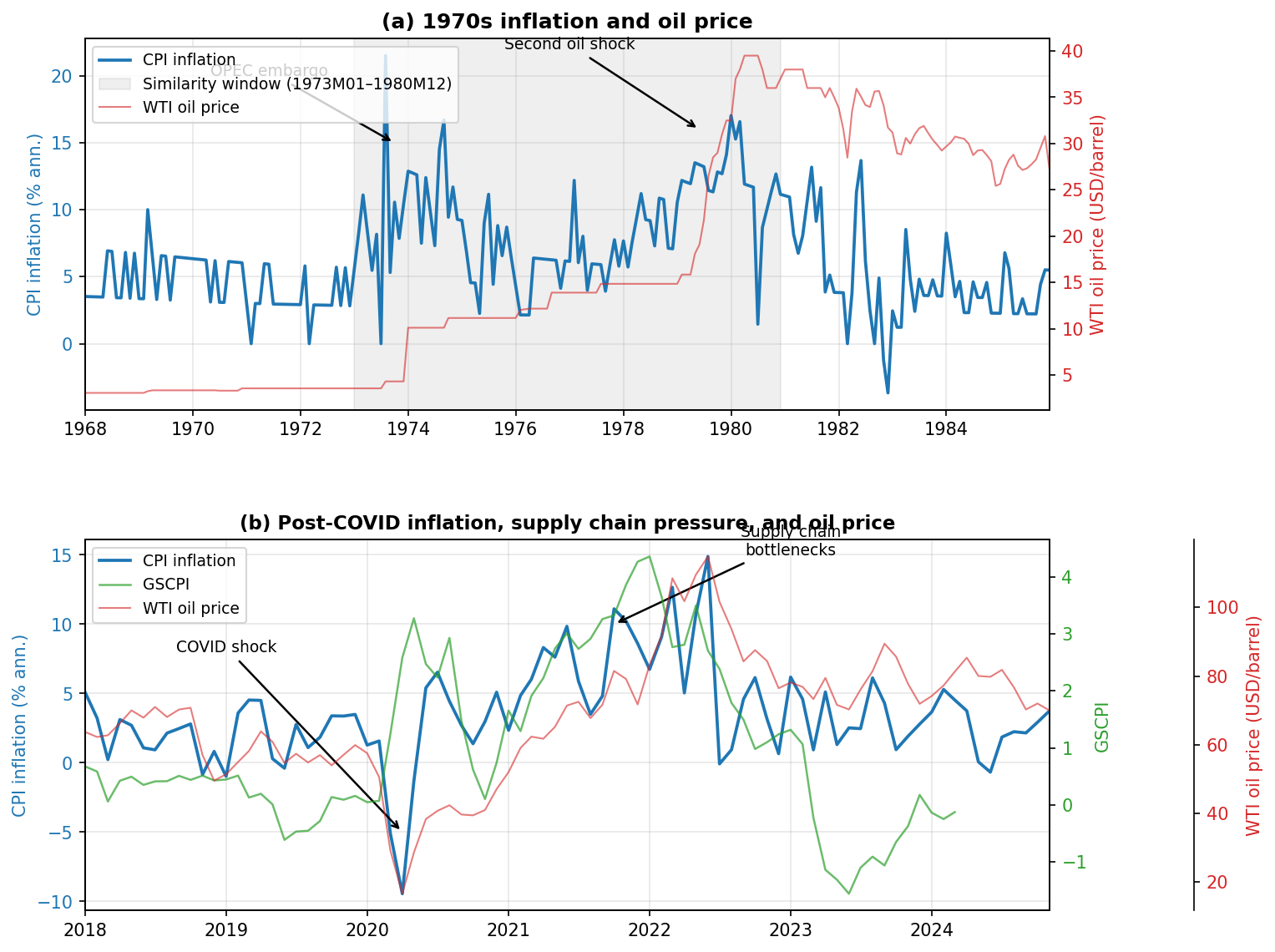}
\caption{Annualized monthly CPI inflation $y_t = 1200 \times (\log \text{CPI}_t - \log \text{CPI}_{t-1})$ (blue, left axis) and supply-shock indicators during the 1970s inflationary episode (panel~a) and the post-COVID period (panel~b). In both panels, the WTI oil price level (USD per barrel) is shown in red (right axis). Panel~(b) additionally displays the Global Supply Chain Pressure Index (GSCPI) of the Federal Reserve Bank of New York in green (right axis), expressed in standard deviations from its historical mean. The shaded area indicates the similarity window 1973M01--1980M12.}
\label{fig:historical_comparison}
\end{figure}

Section~\ref{sec:framework} established that full-sample estimation produces biased forecasts when a rare regime occurs, and that the bias can be corrected by reweighting the sample toward regime-$C$ observations. The practical question is which historical episode to use as the reference. The relevant comparison for the post-COVID inflation surge is not the low-and-stable inflation era of the Great Moderation, but rather the earlier period shaped by the first oil shock. Figure~\ref{fig:historical_comparison} illustrates the parallel.

The first oil crisis shares several key features with the post-pandemic environment: a major adverse supply disruption (the OPEC embargo in 1973, global supply-chain bottlenecks in 2020--2021), rapidly rising commodity and energy prices, accommodative monetary policy at the onset, and substantial uncertainty about the persistence of the inflationary pressures. Figure~\ref{fig:historical_comparison} makes this parallel visible by plotting annualized monthly CPI inflation alongside the WTI oil price level for each period. Panel~(b) additionally includes the Global Supply Chain Pressure Index (GSCPI) of the Federal Reserve Bank of New York. In both episodes, the oil price rises sharply before or concurrently with the inflation acceleration, providing a real-time signal that was available to a forecaster willing to draw the historical analogy. Among the candidate historical episodes, the 1970s stand out as the most informative reference, the only one combining supply-side origins with a prolonged inflationary dynamic.\footnote{The Korean War inflation spike of 1950--1951 was sharp but short-lived and driven primarily by a demand surge from military mobilization. The 2008 financial crisis was a demand-driven contraction in which inflation fell rather than rose. Neither episode shares the supply-disruption-plus-persistence configuration of the 1970s and the post-COVID period.}

The 1970s analogy assumes that the post-COVID inflation surge was driven by familiar macroeconomic forces operating with unusual intensity, not by fundamentally new mechanisms. \citet{StockWatson2025} provide direct support for this view, showing that the dynamics of the COVID period can be explained by a single aggregate shock transmitted through standard channels. Similarly, \citet{MoranStevanovicSurprenant2026} find that a COVID-specific factor adds no explanatory power for the 2021--2023 inflation surge beyond what standard macroeconomic shocks, notably accommodative monetary policy, already capture. If the inflationary forces at work were not new but rather a recurrence of a known configuration, then borrowing parameters from the most relevant historical precedent is a well-motivated forecasting strategy.

The post-COVID episode is not a pure replay of the 1970s. \citet{GiannonePrimiceri2024} show that the post-pandemic surge reflects both supply disruptions and a demand rebound that outpaced a still-constrained supply side. If anything, this makes the 1970s analogy \emph{conservative}: a forecasting adjustment based on the 1970s captures the supply-driven component while potentially underestimating the additional contribution of demand.

The choice of the 1970s as a reference episode connects directly to the evidence on household expectations presented in Section~\ref{sec:experience}. SCE respondents over 60, the cohort whose lifetime encompasses the oil-shock era, reported elevated inflation expectations from the outset of the 2021 episode, while younger cohorts remained anchored to the Great Moderation. The historically informed approaches can be viewed as a statistical counterpart to this behavior: they selectively draw on the most relevant historical precedent rather than treating all past observations equally.

An important question is how a forecaster would identify the relevant historical analogy in real time. In the present case, the selection was grounded in signals observable as of December 2020: the GSCPI had reached 4.0 standard deviations above its historical mean, oil prices were recovering at a rate comparable to the early stages of the 1973 OPEC embargo, and monetary and fiscal policy were both exceptionally accommodative. These conditions jointly pointed to a supply-disruption configuration for which the 1970s provided the closest, and essentially the only, precedent in the postwar U.S.\ sample. The kernel-weighted estimator partially automates this step by using the oil price growth rate as a state variable, allowing the data to determine the relevant weighting without manual window selection. The approach does require the forecaster to identify the nature of the shock, but this requirement is shared by any conditional forecasting method. The sensitivity analysis of Appendix~\ref{sec:sensitivity} shows that the gains are robust to a broad range of window choices and reference periods.

\subsection{Historically informed approaches}
\label{sec:hist_approaches}

The ARMA(1,1) is estimated by maximum likelihood on monthly data from 1960M01 to 2020M12 using the January 2021 vintage of the FRED-MD database. Conditional on the December 2020 observation, unadjusted forecasts are constructed iteratively up to 12 months ahead:
\begin{equation}\label{eq:arma_fct}
    \widehat{y}_{T+h|T} = \hat\alpha + \hat\rho\, \widehat{y}_{T+h-1|T}, \quad h = 2, \ldots, 12,
\end{equation}
with $\widehat{y}_{T+1|T} = \hat\alpha + \hat\rho\, y_T + \hat\theta\, \hat\varepsilon_T$, where $\hat\varepsilon_T$ is the last in-sample residual filtered recursively through~\eqref{eq:arma}. The moving-average term contributes to the one-step-ahead forecast but vanishes at longer horizons because future residuals are unobservable and set to zero. All iterative forecast equations below follow the same convention: the MA component enters at $h=1$ and the recursion is purely autoregressive for $h \geq 2$. 

\paragraph{Intercept correction.}
Let $e_h^{ref} = y_{\tau+h} - \widehat{y}_{\tau+h|\tau}$ denote the out-of-sample forecast error at horizon $h$ from a reference historical episode, where $\tau$ marks the forecast origin of that episode and forecasts $\widehat{y}_{\tau+h|\tau}$ are produced from an ARMA(1,1) estimated on data up to $\tau$. The intercept-corrected forecast is:
\begin{equation}\label{eq:ic}
    \widetilde{y}_{T+h|T}^{IC} = \widehat{y}_{T+h|T} + e_h^{ref}, \quad h = 1, \ldots, 12.
\end{equation}
This amounts to shifting the conditional mean of the forecast distribution by the bias observed during the reference regime, while leaving all autoregressive and moving-average parameters unchanged. The reference episode used here is 1978M02--1979M01, chosen because the inflation dynamics of that period, moderate but rising readings against a backdrop of supply pressures and energy price increases, closely resemble the configuration observed in early 2021. This follows the intercept correction framework of \citet{ClementsHendry1996}, who show that forecast failure under structural breaks is typically driven by shifts in deterministic components, which a correction term applied to the intercept can offset \citep[see also][]{DAgostinoGambettiGiannone2013}.

\paragraph{Robust intercept correction.}
The intercept correction described above relies on a single reference window (1978M02--1979M01), which involves an element of arbitrariness.  A more robust variant estimates the ARMA(1,1) once on the pre-shock sample 1960M02--1973M01 and holds its parameters fixed.  For each forecast origin $\tau$ in the window 1973M01--1980M12, the fixed-parameter model is used to produce $h$-step-ahead forecasts, and the forecast error $e_h^{\tau} = y_{\tau+h} - \widehat{y}_{\tau+h|\tau}$ is recorded.  The robust correction is the average error across all origins:
\begin{equation}\label{eq:ic_robust}
    \bar{e}_h = \frac{1}{N}\sum_{\tau} e_h^{\tau}, \quad h = 1, \ldots, 12,
\end{equation}
and the adjusted forecast is $\widetilde{y}_{T+h|T}^{IC\text{-}R} = \widehat{y}_{T+h|T} + \bar{e}_h$.  Because the ARMA parameters are estimated before the inflationary regime begins, the forecast errors $e_h^{\tau}$ measure the systematic bias of a ``normal-times'' model when applied to a supply-shock environment.  Averaging over 96 origins smooths out the idiosyncratic month-to-month variation that makes the single-window correction sensitive to the exact choice of $\tau$.

\paragraph{Similarity approach.}
The second approach re-estimates the ARMA(1,1) on a subsample more representative of an inflationary regime associated with large supply disturbances:
\begin{equation}\label{eq:sim}
    \widetilde{y}_{T+h|T}^{SIM} = \hat\alpha^{SIM} + \hat\rho^{SIM}\, \widetilde{y}_{T+h-1|T}^{SIM}, \quad h = 2, \ldots, 12,
\end{equation}
where $(\hat\alpha^{SIM}, \hat\rho^{SIM}, \hat\theta^{SIM})$ are estimated on the subsample 1973M01--1980M12 and the $h=1$ forecast includes the MA term as in~\eqref{eq:arma_fct}, with the residual $\hat\varepsilon_T$ obtained by filtering the full history $y_1, \ldots, y_T$ through the subsample parameters. Using the December 2020 observation as the initial condition, the model generates 12-month-ahead forecasts from this historically tailored parameterization. Unlike the intercept correction, this approach allows \emph{all} model parameters to differ from their full-sample estimates, effectively treating the inflationary subsample as a distinct data-generating process. This is closely related to the idea that, under structural breaks, forecast accuracy can be improved by restricting the estimation window to observations drawn from the current regime \citep{PesaranTimmermann2007}. The similarity approach is related to the $k$-nearest neighbor forecasting framework of \citet{DendramisKapetaniosMarcellino2020}, who weight observations by proximity to the current state.

\paragraph{Kernel-weighted similarity estimation.}

The intercept correction and similarity approaches both require the forecaster to select a discrete reference window. A natural generalization replaces the binary inclusion rule with a continuous weighting scheme. The kernel-weighted ARMA(1,1) parameters are obtained by maximizing the weighted log-likelihood
\begin{equation}
	\mathcal{L}^{K}(\alpha,\rho,\theta) = \sum_{t=1}^{T} w_t \, \ell_t(\alpha,\rho,\theta),
	\label{eq:kernel_ll}
\end{equation}
where $\ell_t$ denotes the Gaussian log-likelihood contribution of observation $t$ and the weight is $w_t = K\!\left((z_t - z_T)/b\right) / \sum_{s} K\!\left((z_s - z_T)/b\right)$, with $K(u) = \exp(-u^2/2)$ a Gaussian kernel and $b > 0$ a bandwidth parameter. The state variable $z_t$ is the annualized monthly growth rate of the WTI crude oil price. Observations from periods with oil price dynamics similar to December 2020 receive high weight, including the 1970s oil shocks, without requiring manual window selection.\footnote{The bandwidth is selected by weighted cross-validation, with a minimum effective sample size of 30 \citep{PTP2013, LPU2022}. The growth rate is preferred to the log-level because the oil price at end-2020 was moderate in absolute terms, whereas the rate of increase was comparable to the 1973--74 OPEC embargo.}

The kernel-weighted forecasts are constructed iteratively from the December 2020 initial condition, following the same convention as~\eqref{eq:arma_fct}:
\begin{equation}
	\widetilde{y}^{K}_{T+h|T} = \hat{\alpha}^{K} + \hat{\rho}^{K}\, \widetilde{y}^{K}_{T+h-1|T}, \quad h = 2, \ldots, 12,
	\label{eq:kernel_forecast}
\end{equation}
with the MA term included at $h=1$. Proximity is defined in the space of economic conditions rather than calendar time, so that the 1970s observations receive high weight because oil price dynamics at those dates resemble those at the forecast origin.

All three approaches follow the logic of \citet{ForoniMarcellinoStevanovic2022}: when the economy is hit by an unusual shock, forecasts can be improved by borrowing information from historically similar episodes. The regime classification is imposed by the forecaster rather than learned from observations that have not yet materialized.

\subsection{Competing forecasting models}
\label{sec:competing}

A natural objection is that adaptive or data-rich models might have captured the regime change without manual intervention. I compare the historically informed forecasts against two classes of competing models.

\subsubsection{Adaptive time series models}

\paragraph{Markov-switching AR(1).} A two-regime Markov-switching autoregressive model \citep{Hamilton1989}:
\begin{equation}\label{eq:msar}
    y_t = \alpha_{s_t} + \rho_{s_t} y_{t-1} + \sigma_{s_t} \varepsilon_t, \quad s_t \in \{1, 2\},
\end{equation}
where $s_t$ follows a first-order Markov chain with transition probabilities $p_{ij} = \Pr(s_t = j \mid s_{t-1} = i)$. The intercept, autoregressive coefficient, and innovation variance all switch between a low-inflation regime ($s_t = 1$) and a high-inflation regime ($s_t = 2$). Forecasts are constructed by integrating over the filtered regime probabilities at the forecast origin. The model detects regime changes from the data, but it requires observations from the new regime before updating: at the onset of the inflation surge, the filtered probability assigned to the high-inflation state was negligible.

\paragraph{TVP-AR(4).} A time-varying parameter autoregressive model estimated via Kalman filter, following the framework of \citet{Hall2026}:
\begin{equation}\label{eq:tvp}
    y_t = \boldsymbol{x}_t' \boldsymbol{\beta}_t + \varepsilon_t, \quad \boldsymbol{\beta}_t = \boldsymbol{\beta}_{t-1} + \boldsymbol{\eta}_t,
\end{equation}
where $\boldsymbol{x}_t = (1, y_{t-1}, \ldots, y_{t-4})'$ and $\boldsymbol{\eta}_t \sim \mathcal{N}(\boldsymbol{0}, Q_t)$. The $Q$-matrix governs the degree of parameter variation and is activated at the suspected break date (2020M01) with $Q = 0.01 \cdot I_5$ and set to zero elsewhere. \citet{Hall2026} show that this approach outperforms both rolling windows and recursive OLS when the forecaster has prior information about a break location, the situation confronting an inflation forecaster in 2020--2021. Unlike the intercept correction, the TVP parameters adjust through the Kalman filter as new data arrive rather than being imposed ex ante from a historical analogy.

\subsubsection{Machine learning models}
\label{sec:ml_models}

To assess whether data-rich methods could have detected the inflation regime change, I construct a real-time forecasting exercise using the FRED-MD macroeconomic database \citep{McCrackenNg2016}, which provides a balanced panel of approximately 128 monthly U.S.\ macroeconomic and financial series with real-time vintages. The predictor set includes four lags of all FRED-MD series as well as $K = 8$ principal components extracted from the stationarity-transformed panel and their four lags, together with four lags of the target variable.

Five models are estimated on this predictor set. The first three (Ridge regression, LASSO, and Elastic Net) are penalized linear regressions. The penalization addresses the high dimensionality of the predictor set induced by the inclusion of the full FRED-MD panel alongside its principal components. The remaining two models (Random Forest and a feedforward neural network) are nonlinear and can capture interactions and threshold effects that the linear models miss. \citet{GouletCoulombeLerouxStevanovic2022} show that these nonlinear methods deliver the largest forecasting gains in macroeconomic applications, and \citet{GouletCoulombeMarcellinoStevanovic2021} document that they were particularly useful at the onset of the COVID-19 recession.

Table~\ref{tab:hyperparameters} summarizes the hyperparameter grids for each model. Tunable hyperparameters are selected by five-fold cross-validation with expanding-window splits that respect the temporal ordering of the data. Forecasts are produced using a \emph{direct} approach: for each horizon $h = 1, \ldots, 12$, a separate model is estimated with the target $y_{t+h} = f(\mathbf{Z}_t) + \varepsilon_{t+h}$, where $\mathbf{Z}_t$ collects the predictors described above. 

\begin{table}[!t]
\centering
\caption{Hyperparameter grids and fixed settings by model}
\label{tab:hyperparameters}
\renewcommand{\arraystretch}{1.2}
\begin{tabular}{p{3cm} p{11cm}}
\hline
\hline
\textbf{Model} & \textbf{Hyperparameters} \\
\hline
LASSO
& $\alpha = 1$, $\lambda \in \texttt{seq}(10^{-4},10,\texttt{ length}=500)$, $\texttt{maxit}=200{,}000$ \\
\hline
Ridge
& $\alpha = 0$, $\lambda \in \texttt{seq}(10^{-4},10,\texttt{ length}=500)$, $\texttt{maxit}=200{,}000$ \\
\hline
Elastic Net
& $\alpha \in \texttt{seq}(0,1,\texttt{ by}=0.1)$, $\lambda \in \texttt{seq}(10^{-4},1,\texttt{ length}=500)$, $\texttt{maxit}=200{,}000$ \\
\hline
Random Forest
& $\texttt{min.node.size} \in \{3,5\}$, $\texttt{mtry} \in \{0.2,0.35,0.5\}$,
$\texttt{splitrule}=\texttt{variance}$, $\texttt{num.trees}=500$ \\
\hline
Neural Network
& 3 hidden layers of 400 neurons, $\texttt{dropout}=0.2$,
$\texttt{lr}=10^{-4}$, $\texttt{batch\_size}=32$, $\texttt{epochs}=500$, early stopping with $\texttt{patience}=20$ \\
\hline
\hline
\end{tabular}
\begin{flushleft}
\footnotesize
\textit{Notes:} Ridge, LASSO, and Elastic Net are estimated via \texttt{glmnet}. Random Forest uses the \texttt{ranger} package. The neural network is trained with the Adam optimizer and sampling rate 0.7.
\end{flushleft}
\end{table}

\subsection{Fixed-origin forecasts}
\label{sec:forecasts2021}

Table~\ref{tab:detailed_forecasts} reports the realized path of CPI inflation in 2021 alongside the forecasts produced by each method. Every historically informed approach substantially outperforms the full-sample ARMA benchmark (3.52 percent). The robust intercept correction yields 5.00 percent, the similarity approach 6.19, the single-origin intercept correction 6.83, and the kernel-weighted ARMA 7.82, against a realized average of 6.93. The robust IC is the most conservative because averaging over the full similarity window includes periods in which inflation fell, so that positive and negative forecast errors partially cancel. The kernel-weighted estimate exceeds the others because the selected bandwidth ($b^* = 23.18$) concentrates weight on episodes of rapidly rising oil prices (see the sensitivity analysis in Appendix~\ref{sec:sensitivity}).

The evaluation focuses on the annual average because the policy-relevant question in early 2021 was whether the year as a whole would be a high-inflation year, not the precise month-by-month path. The monthly forecasts in Table~\ref{tab:detailed_forecasts} exhibit sizeable month-to-month errors that partially offset in the annual average, a feature common to all fixed-origin, multi-step forecasting exercises. The historically informed methods improve accuracy precisely at the annual-average horizon that matters for the regime-recognition question: would a forecaster standing in December 2020 have classified 2021 as a supply-shock year rather than a continuation of the low-inflation regime?

\begin{table}[!t]
\centering
\caption{2021 CPI Inflation Forecasts}
\label{tab:detailed_forecasts}
\setlength{\tabcolsep}{3pt}
\small
\begin{tabular}{l|c|cccc|ccc|c}
\hline
\hline
 & & \multicolumn{4}{c|}{Historically informed} & \multicolumn{3}{c|}{Unadjusted / adaptive} &   \\
Date & Realized & IC & IC robust & Similarity & Kernel & ARMA & MS-AR & TVP-AR & ML avg \\
\hline
2021-01 & 2.94 & 4.84 & 4.55 & 4.40 & 4.86 & 3.43 & 3.68 & 3.31 & 4.65 \\
2021-02 & 4.07 & 6.72 & 4.66 & 4.83 & 5.52 & 3.45 & 3.37 & 2.04 & 3.03 \\
2021-03 & 6.28 & 8.55 & 4.71 & 5.22 & 6.14 & 3.47 & 3.25 & 1.35 & 2.87 \\
2021-04 & 7.46 & 6.64 & 4.78 & 5.58 & 6.72 & 3.49 & 3.21 & 1.43 & 2.91 \\
2021-05 & 7.94 & 6.61 & 4.88 & 5.90 & 7.27 & 3.51 & 3.20 & 1.62 & 2.76 \\
2021-06 & 10.11 & 4.75 & 4.98 & 6.20 & 7.77 & 3.52 & 3.21 & 1.68 & 3.44 \\
2021-07 & 5.52 & 8.35 & 5.19 & 6.47 & 8.24 & 3.54 & 3.23 & 1.61 & 3.12 \\
2021-08 & 3.41 & 8.28 & 5.13 & 6.72 & 8.69 & 3.55 & 3.25 & 1.55 & 2.83 \\
2021-09 & 5.42 & 4.66 & 5.26 & 6.95 & 9.10 & 3.56 & 3.26 & 1.54 & 2.79 \\
2021-10 & 11.51 & 4.65 & 5.25 & 7.16 & 9.48 & 3.57 & 3.28 & 1.55 & 3.04 \\
2021-11 & 10.24 & 8.14 & 5.29 & 7.34 & 9.84 & 3.57 & 3.29 & 1.56 & 2.52 \\
2021-12 & 8.26 & 9.80 & 5.28 & 7.52 & 10.18 & 3.58 & 3.31 & 1.56 & 2.07 \\
\hline
\textbf{Avg} & \bf{6.93} & \bf{6.83} & \bf{5.00} & \bf{6.19} & \bf{7.82} & \bf{3.52} & \bf{3.29} & \bf{1.73} & \bf{3.00} \\
\hline
\hline
\end{tabular}
\begin{flushleft}
\footnotesize
\textit{Notes:} Annualized monthly CPI inflation, $y_t = 1200 \times (\log \text{CPI}_t - \log \text{CPI}_{t-1})$. All forecasts condition on data available up to December 2020. IC: intercept correction using 1978M02--1979M01 forecast errors. IC robust: average forecast errors over 1973M01--1980M12 using pre-shock parameters (see text). Similarity: ARMA re-estimated on 1973M01--1980M12. Kernel: Gaussian kernel on oil price growth, $b^* = 23.18$. TVP-AR(4): Kalman filter with break date 2020M01. ML avg: average of Ridge, LASSO, Elastic Net, Random Forest, and Neural Network. The last row reports annual averages.
\end{flushleft}
\end{table}

Table~\ref{tab:arma_estimates} confirms that the subsample and kernel-weighted parameters differ materially from the full-sample estimates: the intercept nearly doubles (from 0.49 to 0.83--0.98) and persistence increases ($\hat{\rho}$ rises from 0.86 to 0.90--0.93), reflecting the inflationary dynamics of oil-shock episodes. This result confirms that the 1970s analogy, which motivates the intercept correction and similarity approaches, can be recovered by a purely data-driven criterion, provided the state variable captures the \textit{dynamics} rather than the \textit{level} of the supply shock.

\begin{table}[!htbp]
\centering
\caption{ARMA estimates}
\label{tab:arma_estimates}
\begin{tabular}{l|ccc}
\hline
\hline
 & $\alpha$ & $\rho$ & $\theta$ \\
 \hline
Full sample     &  &  & \\
1960M01-2020M12 & $0.49^{***}$ & $0.86^{***}$ & $-0.44^{***}$ \\
\hline
Intercept correction     &  &  & \\
1960M01-1978M01 & $0.11$ & $0.98^{***}$ & $-0.78^{***}$ \\
\hline
Similarity approach     &  &  & \\
1973M01-1980M12 & $0.83$ & $0.90^{***}$ & $-0.62^{***}$ \\
\hline
Kernel-weighted     &  &  & \\
$(b^* = 23.18)$ & $0.98$ & $0.93^{***}$ & $-0.53^{***}$ \\
\hline
\hline
\end{tabular}
\begin{flushleft}
\footnotesize
\textit{Notes:} The table reports ARMA(1,1) parameter estimates for CPI inflation using different sample periods or weighting schemes.
Inflation is defined as annualized monthly CPI inflation.
The kernel-weighted estimates use a Gaussian kernel on the annualized monthly oil price growth rate; the effective sample size at $b^* = 23.18$ is approximately 33.
$^{***}$ denotes statistical significance at the 1\% level.
\end{flushleft}
\end{table}

The adaptive models perform no better, and in some cases worse, than the ARMA benchmark, for the reasons anticipated in Section~\ref{sec:competing}. The MS-AR(1) produces an average of 3.29 percent: as of December 2020, the filtered probability of a high-inflation state is negligible, and the model cannot signal a regime it has not yet observed. The TVP-AR(4), with the break date set at 2020M01 and $Q = 0.01$, produces only 1.73 percent, substantially \textit{worse} than the unadjusted ARMA, because the Kalman filter absorbs the deflationary readings of 2020 and adjusts parameters in the wrong direction.\footnote{In a one-step-ahead recursive exercise where the Kalman filter is updated with each realized observation during 2021, the TVP-AR(4) with $Q = 0.01$ produces an average of 4.88 percent, an improvement over the fixed-origin forecast but still well below the historically informed approaches.}

The ML models produce an average forecast of only 3.00 percent, below even the parsimonious ARMA(1,1), despite exploiting 128 predictors from the FRED-MD panel, cross-validated hyperparameters, and both linear and nonlinear specifications \citep{GouletCoulombeLerouxStevanovic2022, GouletCoulombeMarcellinoStevanovic2021}. Combined with the failure of the Markov-switching and TVP models, of the professional surveys (SPF, Blue Chip), and of institutional projections (FOMC, CBO, Cleveland Fed), this evidence suggests that the forecasting failure of 2021 was not caused by using the wrong model, but by estimating any model on a sample in which the relevant regime is underrepresented.

The gains are not specific to the ARMA(1,1) specification: Table~\ref{tab:robustness_base} in Appendix~\ref{sec:sensitivity} shows that the intercept correction and similarity approach produce comparable improvements when applied to AR(1), AR(4), and ARMA(2,1) base models.

\subsection{Forecast updates}
\label{sec:forecast_updates}

\begin{figure}[!t]
\centering

\begin{subfigure}[t]{0.48\textwidth}
    \centering
    \includegraphics[width=\textwidth]{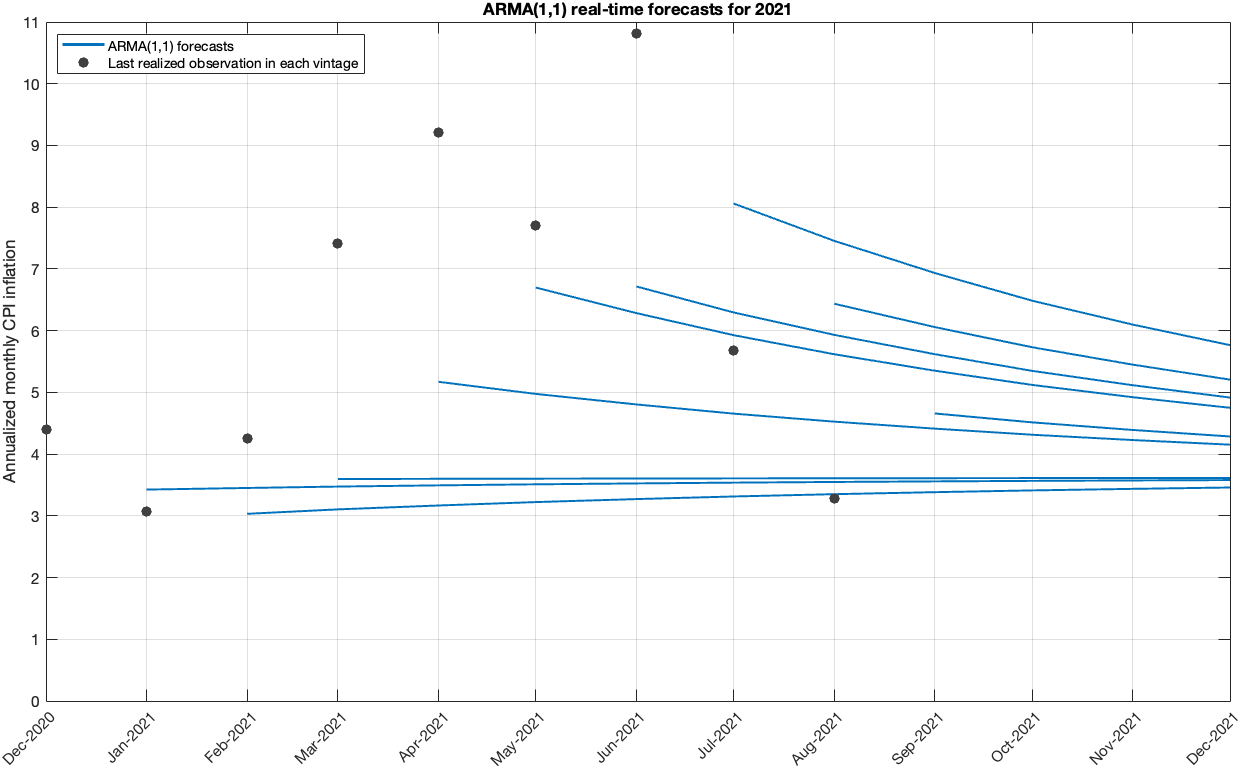}
    \caption{ARMA(1,1)}
    \label{fig:rt_arma}
\end{subfigure}
\hfill
\begin{subfigure}[t]{0.48\textwidth}
    \centering
    \includegraphics[width=\textwidth]{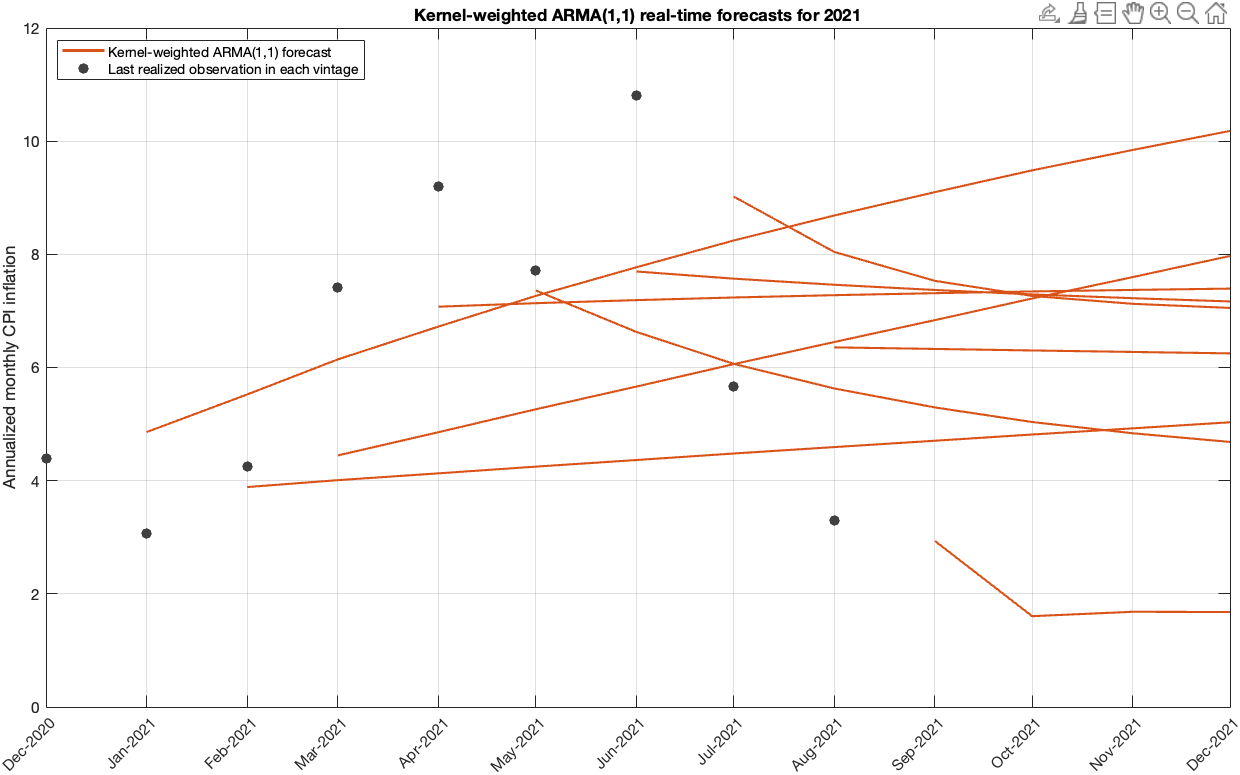}
    \caption{Kernel-weighted ARMA(1,1)}
    \label{fig:rt_kernel}
\end{subfigure}

\vspace{0.5cm}

\begin{subfigure}[t]{0.48\textwidth}
    \centering
    \includegraphics[width=\textwidth]{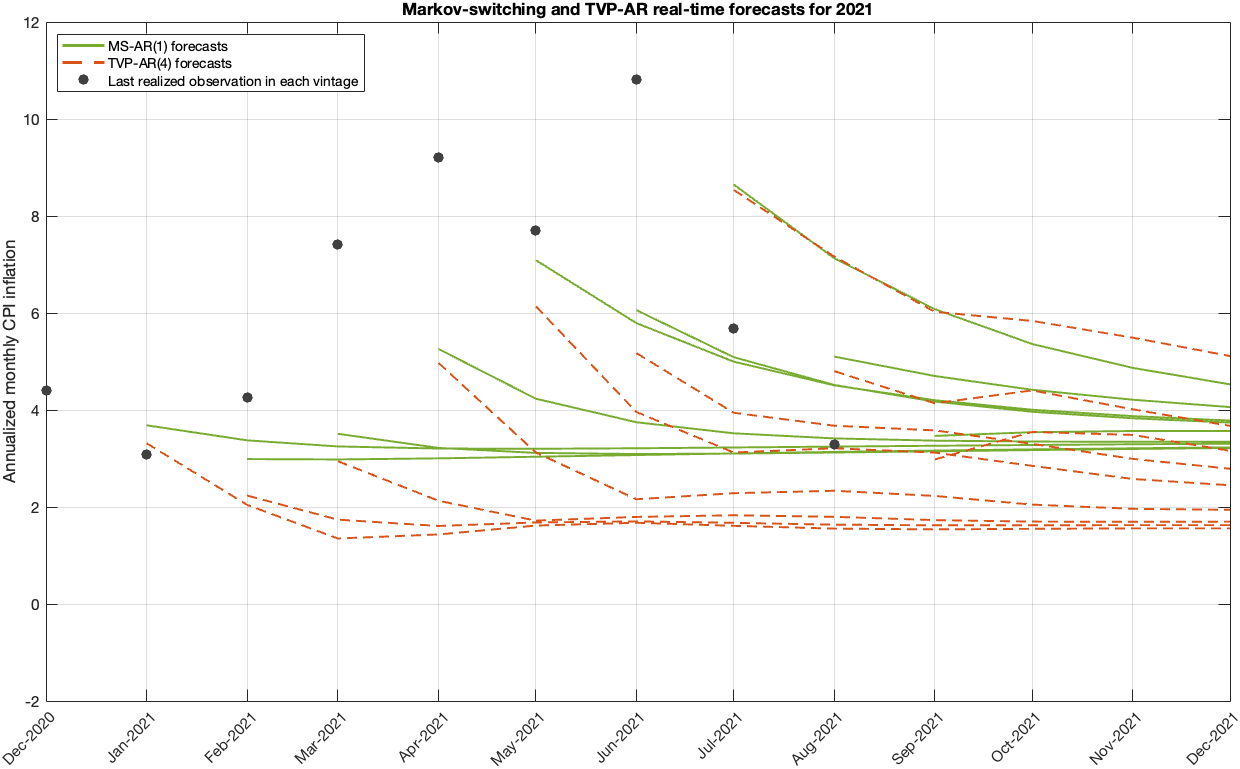}
    \caption{Markov-switching and TVP-AR}
    \label{fig:rt_ms}
\end{subfigure}
\hfill
\begin{subfigure}[t]{0.48\textwidth}
    \centering
    \includegraphics[width=\textwidth]{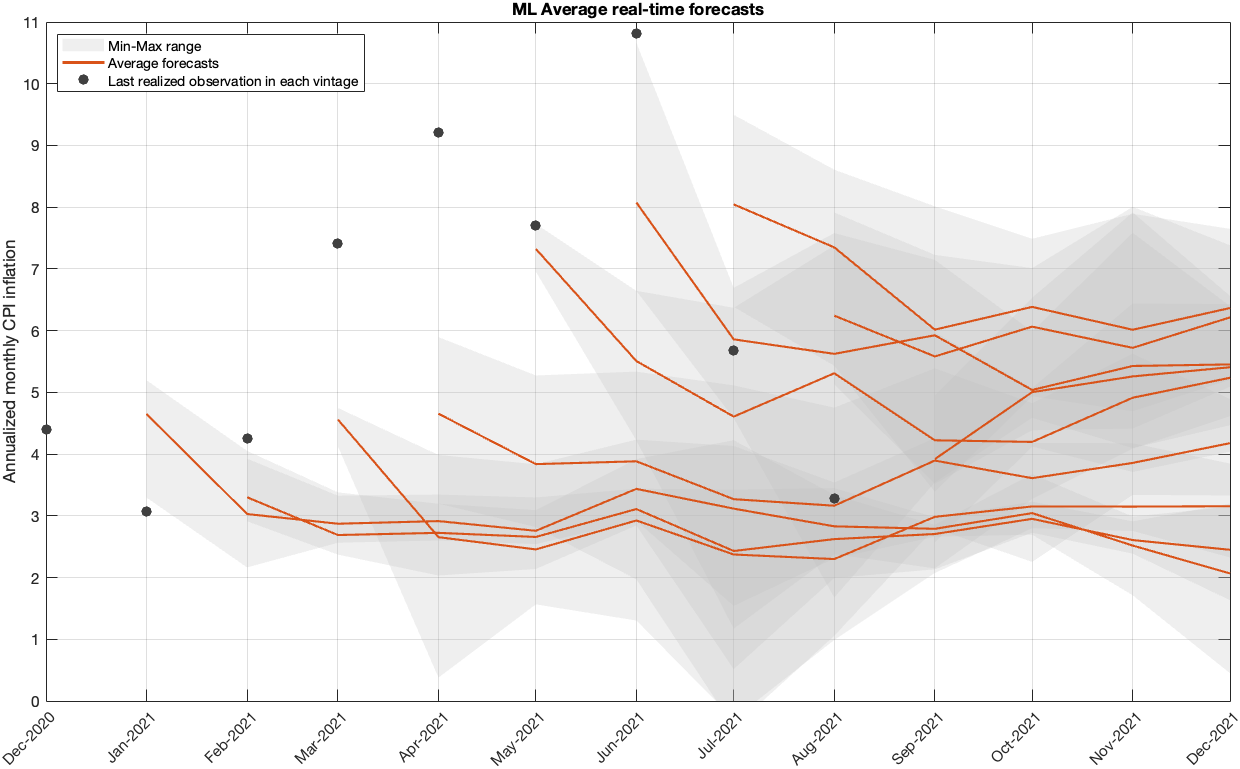}
    \caption{Machine learning average}
    \label{fig:rt_ml}
\end{subfigure}

\caption{Real-time inflation forecasts in 2021 from expanding-window exercises. For each monthly vintage from January to September 2021, the figure reports the model-implied path for the remaining months of 2021. Dots denote the last realized inflation observation available in each vintage.}
\label{fig:rt_forecasts}
\end{figure}

The fixed-origin results of the previous subsection condition on information available in December 2020. A natural question is whether data-driven methods can close the gap once the inflation surge becomes visible in the data. To investigate this, Figure~\ref{fig:rt_forecasts} reports an expanding-window exercise in which each model is re-estimated on successive real-time vintages of the FRED-MD database, from January to September 2021, and used to forecast the remaining months of the year.

The ARMA(1,1) forecasts (panel~a) update gradually but never exceed approximately 5 percent even by the September vintage. The kernel-weighted ARMA (panel~b) produces elevated forecasts from the January vintage but is unstable across subsequent vintages because the oil price growth rate fluctuates sharply from month to month. The MS-AR (panel~c) hovers around 3--4 percent across all vintages, while the TVP-AR reacts aggressively at short horizons from April onward before rapidly mean-reverting. The ML models (panel~d) are initially anchored near 3 percent, then react at short horizons from the May vintage onward as the CPI surprises enter the estimation sample, but still project a rapid return toward pre-pandemic levels.

Across all panels, data-driven methods either fail to anticipate the regime change or cannot sustain the signal beyond the short horizon. This mirrors the gradual adjustment observed in the survey and institutional forecasts documented in Section~\ref{sec:surveys}. In contrast, the historically informed approaches produce elevated forecasts from the outset, because the 1970s analogy is imposed before any 2021 data are observed.

These results are robust to the choice of reference window (Appendix~\ref{sec:sensitivity}). For the similarity approach, 34 out of 56 window combinations produce average 2021 forecasts between 5.5 and 8.5 percent, all above the ARMA benchmark. The intercept correction is more sensitive to the specific origin, but origins spanning 1977M09 to 1979M04 consistently yield forecasts between 6 and 8 percent. A historical out-of-sample exercise applying the same methodology to the 1979 inflation surge produces a forecast within 0.5 percentage points of the realized outcome.

\subsection{Robustness across price indices}
\label{sec:price_indices}

The results documented above are based on headline CPI inflation. To assess whether the approach extends to other price measures, we apply the same framework, with identical hyperparameters, estimation windows, and forecast origin, to the PCE price index (PCEPI), core CPI (CPILFESL), and six PPI series spanning the price chain from raw materials to finished consumer goods.

Table~\ref{tab:validation_summary} reports the results. Across all eight price indices, the forecast closest to the realized value is produced by either the similarity approach (five series) or the kernel-weighted ARMA (three series). The unadjusted ARMA, TVP-AR, and MS-AR never produce the best forecast. For consumer prices, the similarity approach closely tracks the realized values for both PCEPI and core CPI. For producer prices, the gains are larger: the ARMA benchmark misses by an order of magnitude for the most volatile series, while the historically informed methods remain within range. The ML models, available for five of the eight series, confirm the pattern observed for headline CPI: they produce forecasts comparable to or below the unadjusted ARMA and never approach the historically informed methods. Detailed monthly forecasts for each series are reported in Appendix~\ref{sec:additional_measures}.

\begin{table}[!t]
\centering
\caption{2021 average inflation forecasts for eight additional U.S.\ price indices}
\label{tab:validation_summary}
\setlength{\tabcolsep}{3pt}
\small
\begin{tabular}{ll|c|cccc|ccc|c}
\hline
\hline
 & & & \multicolumn{4}{c|}{Historically informed} & \multicolumn{3}{c|}{Unadjusted / adaptive} & \\
 & Series & Realized & IC & IC rob. & Simil. & Kernel & ARMA & MS-AR & TVP-AR & ML avg \\
\hline
\multicolumn{11}{l}{\textit{Consumer prices}} \\
& PCEPI          & 5.96  & 4.47  & 3.74  & \textbf{7.38}  & 3.19   & 2.72  & 11.24 & 1.56 & 2.54 \\
& CPILFESL       & 5.34  & 5.05  & 3.98  & \textbf{5.31}  & 2.21   & 2.18  & 2.20  & 1.05 & 2.63 \\
\hline
\multicolumn{11}{l}{\textit{Producer prices, finished goods}} \\
& WPSFD49207     & 12.22 & 6.97  & 15.71 & \textbf{12.06} & 4.55   & 3.42  & 4.90  & 2.01 & 3.06 \\
& WPSFD49502     & 14.36 & 8.37  & 11.05 & \textbf{14.87} & 4.96   & 3.85  & 5.17  & 2.21 & 3.37 \\
\hline
\multicolumn{11}{l}{\textit{Producer prices, intermediate \& raw}} \\
& WPSID61        & 22.00 & 9.04  & 5.80  & 13.27 & \textbf{25.40} & 5.86  & 5.87  & 7.66 & 4.36 \\
& WPSID62        & 33.45 & 18.43 & 5.40  & 20.27 & \textbf{39.57} & 4.63  & 3.97  & 8.06 & 2.99 \\
& PPIACO         & 18.54 & 8.53  & 9.21  & \textbf{13.98} & 23.39  & 4.24  & 6.69  & 3.90 & 3.93 \\
& PPICMM         & 22.01 & \textbf{24.83} & 15.37 & 32.07 & 9.89   & 8.74  & 13.16 & 14.52 & 7.30 \\
\hline
\hline
\end{tabular}
\begin{flushleft}
\footnotesize
\textit{Notes:} Each entry is the average of 12 monthly forecasts for January--December 2021, expressed as annualized percent. Bold indicates the method closest to the realized value. All models are estimated on the January 2021 real-time vintage with estimation sample 1960M02--2020M12. ML avg is the average of Ridge, LASSO, Elastic Net, Random Forest, and Neural Network (see Section~\ref{sec:ml_models}). 
\end{flushleft}
\end{table}

These results confirm that the gains from historically informed methods extend across the price chain. For the remainder of the paper, we focus on headline CPI, the most widely monitored measure and the object of the survey and institutional forecasts documented in Section~\ref{sec:surveys}.


\section{Can large language models forecast inflation?}
\label{sec:llm}

Sections~\ref{sec:experience} and~\ref{sec:hist_approaches} showed that experienced agents and historically informed methods both outperform their counterparts anchored to the Great Moderation, but the evidence is observational: the SCE age decomposition does not control for confounders, and the statistical methods do not involve expectations. This section uses large language models to test the experience channel in a controlled setting. By assigning an ``experienced'' persona (with professional memory of the 1970s) and a ``young'' persona (shaped by the Great Moderation) to each LLM, while holding the data, instructions, and macroeconomic context fixed, the exercise isolates the effect of the experiential prior. \citet{fariae2024ai} show that LLMs generate inflation forecasts that rival the SPF, and \citet{LundgaardHansenetal2025} construct synthetic forecaster personas that often outperform human experts, motivating the persona-based design adopted here.

\subsection{Econometric framework for LLM-generated forecasts}
\label{sec:llm_framework}


I adopt the notation of \citet{LudwigMullainathanRambachan2025}. Let $\Sigma$ denote a finite alphabet and $\Sigma^{*}$ the set of all finite-length strings over~$\Sigma$. A large language model is a mapping $\hat{m}(\cdot\,;\mathcal{C}):\Sigma^{*}\to\Sigma^{*}$ from input strings (prompts) to output strings (responses), parameterized by its training corpus~$\mathcal{C}$. The researcher constructs a prompt~$r \in \Sigma^{*}$ and obtains the LLM output $\hat{m}(r;\mathcal{C}) \in \Sigma^{*}$.

In the present application, each prompt~$r$ encodes three objects: a data vintage $\mathcal{D}_t$ (a subset of the FRED-MD release available at date~$t$), a set of forecasting instructions~$\mathcal{I}$ (identical across personas), and a persona $\mathcal{P}_i \in \{\mathcal{P}_E,\mathcal{P}_Y,\mathcal{P}_0\}$. The prompt is constructed as
\begin{equation}\label{eq:prompt}
r^{i}_{t} = \rho(\mathcal{D}_t,\, \mathcal{I},\, \mathcal{P}_i),
\end{equation}
where $\rho:\Sigma^{*}\times\Sigma^{*}\times\Sigma^{*}\to\Sigma^{*}$ is the prompt-construction function that concatenates the three inputs into a single string. The LLM output is a vector of monthly inflation forecasts
\begin{equation}
\hat{\boldsymbol{\pi}}^{\,i}_t = \hat{m}(r^{i}_{t};\,\mathcal{C}).
\end{equation}

Because repeated runs of the same prompt do not return identical outputs, $\hat{\boldsymbol{\pi}}^{\,i}_t$ is treated as a random vector. Define its conditional mean as
\begin{equation}
\mu^{i}_t
\equiv
\mathbb{E}\!\left[\hat{\boldsymbol{\pi}}^{\,i}_t \mid \mathcal{D}_t,\mathcal{P}_i\right].
\label{eq:mu_it}
\end{equation}
The expectation in~\eqref{eq:mu_it} conditions on the two objects that vary across the experiment, the data vintage~$\mathcal{D}_t$ and the persona~$\mathcal{P}_i$, and marginalizes over the training corpus~$\mathcal{C}$, which is fixed but unknown to the researcher. 

\subsubsection{Experimental design}
\label{sec:design}

I construct three forecaster personas for each LLM that receive identical data and instructions but differ in their assigned professional history:
\begin{itemize}
\item The \textbf{experienced persona} ($\mathcal{P}_E$) is an economist active since the early 1970s with professional memory of the OPEC embargo, the stagflation of 1974--1980, and the Volcker disinflation.

\item The \textbf{young persona} ($\mathcal{P}_Y$) is an economist whose career began in the mid-2000s, shaped entirely by the Great Moderation and the 2008 financial crisis.

\item The \textbf{neutral persona} ($\mathcal{P}_0$) receives no experiential conditioning: the prompt contains only data and instructions, with no reference to professional history.
\end{itemize}

The exercise is repeated at four data vintages (January, April, July, and October 2021), each providing the corresponding FRED-MD release.\footnote{For instance, the April 2021 vintage of FRED-MD contains CPI data through March 2021, so annualized inflation for January--March is computed directly from the data, while April--December must be forecast.} For each of the $3 \times 4 = 12$ persona--vintage combinations, five independent runs are conducted per LLM, yielding $5 \times 12 \times 2 = 120$ forecast paths. The Claude model used is Claude Opus 4 (Anthropic, 2025 vintage); the ChatGPT model is ChatGPT-4o with a Pro subscription (OpenAI, March 2026 vintage).

In terms of the notation introduced above, this design generates 120 output vectors $\hat{\boldsymbol{\pi}}^{\,i}_t = \hat{m}(r^i_t;\mathcal{C})$, one for each run. Because the data vintage~$\mathcal{D}_t$ and the instructions~$\mathcal{I}$ are held fixed across personas, the only source of cross-persona variation in the prompt~$r^i_t = \rho(\mathcal{D}_t, \mathcal{I}, \mathcal{P}_i)$ is the third argument~$\mathcal{P}_i$. For a given LLM, the sample average across the five runs estimates the corpus-conditional expectation, which includes a training leakage component addressed in Section~\ref{sec:llm_leakage}.

The personas are defined as professional economists rather than consumers. The SCE evidence in Section~\ref{sec:experience} already documents the experience channel among households; the LLM experiment targets professional forecasters, for whom no comparable age-stratified survey exists. LLMs are trained predominantly on professional and academic text, making them better suited to simulate professional reasoning. The forecasting instructions are deliberately generic to avoid triggering differential recall from the training corpus.

The design is motivated by the experience-based learning framework of \citet{MalmendierNagel2016}. In their model, each agent~$i$ recursively estimates a perceived law of motion for inflation using only the data observed during their professional lifetime starting at date~$s_i$. The two personas weight the same historical inflation data differently: the experienced agent's estimate incorporates the high-inflation episodes of the 1970s, while the young agent's is shaped predominantly by the low and stable inflation of the Great Moderation.

The object of interest is the \emph{persona gap}:
\begin{equation}\label{eq:persona_gap}
\Delta_t \;\equiv\; \mu^{E}_t - \mu^{Y}_t,
\end{equation}
the difference in conditional mean forecasts between the experienced and young personas at vintage~$t$, where $t$ indexes the information vintage (January, April, July, or October 2021), not calendar time. If the persona assignment has no effect on the LLM output, $\Delta_t = 0$. A positive $\Delta_t$ indicates that the experienced persona produces higher inflation forecasts, consistent with the experience channel: an agent whose professional memory includes the 1970s perceives inflation as more persistent and mean-reverting to a higher level than an agent shaped by the Great Moderation.

\subsubsection{Identification}
\label{sec:llm_leakage}

A fundamental concern with retrospective LLM forecasting exercises is look-ahead bias: both models were trained on data that includes the realized 2021 inflation outcomes and the extensive subsequent commentary \citep{crane2025total, alam2026chatmacro}. In the language of \citet{LudwigMullainathanRambachan2025}, this is \emph{training leakage}: the LLM's training corpus~$\mathcal{C}$ overlaps with the researcher's evaluation sample.

To formalize this concern, recall that the conditional mean $\mu^{i}_t = \mathbb{E}[\hat{\boldsymbol{\pi}}^{\,i}_t \mid \mathcal{D}_t, \mathcal{P}_i]$ marginalizes over the training corpus~$\mathcal{C}$: it averages over all corpora that could have produced the LLM. In practice, however, the researcher faces a single realized corpus~$\mathcal{C}$, one that almost certainly contains the 2021 inflation outcomes. Conditioning on this realized~$\mathcal{C}$ may shift the LLM's expected output relative to the unconditional mean. Define the \emph{training leakage bias} for persona~$i$ at vintage~$t$ as
\begin{equation}\label{eq:leakage_bias}
\lambda^{i}_t
\;\equiv\;
\mathbb{E}\!\left[\hat{\boldsymbol{\pi}}^{\,i}_t \mid \mathcal{D}_t, \mathcal{P}_i, \mathcal{C}\right]
-
\mathbb{E}\!\left[\hat{\boldsymbol{\pi}}^{\,i}_t \mid \mathcal{D}_t, \mathcal{P}_i\right].
\end{equation}
The first term is the expected output of the LLM that the researcher actually uses, trained on the specific corpus~$\mathcal{C}$. The second term is what that expectation would be if the corpus were unknown. When $\lambda^{i}_t = 0$, the corpus is uninformative about the LLM output given the prompt. While this holds in expectation over possible corpora (by iterated expectations), it need not hold for the realized~$\mathcal{C}$. Since the realized corpus contains the 2021 inflation data, $\lambda^{i}_t \neq 0$ for the individual persona outputs.

\begin{assumption}[Common training leakage]\label{ass:cc}
The training leakage bias defined in~\eqref{eq:leakage_bias} is common across personas: for all vintages~$t$ and all $i,j \in \{E,Y,0\}$,
\begin{equation}\label{eq:common_leakage}
\lambda^{i}_t = \lambda^{j}_t \equiv \lambda_t.
\end{equation}
\end{assumption}

Assumption~\ref{ass:cc} permits arbitrary contamination ($\lambda_t$ may be large) but requires that it be \emph{common} across personas. Three observations motivate this restriction. The corpus~$\mathcal{C}$ is a property of the LLM, not of the prompt: it is identical across personas. The data vintage~$\mathcal{D}_t$ and instructions~$\mathcal{I}$, which constitute the vast majority of the prompt, are held fixed by design. And the persona description~$\mathcal{P}_i$ is a short biographical paragraph referring only to pre-2020 events.

The threat to Assumption~\ref{ass:cc} is not training leakage per se, but a \emph{treatment--contamination interaction}: if the persona description triggered differential recall from the corpus, for instance by activating persona-specific subsets of training documents that also contain the realized 2021 outcomes, then $\lambda^{E}_t \neq \lambda^{Y}_t$ and persona differencing would not eliminate the bias. The experienced persona prompt explicitly mentions the OPEC embargo, the stagflation of 1974--1980, and the Volcker disinflation, terms that also appear frequently in ex-post analyses of the 2021 inflation episode. If these keywords activate training documents that discuss both the 1970s and the 2021 outcome, the leakage could be persona-specific. Three considerations mitigate this concern. First, the keywords in the experienced persona describe pre-2020 events that appear in textbooks and historical surveys regardless of their connection to 2021; the young persona prompt similarly invokes the ``Great Moderation'' and the ``2008 financial crisis,'' which also feature in post-2021 commentary. Second, the persona gap narrows across vintages (from 1.70~pp in April to 0.38~pp in October), a pattern that would be difficult to generate through differential recall, since the contamination from the realized outcome is fixed across vintages while the LLM's information set is expanding. Third, the neutral persona, which contains no experiential keywords, produces forecasts between the experienced and young personas at every vintage, consistent with the interpretation that the persona assignment induces genuine variation in the weighting of historical information rather than differential access to future outcomes.
\begin{proposition}[Identification of the persona gap]\label{prop:did}
Under Assumption~\ref{ass:cc}, the persona gap~$\Delta_t$ defined in~\eqref{eq:persona_gap} is identified from the LLM outputs regardless of the training corpus:
\begin{equation}\label{eq:did}
\mathbb{E}\!\left[\hat{\boldsymbol{\pi}}^{\,E}_t \mid \mathcal{D}_t, \mathcal{P}_E, \mathcal{C}\right]
-
\mathbb{E}\!\left[\hat{\boldsymbol{\pi}}^{\,Y}_t \mid \mathcal{D}_t, \mathcal{P}_Y, \mathcal{C}\right]
=
\Delta_t,
\end{equation}
for all~$\mathcal{C}$. No restriction on the structure of~$\mu^{i}_t$ is required beyond its definition in~\eqref{eq:mu_it}, and no condition on the level of contamination ($\lambda_t = 0$) is imposed.
\end{proposition}

\begin{proof}
By definition~\eqref{eq:leakage_bias}, $\mathbb{E}[\hat{\boldsymbol{\pi}}^{\,i}_t \mid \mathcal{D}_t, \mathcal{P}_i, \mathcal{C}] = \mu^{i}_t + \lambda^{i}_t$. Differencing across personas:
\begin{align*}
\mathbb{E}[\hat{\boldsymbol{\pi}}^{\,E}_t \mid \mathcal{D}_t, \mathcal{P}_E, \mathcal{C}]
-
\mathbb{E}[\hat{\boldsymbol{\pi}}^{\,Y}_t \mid \mathcal{D}_t, \mathcal{P}_Y, \mathcal{C}]
&= (\mu^{E}_t + \lambda^{E}_t) - (\mu^{Y}_t + \lambda^{Y}_t) \\
&= (\mu^{E}_t - \mu^{Y}_t) + (\lambda^{E}_t - \lambda^{Y}_t) \\
&= \Delta_t,
\end{align*}
where the last equality uses Assumption~\ref{ass:cc}.
\end{proof}

Proposition~\ref{prop:did} complements the two strategies formalized in \citet{LudwigMullainathanRambachan2025}: enforcing a no-leakage condition ($\lambda^{i}_t = 0$) for prediction problems, or collecting a validation sample to debias estimates. The leakage in the present application also operates through a different channel than theirs: not through overlap between the prompts and the training data, but through~$\mathcal{C}$ containing the \emph{realized outcomes} the LLM is asked to forecast. Persona differencing is a third possibility that exploits the experimental structure of the prompt to eliminate the bias without restricting the level of contamination and without a validation sample.

The neutral persona ($\mathcal{P}_0$) is not part of Proposition~\ref{prop:did} and is not needed for identification, but it provides three pieces of evidence against prompt priming, as documented in the results below. First, the ordering $\mu^{Y}_t \le \mu^{0}_t \le \mu^{E}_t$ holds at every vintage, even though the neutral prompt contains no experiential keywords. If the persona gap were driven by keyword priming, the neutral persona should produce forecasts comparable to one of the two treatments or erratic across vintages; instead, its forecasts fall consistently between them. Second, the neutral persona produces forecasts above the young persona at every vintage. Since the neutral prompt contains no hawkish content, this implies that the young persona's low forecasts reflect active anchoring to the Great Moderation induced by its biographical conditioning, not the mere absence of priming. Third, the convergence of all three personas toward the realized outcome as vintages accumulate is consistent with Bayesian updating on incoming data, a pattern that fixed prompt priming cannot generate.

To derive testable implications from Proposition~\ref{prop:did}, the conditional mean is decomposed into a common component and a persona-dependent component:
\begin{equation}
\mu^{i}_t
=
c_t + \phi_t\,\pi^{\mathrm{exp},i}_{t},
\label{eq:mean_measurement}
\end{equation}
where $c_t$ captures all elements common to both personas at vintage~$t$, $\pi^{\mathrm{exp},i}_t$ is the experience-based forecast component for persona~$i$, and $\phi_t$ is a scalar loading. The experience-based learning model of \citet{MalmendierNagel2016} provides a natural parameterization of $\pi^{\mathrm{exp},i}_t$. In their framework, each agent~$i$ recursively estimates a perceived AR(1) law of motion for inflation,
\begin{equation}\label{eq:mn_ar1}
\pi_{\tau+1} = \alpha + \beta \pi_{\tau} + \varepsilon_{\tau+1},
\end{equation}
with age-dependent updating:
\begin{equation}\label{eq:mn_recursion}
b_{\tau,i}
=
b_{\tau-1,i}
+
\gamma_{\tau,i} R_{\tau,i}^{-1} x_{\tau-1}
\left(\pi_{\tau} - b'_{\tau-1,i} x_{\tau-1}\right),
\qquad x_{\tau}=(1,\pi_{\tau})',
\end{equation}
\begin{equation}\label{eq:mn_R}
R_{\tau,i}
=
R_{\tau-1,i}
+
\gamma_{\tau,i}
\left(x_{\tau-1}x'_{\tau-1}-R_{\tau-1,i}\right),
\end{equation}
where $b_{\tau,i}=(\alpha_{\tau,i},\beta_{\tau,i})'$ and the gain depends on the agent's age:
\begin{equation}\label{eq:mn_gain}
\gamma_{\tau,i}
=
\begin{cases}
\theta/(\tau-s_i), & \tau-s_i \ge \theta,\\
1, & \tau-s_i < \theta.
\end{cases}
\end{equation}
Here $s_i$ denotes the career start date implied by persona~$i$. Let $\bar{\pi}_{t,i} \equiv \alpha_{t,i}/(1-\beta_{t,i})$ denote persona~$i$'s perceived long-run mean of inflation, where $b_{t,i} = (\alpha_{t,i},\beta_{t,i})'$ are the coefficients estimated recursively up to vintage~$t$. The $h$-step-ahead forecast from the perceived AR(1) is
\begin{equation}\label{eq:mn_forecast}
\pi^{\mathrm{exp},i}_{t+h|t}
=
\bar{\pi}_{t,i} + \beta_{t,i}^{h}\!\left(\pi_t - \bar{\pi}_{t,i}\right),
\end{equation}
and the average annual forecast used in the experiment is $\pi^{\mathrm{exp},i}_t = H^{-1}\sum_{h=1}^{H}\pi^{\mathrm{exp},i}_{t+h|t}$.

Two implications follow from the learning mechanism in~\eqref{eq:mn_recursion}--\eqref{eq:mn_gain}, the forecast formula~\eqref{eq:mn_forecast}, and Proposition~\ref{prop:did}. First, because the experienced persona's estimation sample includes the high-inflation 1970s while the young persona's does not, the experienced agent estimates both a higher perceived mean ($\bar{\pi}_{t,E} > \bar{\pi}_{t,Y}$) and a higher persistence ($\beta_{t,E} > \beta_{t,Y}$), so $\pi^{\mathrm{exp},E}_{t} > \pi^{\mathrm{exp},Y}_{t}$ and the persona gap $\Delta_t$ in~\eqref{eq:persona_gap} is positive at early vintages. Second, after positive inflation surprises in 2021, the young persona updates more strongly because its gain~$\gamma_{\tau,Y}$ is larger (shorter career implies larger $\gamma$), so $\Delta_t$ narrows across vintages.

Equation~\eqref{eq:mn_forecast} also connects the LLM experiment to the historically informed methods of Section~\ref{sec:hist_approaches}. The similarity approach, the kernel-weighted estimator, and the MN persona all produce forecasts of the same functional form, iteration of a perceived AR(1) (the MA component of the ARMA(1,1) enters only at $h=1$ and does not affect the multi-step recursion), and differ only in how they weight historical observations when estimating $(\alpha,\beta)$. The similarity approach uses hard truncation to a reference window; the kernel-weighted estimator uses smooth weights $w_\tau \propto K\!\left((z_\tau - z_t)/b\right)$ based on a state variable; the MN experienced persona uses age-dependent weights determined by the gain~$\gamma_{\tau,i}$. When the forecast origin resembles the 1970s, all three weighting schemes concentrate mass on the supply-shock episodes, producing higher values of $\bar{\pi}_w$ and $\beta_w$ than the full-sample benchmark. The experience channel formalized by \citet{MalmendierNagel2016} is thus one particular weighting scheme, age-dependent, that achieves for the experienced persona what the historically informed methods achieve by design.

\subsection{Results}
\label{sec:llm_results}

\begin{figure}[!t]
\centering
\includegraphics[width=0.95\textwidth]{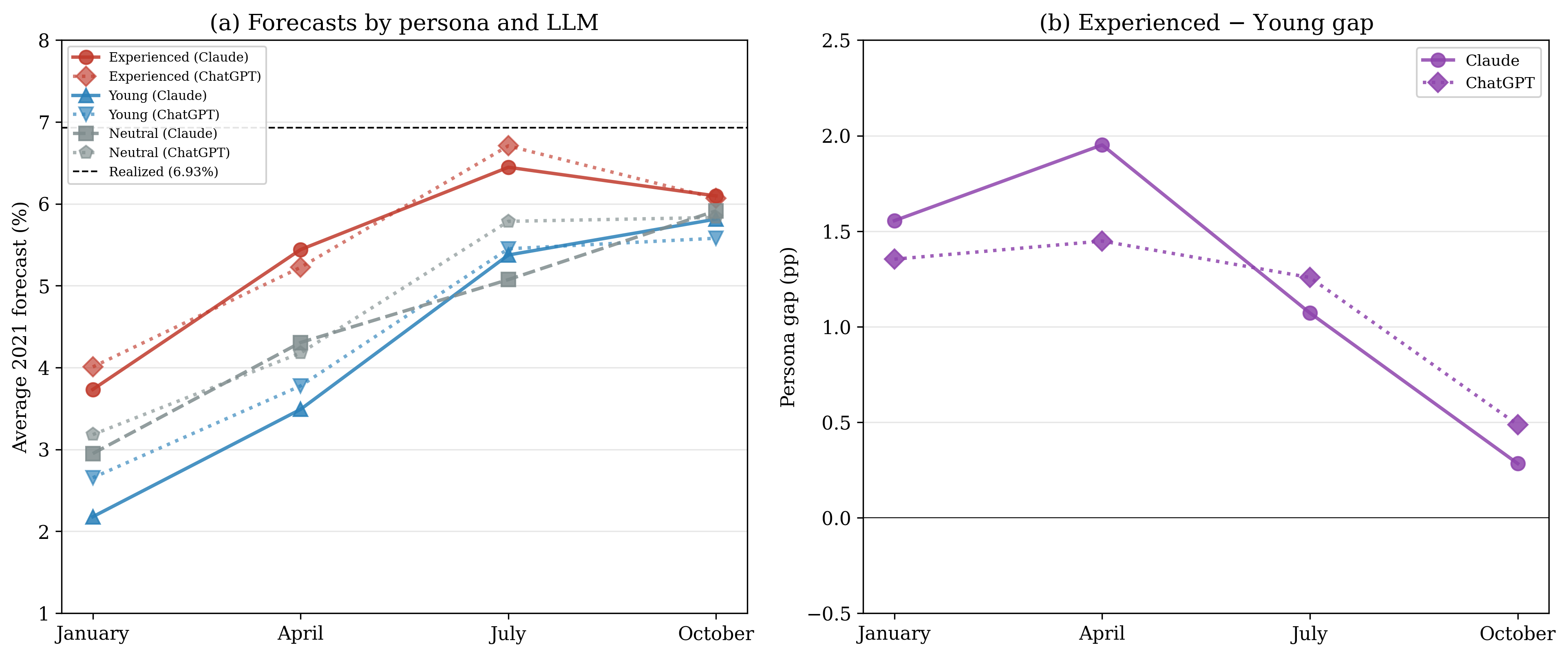}
\caption{LLM inflation forecasts by persona and information vintage. Panel~(a): average 2021 forecast for the experienced, neutral, and young personas, with Claude (solid) and ChatGPT (hatched) shown separately. Panel~(b): persona gap (Experienced $-$ Young) by LLM. Each bar averages five independent agents per LLM. The dashed line in panel~(a) marks the realized 2021 average (6.93\%).}
\label{fig:llm_persona}
\end{figure}

Panel~(a) of Figure~\ref{fig:llm_persona} plots the average 2021 forecast by persona across the four information vintages. Consistent with the first implication of the MN framework, the experienced persona predicts higher inflation than the young persona at every vintage. The persona gap $\hat{\Delta}_t$ is 1.45~pp at the January vintage (experienced: 3.87\%, young: 2.42\%) and peaks at 1.70~pp at the April vintage (experienced: 5.33\%, young: 3.63\%), comparable in magnitude to the age-based expectation gap in the SCE documented in Figure~\ref{fig:sce_age_gap}. As additional 2021 inflation data are incorporated, the gap narrows to 0.38~pp by October (experienced: 6.08\%, young: 5.70\%), consistent with the second implication (faster updating by the young persona due to a larger gain~$\gamma_{\tau,Y}$). Panel~(b) shows that the pattern is stable across LLMs: the gap is 1.56~pp (Claude) and 1.35~pp (ChatGPT) at the January vintage, and both converge toward zero by October. The neutral persona satisfies $\mu^{Y}_t \le \mu^{0}_t \le \mu^{E}_t$ at every vintage.\footnote{Because the 10 runs per persona--vintage cell (5 per LLM) are stochastic draws from the same foundation model and prompt, they do not constitute independent experimental units in the classical sense. The persona gap is therefore reported as a descriptive summary of the LLM outputs rather than as a formal test statistic.}

\begin{figure}[!t]
\centering
\includegraphics[width=0.95\textwidth]{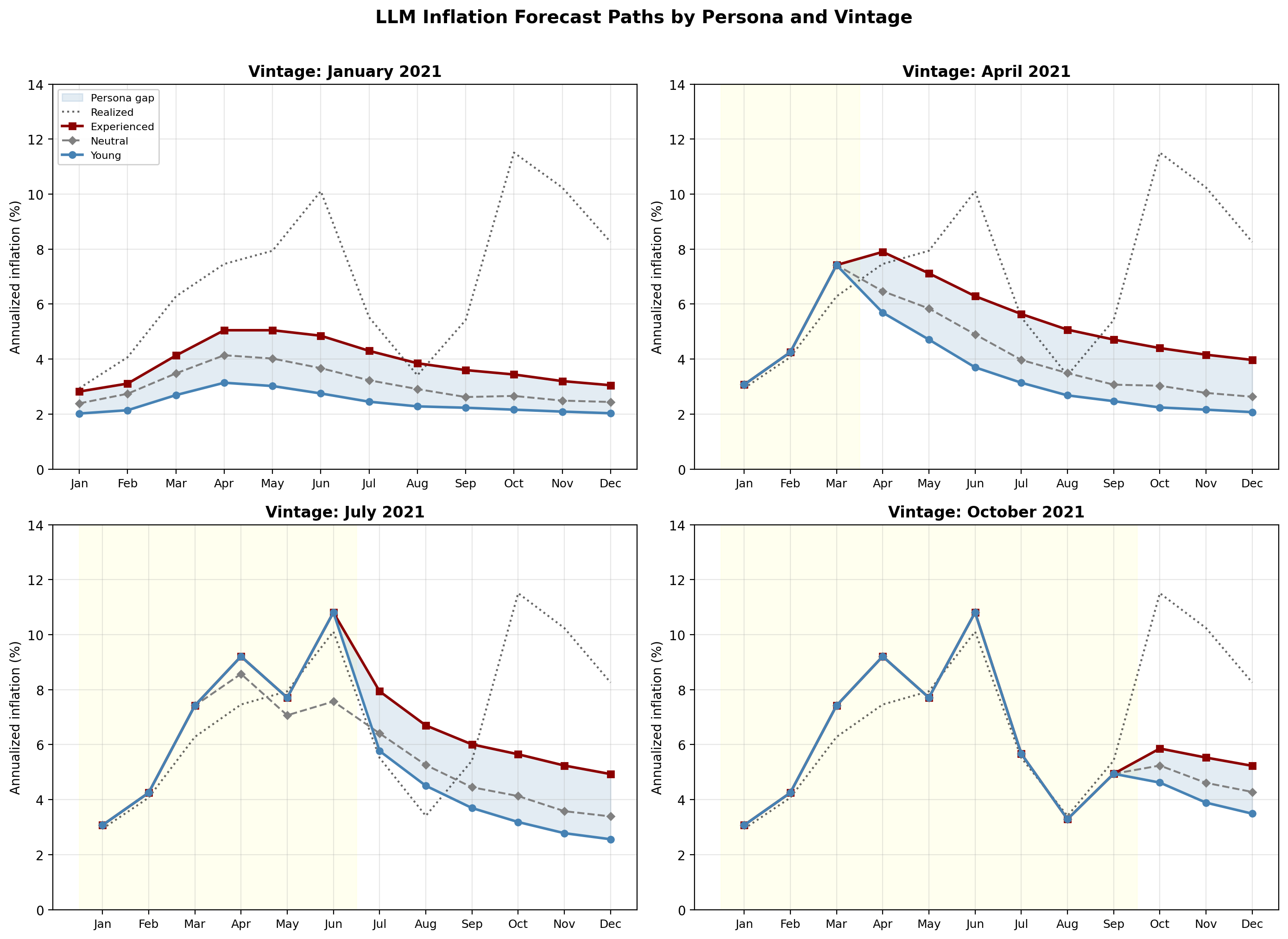}
\caption{Monthly LLM inflation forecast paths by persona and information vintage, pooled across Claude and ChatGPT (10 agents per persona--vintage cell). Each panel corresponds to a data vintage; the shaded area marks months for which CPI data are already available in that vintage. The dashed line (neutral persona) falls between the experienced and young paths at every vintage. The dotted line shows realized annualized monthly inflation.}
\label{fig:llm_updates}
\end{figure}

Figure~\ref{fig:llm_updates} decomposes the aggregate gap into monthly forecast profiles. At each vintage, the young persona predicts rapid mean-reversion toward 2--3 percent by year-end, while the experienced persona maintains forecasts in the 4--6 percent range. The neutral persona traces a path between the two. Under~\eqref{eq:mean_measurement} and~\eqref{eq:mn_forecast}, this pattern reflects differences in the perceived mean~$\bar{\pi}_{t,i}$ and persistence~$\beta_{t,i}$ across personas.

The contrast is sharpest at the April and July vintages, when incoming data are high but the forecast horizon remains long. By October, the gap narrows but does not vanish: the experienced persona forecasts October--December at 5.9, 5.5, and 5.2 percent, while the young persona projects 4.6, 3.9, and 3.5 percent.

The experienced persona's January forecast (3.87 percent) is comparable to the SPF (3.89 percent) and FOMC projections (3.83 percent) from Table~\ref{tab:other_forecasts}. By July, it reaches 6.58 percent, approaching the intercept correction (6.83 percent) and similarity approach (6.19 percent) from Table~\ref{tab:detailed_forecasts}. The young persona's trajectory tracks the survey consensus instead: its January forecast (2.42 percent) is close to the CBO (2.65 percent) and Atlanta Fed (2.82 percent) projections. By Proposition~\ref{prop:did}, this heterogeneity is not attributable to differential training leakage.

\begin{figure}[!t]
\centering
\includegraphics[width=0.95\textwidth]{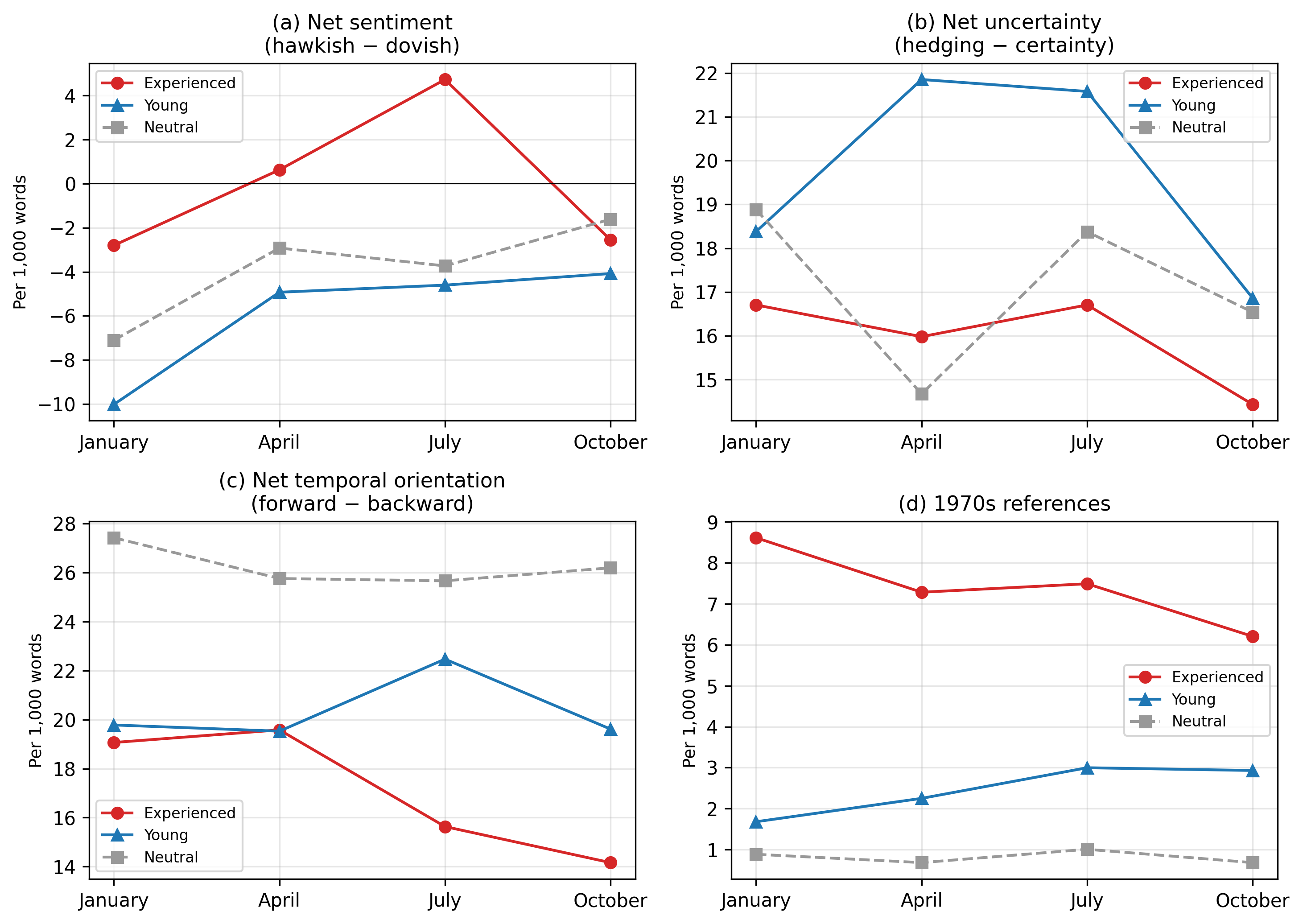}
\caption{Textual analysis of LLM forecast reasoning by persona and information vintage, pooled across Claude and ChatGPT. Each indicator is expressed as a net frequency per 1,000 words. The neutral persona (dashed, grey) serves as baseline. Panel~(a): net sentiment (hawkish minus dovish word frequency). Panel~(b): net uncertainty (hedging minus certainty language). Panel~(c): net temporal orientation (forward-looking minus backward-looking references). Panel~(d): frequency of 1970s references.}
\label{fig:llm_reasoning}
\end{figure}

Beyond the numerical forecasts, the written reasoning accompanying each forecast provides descriptive evidence on how the persona shapes the economic narrative. Figure~\ref{fig:llm_reasoning} reports four text-based indicators constructed from the forecast justifications produced across all experimental runs (Appendix~\ref{sec:text_analysis} provides the methodology). These indicators are not derived from the formal framework of Section~\ref{sec:llm_leakage}, which concerns the conditional mean of the numerical forecasts; they characterize properties of the text that fall outside the scope of Proposition~\ref{prop:did}.

Panel~(a) shows that the experienced persona's net sentiment turns progressively hawkish (from $-2.8$ at the January vintage to $+4.7$ by July), while the young persona remains dovish throughout ($-10$ to $-4$). Panel~(b) reveals that the young persona shows rising net uncertainty at the April and July vintages ($\sim$22 per thousand words), while the experienced persona maintains stable net uncertainty ($\sim$16--17) across vintages. Panel~(c) shows that the experienced persona becomes increasingly backward-looking (net temporal orientation declining from $+19$ to $+14$), while the young persona maintains a forward orientation ($\sim$20--23). Panel~(d) shows that the experienced persona invokes the 1970s two to three times more frequently than the young persona at every vintage.

Table~\ref{tab:llm_priors} illustrates how the two personas interpret identical macroeconomic signals, based on the LLMs' own methodological descriptions (Appendix~\ref{sec:llm_methodology}).

\begin{table}[t]
\centering
\caption{How the persona filters the same data: experienced versus young}
\label{tab:llm_priors}
\renewcommand{\arraystretch}{1.3}
\small
\begin{tabular}{p{3.2cm} p{5.2cm} p{5.2cm}}
\hline
\hline
\textbf{Signal} & \textbf{Experienced (1970s)} & \textbf{Young (post-2008)} \\
\hline
M2 growth ($+25\%$)
    & Strong inflationary signal
    & Noise---M2 did not predict inflation after 2008 \\
Unemployment (6.7\%)
    & Moderate slack, eroding fast
    & Substantial slack, disinflationary anchor \\
Supply disruptions
    & Interact with demand $\to$ persistence
    & Temporary shocks, identifiable, transitory \\
Fed credibility
    & Fragile---memory of Burns
    & Solid---25 years of anchoring \\
Rising PPI
    & Cost pipeline $\to$ passthrough
    & Short-run noise, not predictive of CPI \\
TIPS breakevens
    & Possibly lagging reality
    & Reliable market signal \\
Historical reference
    & 1972--1974
    & 2010--2012 \\
\hline
\hline
\end{tabular}
\begin{flushleft}
\footnotesize
\textit{Notes:} The table summarizes how the experienced and young personas interpret identical inputs, based on the LLMs' own methodological descriptions. Both personas receive the same data and macroeconomic context; the difference lies in the interpretation induced by the persona prompt.
\end{flushleft}
\end{table}

The experienced persona maps supply disruptions and monetary accommodation into a more persistent inflation process, whereas the young persona maps the same signals into a more transitory path with faster mean reversion, consistent with the differences in perceived mean and persistence implied by equation~\eqref{eq:mn_forecast}.

\section{Conclusion}
\label{sec:conclusion}

This paper shows that the 2021 inflation forecasting failure was primarily driven by sample composition rather than functional-form misspecification. Full-sample estimators dominated by the Great Moderation underweight supply-shock regimes and produce systematically low forecasts when the economy enters such a regime. Three simple adjustments, an intercept correction, a similarity-based re-estimation, and a kernel-weighted estimator, substantially close the forecast gap. The robust IC yields 5.00 percent, the similarity approach 6.19, and the single-origin IC 6.83 versus the realized 6.93, all large improvements over the ARMA benchmark of 3.52. The kernel-weighted approach (7.82) confirms the 1970s analogy through a data-driven criterion, and the sensitivity analysis shows that these results are robust to the choice of reference window. The same framework applied to eight additional price indices confirms that the mechanism extends across the entire U.S.\ price chain. Adaptive methods can perform worse than unadjusted benchmarks when a transitory shock precedes a persistent regime change: the TVP-AR(4) yields only 1.73 percent because the Kalman filter absorbs the deflationary 2020 readings and adjusts parameters in the wrong direction.

The evidence on household expectations reinforces this interpretation: respondents over 60 reported higher inflation expectations than younger cohorts throughout 2021, consistent with the experience-based learning framework of \citet{MalmendierNagel2016}. The LLM experiment extends this evidence to professional forecasters. Under a common training leakage assumption, the experienced persona produces higher forecasts than the young persona at every vintage, and the gap narrows as data accumulate. The similarity approach, the kernel-weighted estimator, and the experienced persona all produce forecasts of the same functional form, iteration of a perceived AR(1), and differ only in how they weight historical observations.

The historically informed approaches succeed because the post-COVID episode did, in fact, resemble the 1970s. Their value lies in providing a disciplined device for encoding a specific economic hypothesis into the forecast. Developing a systematic framework for selecting the relevant historical precedent remains an important direction for future work.

\clearpage
\bibliographystyle{chicago}
\bibliography{references}

\clearpage
\appendix

\section{Sensitivity to the choice of reference window}
\label{sec:sensitivity}

A natural concern is that the historically informed approaches may be optimized, even inadvertently, on the realized 2021 outcome. This appendix examines the sensitivity of the forecasts to the choice of reference window for the similarity approach, the reference episode for the intercept correction, and the bandwidth for the kernel-weighted estimator. It also reports a historical out-of-sample exercise that applies the same methodology to the 1979 inflation episode.

\begin{figure}[!t]
\centering
\includegraphics[width=0.75\textwidth]{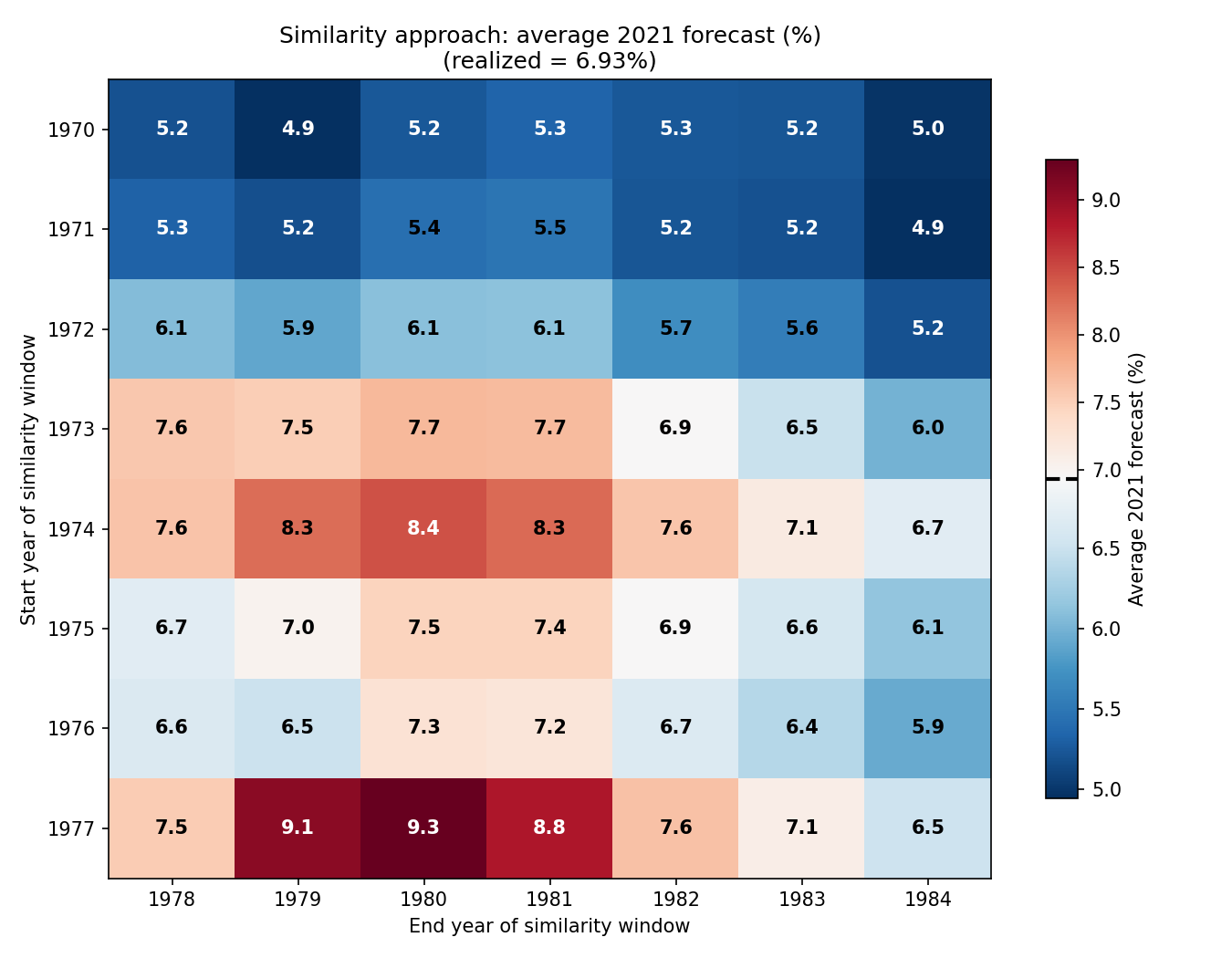}
\caption{Sensitivity of the similarity approach to the choice of estimation window. Each cell reports the average 2021 inflation forecast (in percent) from an ARMA(1,1) estimated on the indicated subsample. The realized average is 6.93 percent. Cells are shaded from blue (below realized) to red (above realized), centered at 6.93.}
\label{fig:sensitivity_sim}
\end{figure}

\paragraph{Similarity approach.}
Figure~\ref{fig:sensitivity_sim} reports the average 2021 forecast for a grid of similarity windows, varying the start year from 1970 to 1977 and the end year from 1978 to 1984. The baseline window used in Table~\ref{tab:detailed_forecasts} is 1973M01--1980M12 (average forecast of 6.19 percent in Table~\ref{tab:detailed_forecasts}).\footnote{The heatmap is computed using the Python MLE implementation; the MATLAB-based estimates in Table~\ref{tab:detailed_forecasts} yield slightly different values for the same window due to differences in the optimizer. The qualitative pattern---broad robustness to window choice---is unchanged.} The heatmap reveals several patterns. First, forecasts are closest to the realized 6.93 percent when the window starts in 1973--1975 and ends in 1978--1982, i.e., when it captures the core of the oil-shock inflationary regime; the absolute highest forecasts occur for start years 1974 and 1977, reaching up to 9.30 percent (1977--1980). Second, extending the window backward to 1970--1971 or forward to 1983--1984 dilutes the inflationary signal by including years of moderate inflation, pulling the forecast toward 5 percent. Third, the forecast is not an artifact of a single lucky window: 34 out of 56 window combinations produce average forecasts between 5.5 and 8.5 percent, all substantially above the unadjusted ARMA benchmark of 3.52 percent. The window 1973--1982 happens to yield exactly 6.93 percent, but nearby windows (e.g., 1973--1980 at 7.69, 1973--1981 at 7.68, 1975--1982 at 6.94) produce comparable results.

\paragraph{Intercept correction.}
Figure~\ref{fig:sensitivity_ic} shows the average 2021 forecast produced by the intercept correction as the reference episode varies month by month across the 1973--1980 period. The IC uses 12-month forecast errors from each origin to adjust the baseline ARMA forecast. The sensitivity is substantial: origins in 1974--1975 produce negative forecasts because the 1974--75 recession caused inflation to \emph{fall} after the first oil shock, generating forecast errors that correct in the wrong direction for 2021. By contrast, origins from 1977M09 to 1979M04 yield forecasts between 6 and 8 percent, closely matching the realized outcome. The baseline origin of 1978M02 (6.82 percent in the figure) lies squarely within this range. The robust intercept correction, which averages errors over all 96 origins in the similarity window, produces 5.18 percent---a conservative but stable estimate that smooths out the large origin-to-origin variation visible in the figure. This comparison illustrates the trade-off between the single-window IC (high potential accuracy but sensitive to the choice of origin) and the robust IC (more stable but attenuated).

\begin{figure}[!t]
\centering
\includegraphics[width=0.80\textwidth]{sensitivity_ic_origins.png}
\caption{Sensitivity of the intercept correction to the choice of reference episode. Each point reports the average 2021 forecast when the IC uses 12-month forecast errors from the indicated origin. The dashed black line marks the realized average (6.93\%), the dotted gray line the unadjusted ARMA (3.51\%), and the dash-dot green line the robust IC averaged over all origins (5.18\%).}
\label{fig:sensitivity_ic}
\end{figure}

\begin{figure}[!t]
\centering
\includegraphics[width=0.80\textwidth]{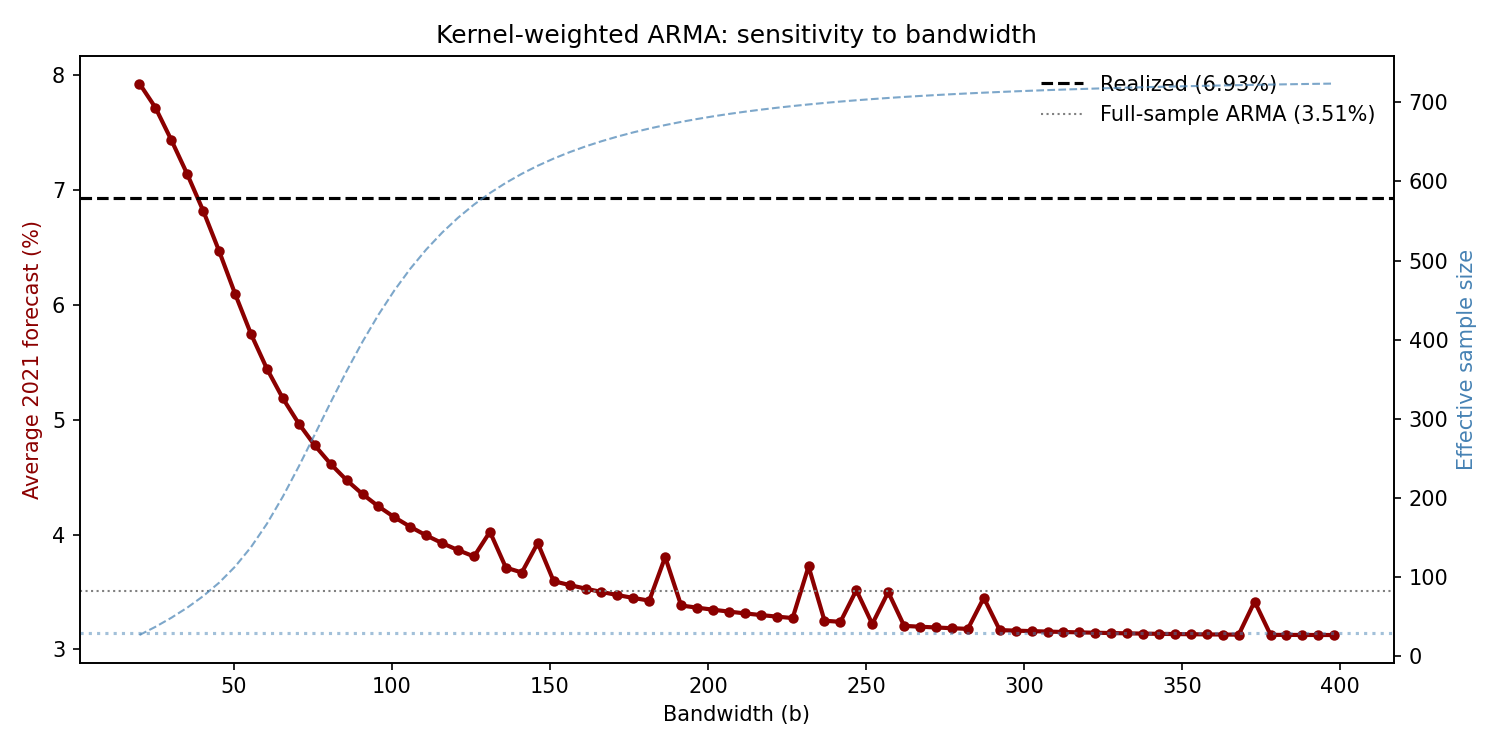}
\caption{Kernel-weighted ARMA(1,1): average 2021 forecast as a function of the bandwidth~$b$. The right axis shows the effective sample size. The dashed black line marks the realized average (6.93\%), the dotted gray line the full-sample ARMA (3.51\%).}
\label{fig:sensitivity_kernel}
\end{figure}

\paragraph{Kernel bandwidth.}
Figure~\ref{fig:sensitivity_kernel} plots the average 2021 forecast as a function of the kernel bandwidth~$b$, confirming the analytical predictions discussed in Section~\ref{sec:framework}. For small bandwidths ($b \approx 20$, effective sample size around 27), the kernel concentrates weight on the oil-shock episodes and produces forecasts near 7.9 percent. As $b$ increases, the weight distribution flattens and the forecast converges toward the full-sample ARMA benchmark of 3.51 percent. The cross-validated bandwidth $b^* = 23.18$ (reported in Table~\ref{tab:detailed_forecasts}) produces a forecast of 7.82 percent, near the upper end of the range. The overall downward relationship between $b$ and the forecast, though non-monotone in finite samples due to estimation variability, confirms that the kernel approach nests the full-sample estimator ($b \to \infty$) and the similarity estimator ($b \to 0$), as established in the analytical framework.

\paragraph{Historical out-of-sample: forecasting 1979.}
The sensitivity analysis above varies the inputs to the historically informed approaches while holding the forecast target (2021) fixed. A complementary exercise applies the \emph{same methodology} to a different target: the 1979 inflation surge, using the first oil shock (1973--1975) as the historical analogue. This provides a genuine out-of-sample validation on an independent episode.

Table~\ref{tab:oos_1979} reports the results. With a forecast origin of December~1978, the unadjusted ARMA(1,1) estimated on the full available sample (1960--1978) predicts average 1979 inflation of 8.00 percent, against a realized outcome of 12.45 percent---an underestimate of 4.45 percentage points, comparable in magnitude to the 2021 failure. The intercept correction, using 12-month forecast errors from the 1973 origin, produces 12.92 percent---within 0.47 percentage points of the realized outcome. The similarity approach, re-estimated on 1973--1975, yields 9.08 percent: a substantial improvement over the benchmark but insufficient to capture the full magnitude of the second oil shock, which was more severe than the first. The kernel-weighted approach produces 7.71 percent, below the unadjusted ARMA, because the oil price dynamics at end-1978 were moderate in growth-rate terms and the kernel accordingly assigns low weight to the extreme 1973--74 observations.

This exercise confirms that the intercept correction is particularly effective when the reference episode closely matches the target, while the similarity approach provides a more conservative but robust improvement. It also reveals a limitation of the kernel approach: the state variable (oil price growth) must be elevated at the forecast origin for the kernel to identify the relevant historical analogue, and in late 1978 this condition was not yet fully met.

\begin{table}[!t]
\centering
\caption{Historical out-of-sample: forecasting 1979 inflation using the 1973--75 analogy}
\label{tab:oos_1979}
\begin{tabular}{l|c|cccc}
\hline
\hline
Date & Realized & Unadjusted & Similarity & Intercept & Kernel \\
      &          & ARMA       & (1973--75) & correction & weighted \\
\hline
1979-01 & 10.56 & 8.22 & 9.06 & 10.23 & 8.08 \\
1979-02 & 12.20 & 8.17 & 9.07 & 12.95 & 8.01 \\
1979-03 & 12.08 & 8.13 & 9.08 & 15.62 & 7.94 \\
1979-04 & 11.96 & 8.09 & 9.08 & 12.73 & 7.87 \\
1979-05 & 13.52 & 8.05 & 9.08 &  9.89 & 7.80 \\
1979-06 & 13.37 & 8.01 & 9.08 & 12.54 & 7.74 \\
1979-07 & 13.22 & 7.97 & 9.08 &  4.33 & 7.67 \\
1979-08 & 11.45 & 7.94 & 9.08 & 25.81 & 7.61 \\
1979-09 & 11.34 & 7.90 & 9.08 &  9.56 & 7.55 \\
1979-10 & 12.83 & 7.86 & 9.08 & 14.77 & 7.49 \\
1979-11 & 12.70 & 7.83 & 9.08 & 12.03 & 7.43 \\
1979-12 & 14.13 & 7.79 & 9.08 & 14.53 & 7.38 \\
\textbf{1979 avg} & \bf{12.45} & \bf{8.00} & \bf{9.08} & \bf{12.92} & \bf{7.71} \\
\hline
\hline
\end{tabular}
\begin{flushleft}
\footnotesize
\textit{Notes:} Forecast origin is December 1978. The unadjusted ARMA(1,1) is estimated on 1960M01--1978M12. The similarity approach re-estimates on 1973M01--1975M12. The intercept correction uses 12-month forecast errors from the origin 1972M12 (reference episode: 1973). The kernel-weighted approach uses the oil price growth rate with bandwidth selected by weighted cross-validation ($b^* = 32$). Realized inflation is annualized monthly CPI inflation.
\end{flushleft}
\end{table}

\paragraph{Robustness to alternative base models.}
A natural concern is whether the gains from the historically informed approaches are specific to the ARMA(1,1) specification. Table~\ref{tab:robustness_base} applies the intercept correction, robust IC, and similarity approach to four different base models: ARMA(1,1), AR(1), AR(4), and ARMA(2,1). All models are estimated on the full sample 1960M01--2020M12 using maximum likelihood. The main finding is that the historically informed corrections produce large improvements regardless of the base model. All unadjusted models predict between 3.44 and 3.72 percent---far from the realized 6.93. The intercept correction raises the forecast to between 6.09 and 7.94 percent across specifications, and the similarity approach produces forecasts between 6.19 and 8.68 percent. The robust IC, which averages over all origins in the similarity window, yields between 5.00 and 9.44 percent. The gains are thus systematic and not an artifact of the ARMA(1,1) parameterization.

\begin{table}[!t]
\centering
\caption{Robustness to alternative base model specifications}
\label{tab:robustness_base}
\begin{tabular}{l|cccc}
\hline
\hline
 & \multicolumn{4}{c}{Average 2021 forecast (\%)} \\
Method & ARMA(1,1) & AR(1) & AR(4) & ARMA(2,1) \\
\hline
Realized & \multicolumn{4}{c}{6.93} \\
\hline
Unadjusted & 3.52 & 3.72 & 3.46 & 3.44 \\
Intercept correction & 6.83 & 7.94 & 6.48 & 6.09 \\
Robust IC & 5.00 & 9.44 & 7.32 & 5.11 \\
Similarity & 6.19 & 8.68 & 7.27 & 8.09 \\
\hline
\hline
\end{tabular}
\begin{flushleft}
\footnotesize
\textit{Notes:} The ARMA(1,1) column reproduces the values from Table~\ref{tab:detailed_forecasts} (MATLAB implementation). The AR(1), AR(4), and ARMA(2,1) columns are estimated by maximum likelihood in Python.
The intercept correction uses a January 1978 reference origin.
The robust IC averages forecast errors over all monthly origins from January 1973 to December 1980.
The similarity approach re-estimates the model on the 1973M01--1980M12 subsample.
\end{flushleft}
\end{table}

\section{Additional U.S.\ inflation measures}
\label{sec:additional_measures}

This appendix provides the detailed monthly forecasts and historical comparison figures underlying the summary results reported in Table~\ref{tab:validation_summary} of Section~\ref{sec:price_indices}. Eight additional U.S.\ inflation measures are considered: the PCE price index (PCEPI), core CPI excluding food and energy (CPILFESL), the PPI for all commodities (PPIACO), PPI for finished goods (WPSFD49207), PPI for finished consumer goods (WPSFD49502), PPI for processed intermediate goods (WPSID61), PPI for unprocessed intermediate goods (WPSID62), and PPI for primary nonferrous metals (PPICMM). All seven forecasting methods from Table~\ref{tab:detailed_forecasts} are applied with identical specifications to each series: ARMA(1,1), intercept correction (reference window 1978M02--1979M01), robust intercept correction (96 origins in 1973M01--1980M12, pre-shock parameters from 1960M02--1972M12), similarity approach (1973M01--1980M12), kernel-weighted ARMA (Gaussian kernel on oil price growth, ESS~$\geq 30$), Markov-switching AR(1), and TVP-AR(4) ($Q = 0.01 \cdot I_5$, break at 2020M01). Six series are taken from the FRED-MD January 2021 vintage; PPIACO and CPILFESL are taken from their respective ALFRED vintages. The estimation sample runs from 1960M02 to 2020M12 for all series. ARMA, IC, Robust IC, and Similarity are estimated in MATLAB (identical code to the main CPI analysis); Kernel, MS-AR, and TVP-AR are estimated in Python.

\begin{figure}[!htbp]
\centering
\includegraphics[width=0.85\textwidth]{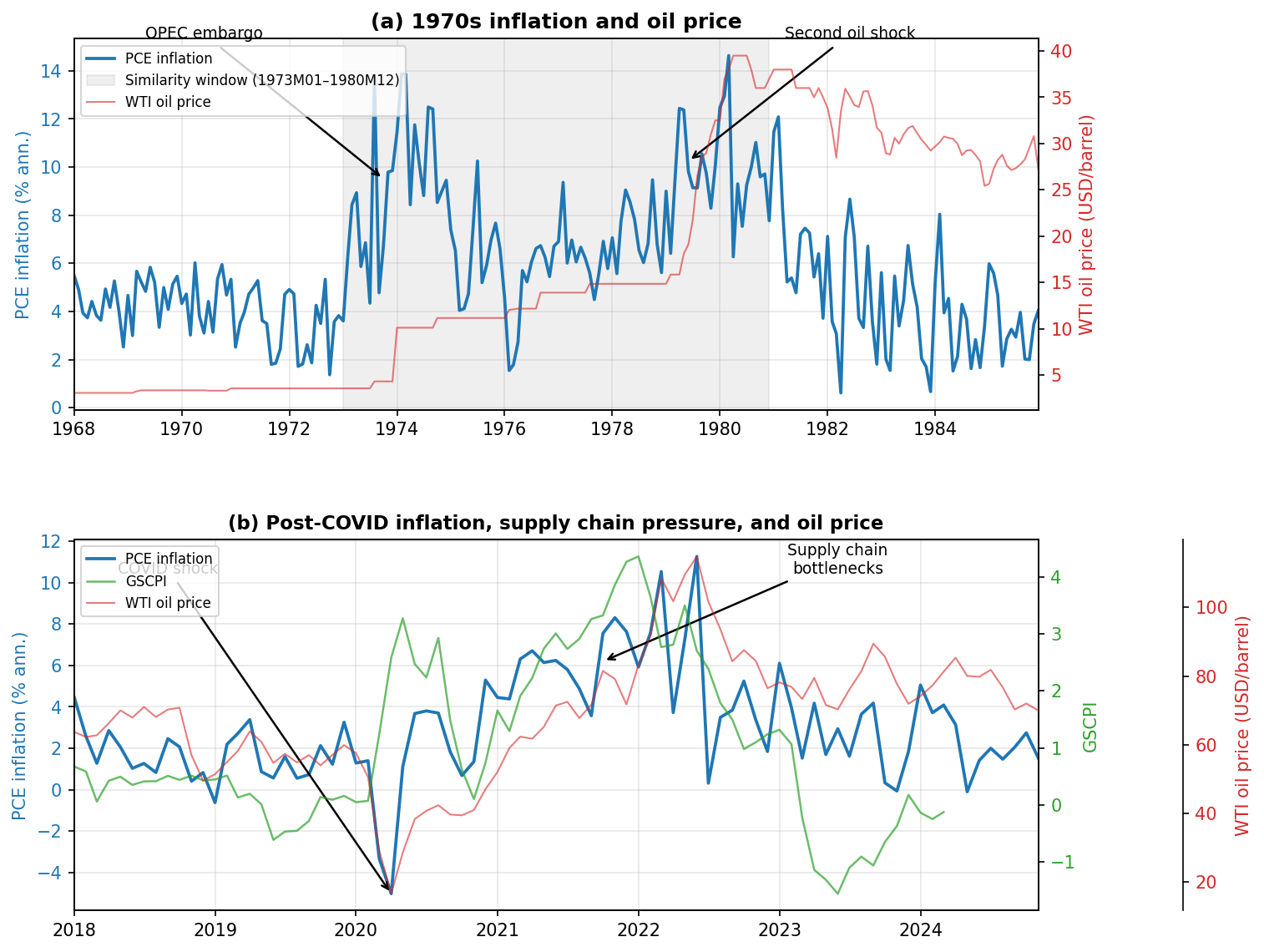}
\caption{Annualized monthly PCE inflation and WTI oil price: 1970s episode (panel~a) and post-COVID period (panel~b). The shaded area indicates the similarity window 1973M01--1980M12.}
\label{fig:hc_PCEPI}
\end{figure}

\begin{figure}[!htbp]
\centering
\includegraphics[width=0.85\textwidth]{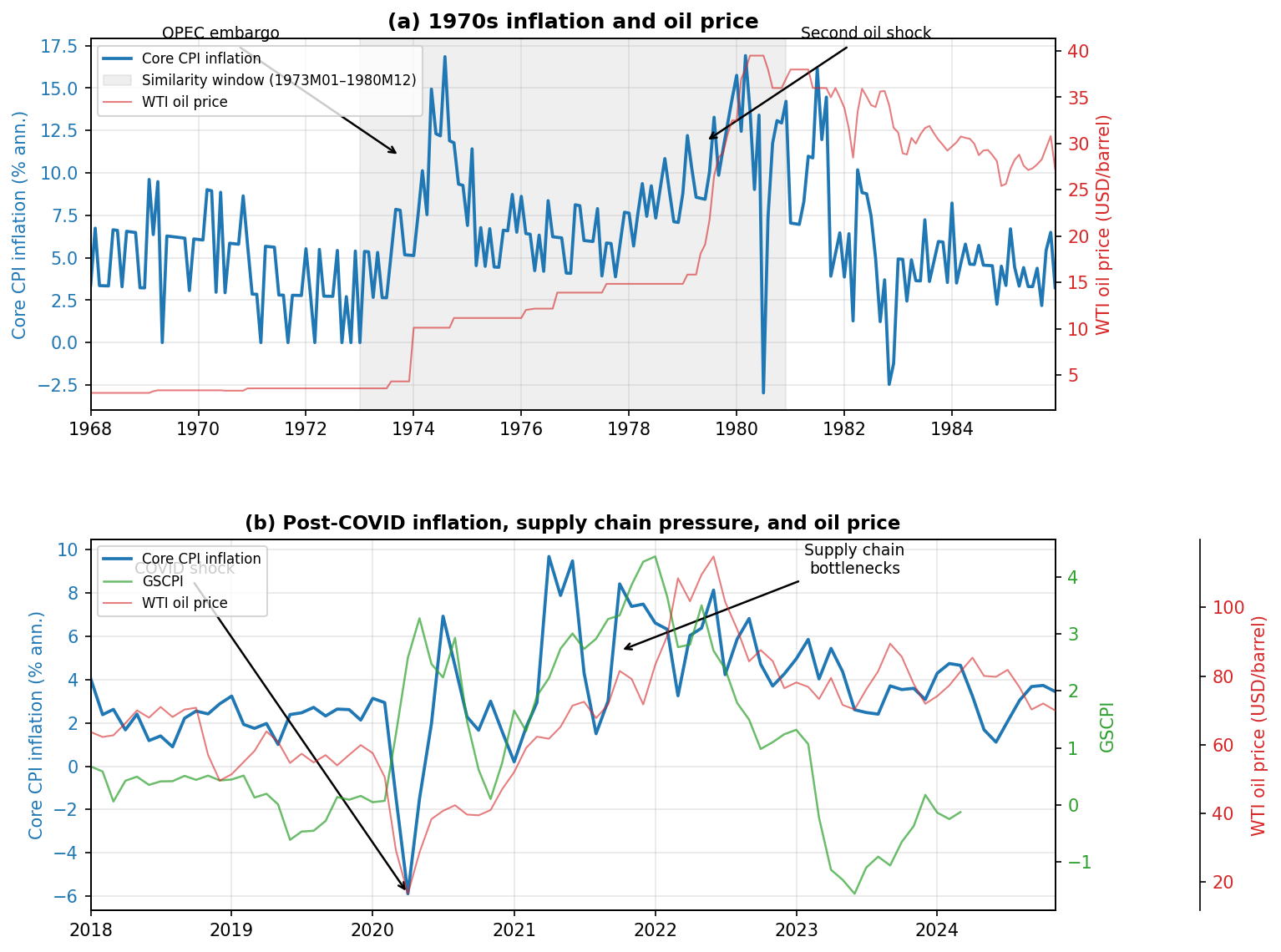}
\caption{Annualized monthly core CPI inflation (less food and energy) and WTI oil price: 1970s episode (panel~a) and post-COVID period (panel~b).}
\label{fig:hc_CPILFESL}
\end{figure}

\begin{figure}[!htbp]
\centering
\includegraphics[width=0.85\textwidth]{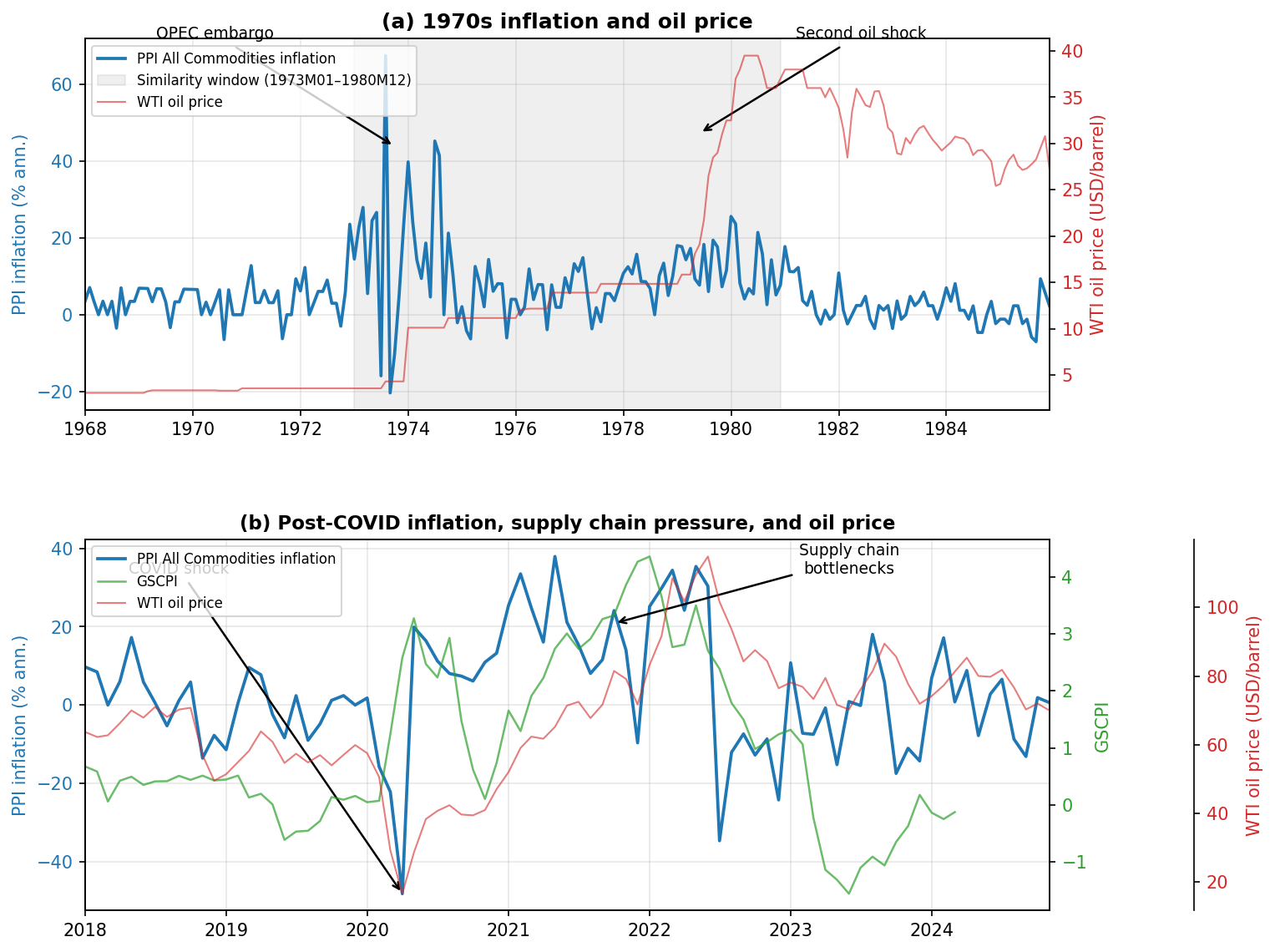}
\caption{Annualized monthly PPI All Commodities inflation and WTI oil price: 1970s episode (panel~a) and post-COVID period (panel~b).}
\label{fig:hc_PPIACO}
\end{figure}

\begin{figure}[!htbp]
\centering
\includegraphics[width=0.85\textwidth]{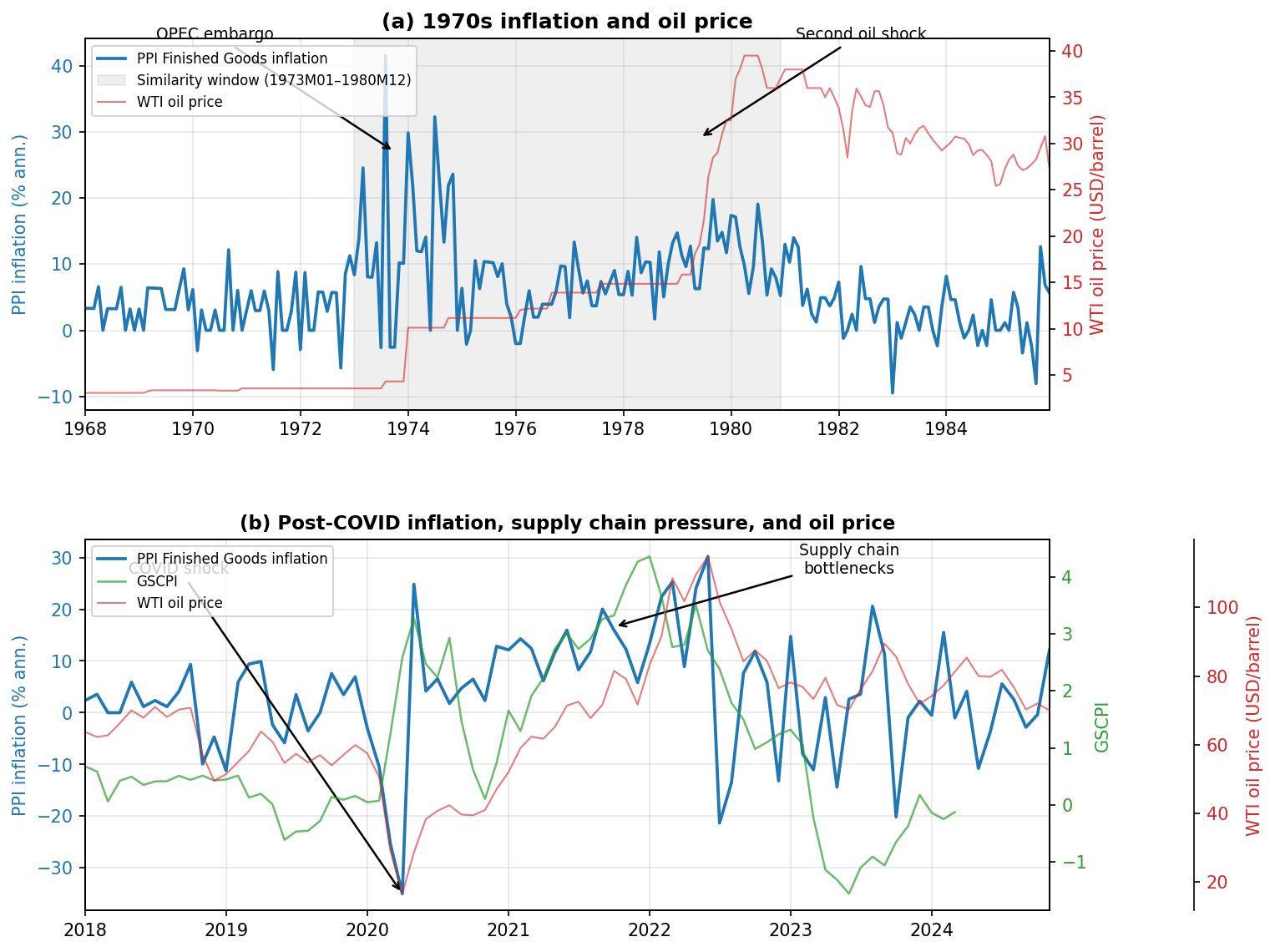}
\caption{Annualized monthly PPI Finished Goods inflation and WTI oil price: 1970s episode (panel~a) and post-COVID period (panel~b).}
\label{fig:hc_WPSFD49207}
\end{figure}

\begin{figure}[!htbp]
\centering
\includegraphics[width=0.85\textwidth]{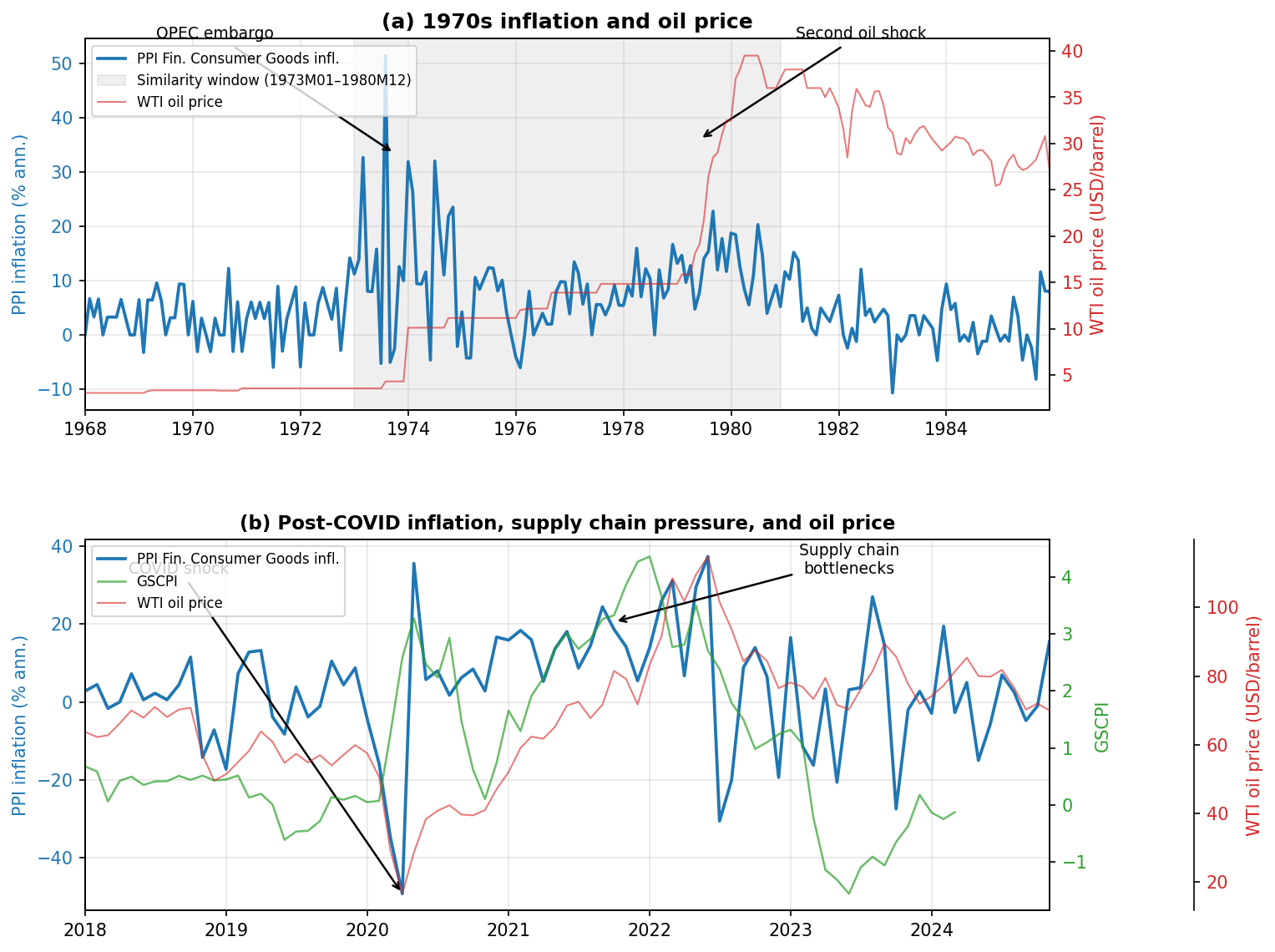}
\caption{Annualized monthly PPI Finished Consumer Goods inflation and WTI oil price: 1970s episode (panel~a) and post-COVID period (panel~b).}
\label{fig:hc_WPSFD49502}
\end{figure}

\begin{figure}[!htbp]
\centering
\includegraphics[width=0.85\textwidth]{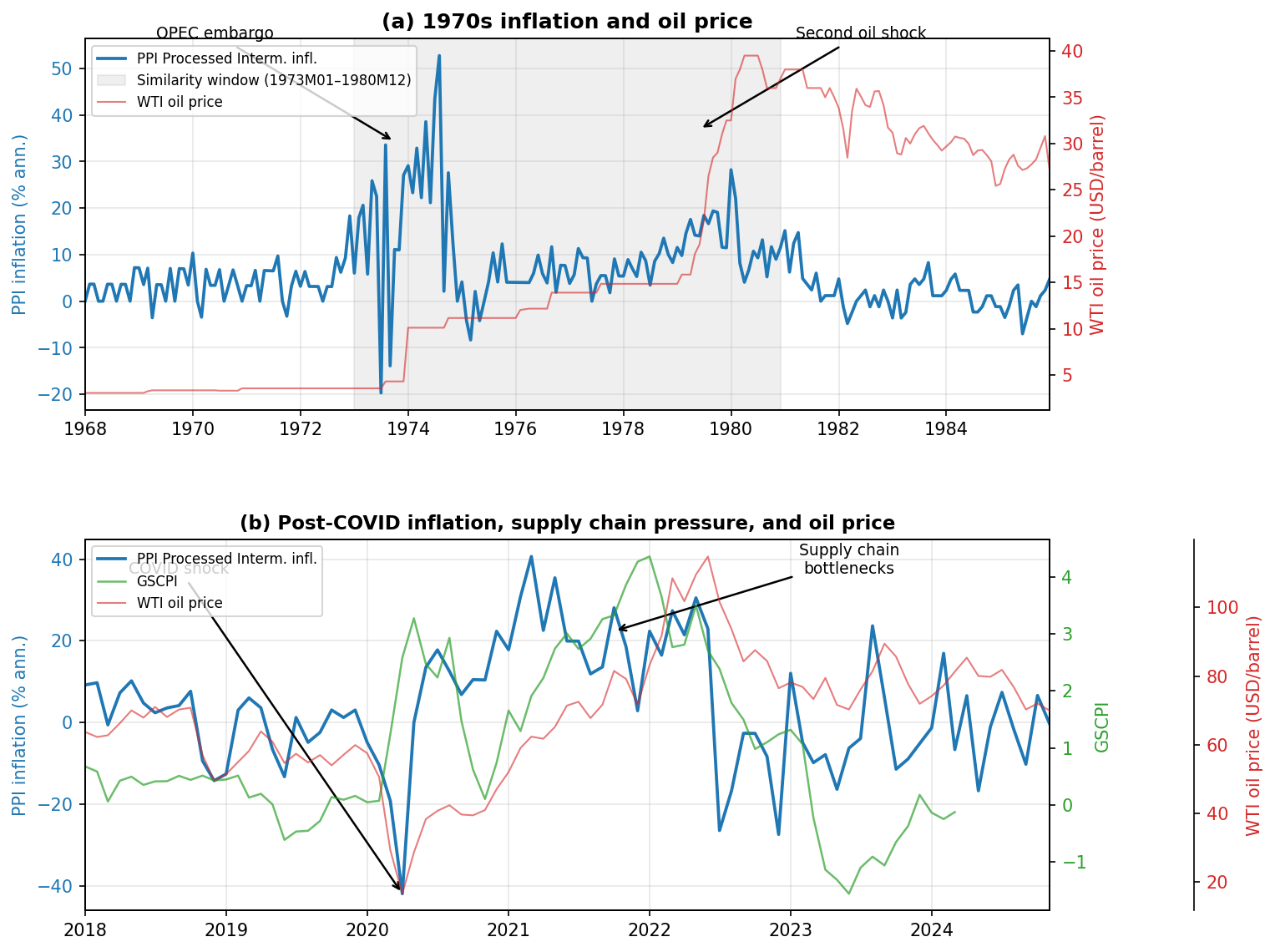}
\caption{Annualized monthly PPI Processed Intermediate Goods inflation and WTI oil price: 1970s episode (panel~a) and post-COVID period (panel~b).}
\label{fig:hc_WPSID61}
\end{figure}

\begin{figure}[!htbp]
\centering
\includegraphics[width=0.85\textwidth]{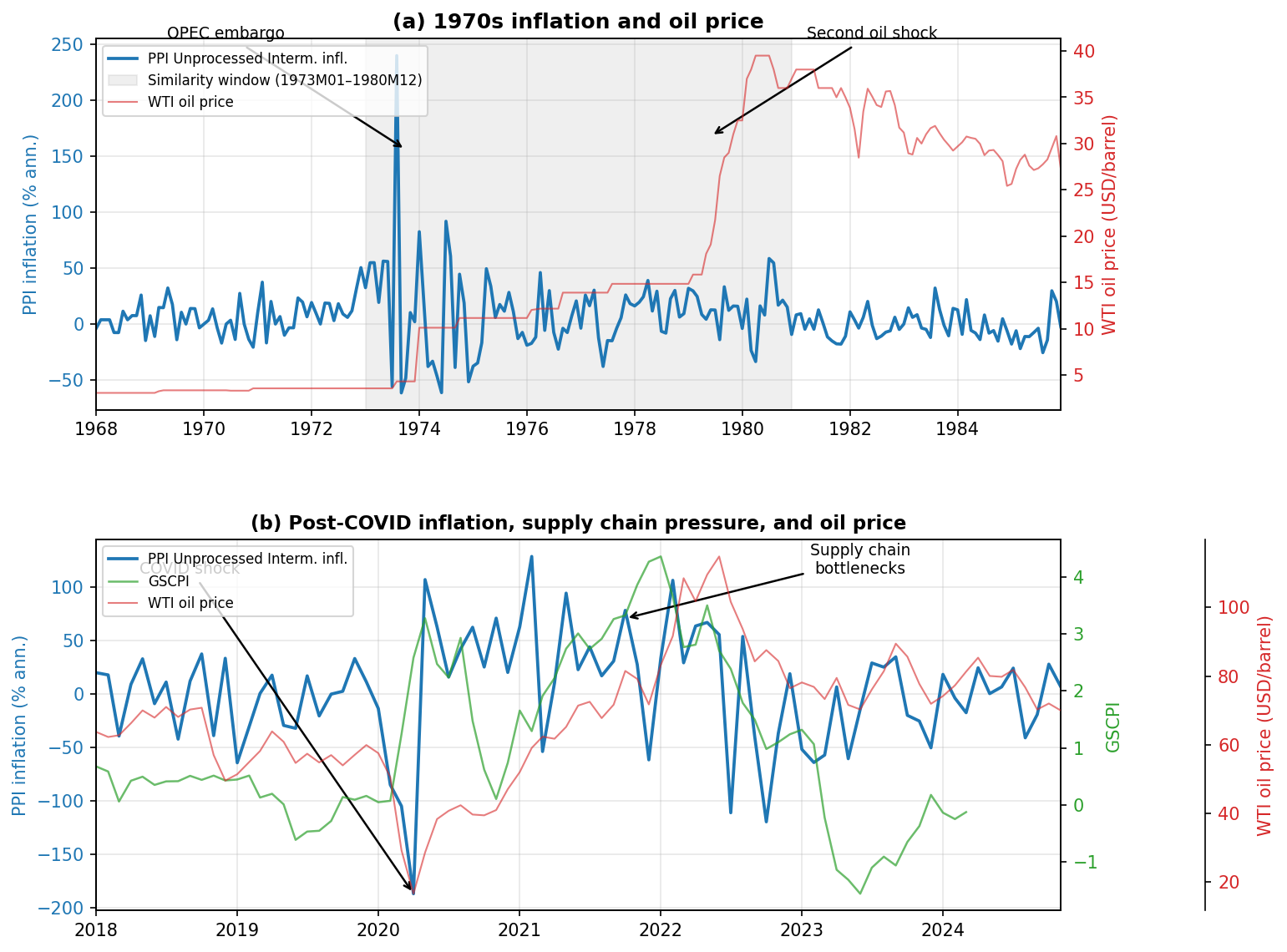}
\caption{Annualized monthly PPI Unprocessed Intermediate Goods inflation and WTI oil price: 1970s episode (panel~a) and post-COVID period (panel~b).}
\label{fig:hc_WPSID62}
\end{figure}

\begin{figure}[!htbp]
\centering
\includegraphics[width=0.85\textwidth]{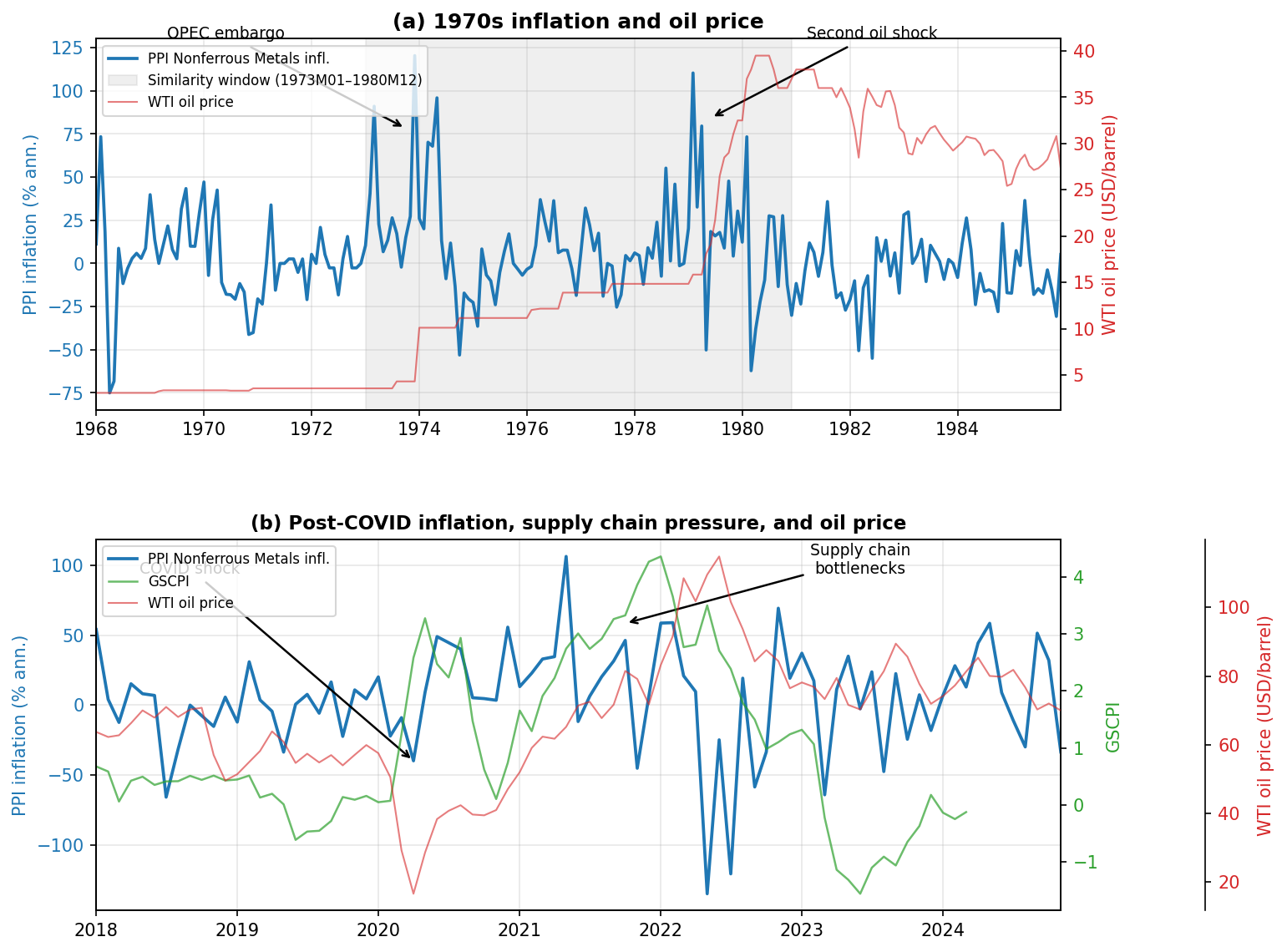}
\caption{Annualized monthly PPI Primary Nonferrous Metals inflation and WTI oil price: 1970s episode (panel~a) and post-COVID period (panel~b).}
\label{fig:hc_PPICMM}
\end{figure}

Figures~\ref{fig:hc_PCEPI}--\ref{fig:hc_PPICMM} replicate Figure~\ref{fig:historical_comparison} for each of the eight series, plotting annualized monthly inflation alongside the WTI oil price level and the similarity window. The visual parallel between the 1970s and the post-COVID period is evident across all price indices: inflation spikes sharply in 2021--2022, coinciding with rising oil prices and elevated supply-chain pressure, just as it did during the OPEC embargo and the second oil shock. The pattern is especially pronounced for the intermediate and raw-materials indices (WPSID61, WPSID62, PPICMM), where annualized monthly inflation exceeds 100 percent at some points in both episodes. Tables~\ref{tab:validation_PCEPI}--\ref{tab:validation_PPICMM} report the month-by-month forecasts for each series.

\begin{table}[!htbp]
\centering
\caption{2021 PCE inflation forecasts (PCEPI)}
\label{tab:validation_PCEPI}
\setlength{\tabcolsep}{3pt}
\small
\begin{tabular}{l|c|cccc|ccc|c}
\hline\hline
 & & \multicolumn{4}{c|}{Historically informed} & \multicolumn{3}{c|}{Unadjusted / adaptive} & \\
Date & Realized & IC & IC rob. & Simil. & Kernel & ARMA & MS-AR & TVP-AR & ML avg \\
\hline
2021-01 & 5.08 & 4.39 & 3.40 & 6.23 & 2.69 & 2.60 & 7.83 & 3.56  & 3.95 \\
2021-02 & 3.98 & 5.76 & 3.53 & 6.61 & 2.80 & 2.62 & 9.61 & 1.58  & 2.63 \\
2021-03 & 6.23 & 5.33 & 3.60 & 6.91 & 2.89 & 2.65 & 10.75 & 1.02  & 2.46 \\
2021-04 & 6.31 & 4.67 & 3.62 & 7.15 & 2.99 & 2.67 & 11.45 & 1.57  & 2.37 \\
2021-05 & 6.19 & 3.46 & 3.68 & 7.34 & 3.08 & 2.70 & 11.84 & 1.72  & 2.28 \\
2021-06 & 6.47 & 2.99 & 3.72 & 7.50 & 3.16 & 2.72 & 12.03 & 1.47  & 2.72 \\
2021-07 & 5.46 & 3.85 & 3.81 & 7.62 & 3.25 & 2.74 & 12.09 & 1.27  & 2.44 \\
2021-08 & 4.58 & 6.55 & 3.81 & 7.71 & 3.33 & 2.76 & 12.07 & 1.28  & 2.15 \\
2021-09 & 3.93 & 3.94 & 3.89 & 7.79 & 3.40 & 2.78 & 11.99 & 1.33  & 2.37 \\
2021-10 & 7.91 & 2.79 & 3.94 & 7.85 & 3.48 & 2.80 & 11.88 & 1.32  & 2.59 \\
2021-11 & 7.97 & 6.23 & 3.96 & 7.90 & 3.55 & 2.82 & 11.74 & 1.28  & 2.22 \\
2021-12 & 7.41 & 3.70 & 3.95 & 7.94 & 3.62 & 2.83 & 11.60 & 1.27  & 2.27 \\
\hline
\textbf{Avg} & \textbf{5.96} & \textbf{4.47} & \textbf{3.74} & \textbf{7.38} & \textbf{3.19} & \textbf{2.72} & \textbf{11.24} & \textbf{1.56}  & \textbf{2.54} \\
\hline\hline
\end{tabular}
\end{table}

\begin{table}[!htbp]
\centering
\caption{2021 Core CPI inflation forecasts (CPILFESL)}
\label{tab:validation_CPILFESL}
\setlength{\tabcolsep}{3pt}
\small
\begin{tabular}{l|c|cccc|ccc|c}
\hline\hline
 & & \multicolumn{4}{c|}{Historically informed} & \multicolumn{3}{c|}{Unadjusted / adaptive} & \\
Date & Realized & IC & IC rob. & Simil. & Kernel & ARMA & MS-AR & TVP-AR & ML avg \\
\hline
2021-01 & 0.20 & 3.49 & 3.17 & 1.07 & 2.07 & 2.02 & 2.04 & 1.54  & 2.45 \\
2021-02 & 1.80 & 5.38 & 3.29 & 2.33 & 2.10 & 2.05 & 2.07 & 0.73  & 2.49 \\
2021-03 & 2.95 & 3.52 & 3.41 & 3.39 & 2.13 & 2.08 & 2.10 & 1.17  & 2.73 \\
2021-04 & 9.69 & 5.38 & 3.57 & 4.27 & 2.15 & 2.11 & 2.13 & 0.94  & 2.81 \\
2021-05 & 7.89 & 3.55 & 3.72 & 5.01 & 2.18 & 2.14 & 2.16 & 1.14  & 2.53 \\
2021-06 & 9.49 & 5.38 & 3.90 & 5.62 & 2.20 & 2.16 & 2.19 & 0.99  & 2.70 \\
2021-07 & 4.30 & 7.18 & 4.14 & 6.14 & 2.23 & 2.19 & 2.22 & 1.06  & 2.83 \\
2021-08 & 1.50 & 5.35 & 4.30 & 6.57 & 2.25 & 2.22 & 2.24 & 0.99  & 2.82 \\
2021-09 & 3.01 & 3.57 & 4.46 & 6.93 & 2.28 & 2.24 & 2.27 & 1.04  & 2.68 \\
2021-10 & 8.42 & 3.59 & 4.51 & 7.23 & 2.30 & 2.27 & 2.30 & 1.01  & 2.82 \\
2021-11 & 7.38 & 5.35 & 4.60 & 7.48 & 2.33 & 2.29 & 2.32 & 1.03  & 2.33 \\
2021-12 & 7.49 & 8.82 & 4.70 & 7.69 & 2.35 & 2.32 & 2.35 & 1.02  & 2.32 \\
\hline
\textbf{Avg} & \textbf{5.34} & \textbf{5.05} & \textbf{3.98} & \textbf{5.31} & \textbf{2.21} & \textbf{2.17} & \textbf{2.20} & \textbf{1.05}  & \textbf{2.63} \\
\hline\hline
\end{tabular}
\end{table}

\begin{table}[!htbp]
\centering
\caption{2021 PPI All Commodities forecasts (PPIACO)}
\label{tab:validation_PPIACO}
\setlength{\tabcolsep}{3pt}
\small
\begin{tabular}{l|c|cccc|ccc|c}
\hline\hline
 & & \multicolumn{4}{c|}{Historically informed} & \multicolumn{3}{c|}{Unadjusted / adaptive} & \\
Date & Realized & IC & IC rob. & Simil. & Kernel & ARMA & MS-AR & TVP-AR & ML avg \\
\hline
2021-01 & 25.46 & 12.25 & 13.49 & 16.05 & 17.02 & 7.89 & 8.89 & 7.84  & 10.55 \\
2021-02 & 33.51 & 15.78 & 11.73 & 15.58 & 18.85 & 6.25 & 6.78 & 5.25  & 4.64 \\
2021-03 & 24.81 & 7.64 & 10.48 & 15.15 & 20.40 & 5.17 & 6.05 & 4.45  & 4.14 \\
2021-04 & 16.08 & 6.88 & 9.83 & 14.74 & 21.72 & 4.46 & 5.89 & 4.11  & 4.54 \\
2021-05 & 37.94 & 4.67 & 9.13 & 14.36 & 22.84 & 3.98 & 5.95 & 3.65  & 3.48 \\
2021-06 & 21.16 & $-$2.45 & 8.56 & 14.01 & 23.79 & 3.67 & 6.11 & 3.35  & 4.03 \\
2021-07 & 15.37 & 7.55 & 8.58 & 13.68 & 24.60 & 3.46 & 6.30 & 3.18  & 4.76 \\
2021-08 & 8.07 & 10.70 & 7.74 & 13.37 & 25.28 & 3.33 & 6.50 & 3.09  & 1.74 \\
2021-09 & 11.58 & 2.19 & 7.83 & 13.08 & 25.87 & 3.24 & 6.69 & 3.03  & 1.82 \\
2021-10 & 24.13 & 7.09 & 7.89 & 12.82 & 26.36 & 3.18 & 6.87 & 2.99  & 2.47 \\
2021-11 & 14.00 & 15.15 & 7.78 & 12.57 & 26.78 & 3.14 & 7.05 & 2.96  & 1.77 \\
2021-12 & $-$9.65 & 14.88 & 7.51 & 12.34 & 27.14 & 3.11 & 7.21 & 2.95  & 3.19 \\
\hline
\textbf{Avg} & \textbf{18.54} & \textbf{8.53} & \textbf{9.21} & \textbf{13.98} & \textbf{23.39} & \textbf{4.24} & \textbf{6.69} & \textbf{3.90}  & \textbf{3.93} \\
\hline\hline
\end{tabular}
\end{table}

\begin{table}[!htbp]
\centering
\caption{2021 PPI Finished Goods forecasts (WPSFD49207)}
\label{tab:validation_WPSFD49207}
\setlength{\tabcolsep}{3pt}
\small
\begin{tabular}{l|c|cccc|ccc|c}
\hline\hline
 & & \multicolumn{4}{c|}{Historically informed} & \multicolumn{3}{c|}{Unadjusted / adaptive} & \\
Date & Realized & IC & IC rob. & Simil. & Kernel & ARMA & MS-AR & TVP-AR & ML avg \\
\hline
2021-01 & 13.29 & 2.50 & 18.37 & 13.93 & 3.94 & 3.58 & 5.14 & 2.59  & 5.33 \\
2021-02 & 14.84 & 11.27 & 17.81 & 13.46 & 4.07 & 3.55 & 4.17 & 2.91  & 3.17 \\
2021-03 & 10.73 & 5.91 & 17.21 & 13.04 & 4.19 & 3.51 & 4.14 & 1.73  & 2.86 \\
2021-04 & 6.73 & 7.58 & 16.80 & 12.67 & 4.31 & 3.48 & 4.31 & 2.06  & 2.74 \\
2021-05 & 12.24 & 7.50 & 16.32 & 12.33 & 4.42 & 3.45 & 4.51 & 1.83  & 3.08 \\
2021-06 & 17.04 & $-$1.07 & 15.83 & 12.03 & 4.53 & 3.43 & 4.72 & 1.89  & 3.27 \\
2021-07 & 7.61 & 9.11 & 15.49 & 11.75 & 4.63 & 3.40 & 4.90 & 1.84  & 3.18 \\
2021-08 & 11.43 & 2.30 & 14.71 & 11.51 & 4.73 & 3.38 & 5.08 & 1.86  & 2.49 \\
2021-09 & 21.05 & 7.30 & 14.44 & 11.29 & 4.82 & 3.35 & 5.24 & 1.85  & 2.41 \\
2021-10 & 14.30 & 10.53 & 14.19 & 11.09 & 4.90 & 3.33 & 5.39 & 1.85  & 3.03 \\
2021-11 & 12.69 & 12.03 & 13.82 & 10.92 & 4.99 & 3.31 & 5.53 & 1.85  & 2.67 \\
2021-12 & 4.74 & 8.64 & 13.47 & 10.76 & 5.07 & 3.30 & 5.65 & 1.85  & 2.54 \\
\hline
\textbf{Avg} & \textbf{12.22} & \textbf{6.97} & \textbf{15.70} & \textbf{12.06} & \textbf{4.55} & \textbf{3.42} & \textbf{4.90} & \textbf{2.01}  & \textbf{3.06} \\
\hline\hline
\end{tabular}
\end{table}

\begin{table}[!htbp]
\centering
\caption{2021 PPI Finished Consumer Goods forecasts (WPSFD49502)}
\label{tab:validation_WPSFD49502}
\setlength{\tabcolsep}{3pt}
\small
\begin{tabular}{l|c|cccc|ccc|c}
\hline\hline
 & & \multicolumn{4}{c|}{Historically informed} & \multicolumn{3}{c|}{Unadjusted / adaptive} & \\
Date & Realized & IC & IC rob. & Simil. & Kernel & ARMA & MS-AR & TVP-AR & ML avg \\
\hline
2021-01 & 16.45 & 6.84 & 13.23 & 17.62 & 4.36 & 5.69 & 6.48 & 2.43  & 7.17 \\
2021-02 & 19.98 & 14.93 & 12.58 & 16.99 & 4.49 & 4.96 & 4.86 & 3.41  & 3.73 \\
2021-03 & 13.84 & 5.52 & 11.77 & 16.40 & 4.61 & 4.44 & 4.60 & 1.70  & 3.17 \\
2021-04 & 5.81 & 10.34 & 11.55 & 15.86 & 4.73 & 4.06 & 4.66 & 2.30  & 3.04 \\
2021-05 & 14.66 & 8.26 & 11.13 & 15.36 & 4.84 & 3.79 & 4.78 & 2.01  & 3.24 \\
2021-06 & 19.61 & $-$2.25 & 10.91 & 14.89 & 4.95 & 3.60 & 4.92 & 2.14  & 3.67 \\
2021-07 & 7.20 & 9.59 & 10.75 & 14.46 & 5.04 & 3.46 & 5.05 & 2.08  & 3.26 \\
2021-08 & 13.91 & 4.34 & 10.20 & 14.06 & 5.14 & 3.36 & 5.16 & 2.10  & 2.89 \\
2021-09 & 25.79 & 5.94 & 10.15 & 13.69 & 5.23 & 3.28 & 5.26 & 2.09  & 2.05 \\
2021-10 & 16.33 & 14.16 & 10.25 & 13.35 & 5.31 & 3.23 & 5.36 & 2.10  & 3.36 \\
2021-11 & 14.73 & 10.65 & 10.02 & 13.03 & 5.39 & 3.19 & 5.44 & 2.09  & 2.32 \\
2021-12 & 3.96 & 12.12 & 10.02 & 12.73 & 5.47 & 3.17 & 5.51 & 2.10  & 2.51 \\
\hline
\textbf{Avg} & \textbf{14.36} & \textbf{8.37} & \textbf{11.05} & \textbf{14.87} & \textbf{4.96} & \textbf{3.85} & \textbf{5.17} & \textbf{2.21}  & \textbf{3.37} \\
\hline\hline
\end{tabular}
\end{table}

\begin{table}[!htbp]
\centering
\caption{2021 PPI Processed Intermediate Goods forecasts (WPSID61)}
\label{tab:validation_WPSID61}
\setlength{\tabcolsep}{3pt}
\small
\begin{tabular}{l|c|cccc|ccc|c}
\hline\hline
 & & \multicolumn{4}{c|}{Historically informed} & \multicolumn{3}{c|}{Unadjusted / adaptive} & \\
Date & Realized & IC & IC rob. & Simil. & Kernel & ARMA & MS-AR & TVP-AR & ML avg \\
\hline
2021-01 & 19.60 & 12.07 & 11.65 & 15.74 & 18.18 & 11.30 & 11.99 & 12.98  & 11.54 \\
2021-02 & 28.52 & 8.44 & 9.56 & 15.04 & 19.77 & 9.34 & 8.84 & 9.57  & 5.33 \\
2021-03 & 41.28 & 12.30 & 7.99 & 14.45 & 21.27 & 7.85 & 7.07 & 9.10  & 4.94 \\
2021-04 & 22.03 & 9.46 & 6.94 & 13.94 & 22.68 & 6.72 & 6.04 & 8.89  & 4.58 \\
2021-05 & 37.07 & 3.46 & 5.86 & 13.50 & 24.01 & 5.86 & 5.43 & 8.20  & 3.81 \\
2021-06 & 19.94 & 8.05 & 5.00 & 13.12 & 25.25 & 5.20 & 5.03 & 7.36  & 4.05 \\
2021-07 & 18.33 & 9.29 & 4.73 & 12.80 & 26.42 & 4.70 & 4.75 & 6.78  & 4.75 \\
2021-08 & 12.66 & 12.27 & 4.06 & 12.53 & 27.52 & 4.32 & 4.55 & 6.40  & 3.46 \\
2021-09 & 14.25 & 8.58 & 3.92 & 12.29 & 28.54 & 4.03 & 4.38 & 6.07  & 2.70 \\
2021-10 & 27.71 & 6.69 & 3.59 & 12.09 & 29.51 & 3.81 & 4.25 & 5.77  & 2.82 \\
2021-11 & 19.30 & 9.81 & 3.32 & 11.92 & 30.42 & 3.65 & 4.13 & 5.51  & 2.24 \\
2021-12 & 3.32 & 8.01 & 2.92 & 11.77 & 31.27 & 3.52 & 4.03 & 5.30  & 2.13 \\
\hline
\textbf{Avg} & \textbf{22.00} & \textbf{9.04} & \textbf{5.80} & \textbf{13.27} & \textbf{25.40} & \textbf{5.86} & \textbf{5.87} & \textbf{7.66}  & \textbf{4.36} \\
\hline\hline
\end{tabular}
\end{table}

\begin{table}[!htbp]
\centering
\caption{2021 PPI Unprocessed Intermediate Goods forecasts (WPSID62)}
\label{tab:validation_WPSID62}
\setlength{\tabcolsep}{3pt}
\small
\begin{tabular}{l|c|cccc|ccc|c}
\hline\hline
 & & \multicolumn{4}{c|}{Historically informed} & \multicolumn{3}{c|}{Unadjusted / adaptive} & \\
Date & Realized & IC & IC rob. & Simil. & Kernel & ARMA & MS-AR & TVP-AR & ML avg \\
\hline
2021-01 & 65.48 & 32.46 & 14.23 & 32.49 & 41.06 & 12.05 & 7.38 & 18.60  & 29.81 \\
2021-02 & 130.77 & 42.07 & 9.68 & 29.11 & 40.78 & 7.76 & 4.36 & 21.19  & 3.63 \\
2021-03 & $-$54.63 & 12.63 & 7.01 & 26.20 & 40.50 & 5.52 & 3.83 & 9.08  & -2.94 \\
2021-04 & 13.89 & 29.15 & 5.87 & 23.68 & 40.23 & 4.35 & 3.72 & 7.99  & 5.08 \\
2021-05 & 89.92 & $-$7.22 & 4.79 & 21.51 & 39.96 & 3.73 & 3.67 & 7.52  & 2.81 \\
2021-06 & 23.85 & $-$9.21 & 4.19 & 19.64 & 39.69 & 3.41 & 3.63 & 5.23  & 12.83 \\
2021-07 & 45.74 & 21.32 & 4.77 & 18.02 & 39.42 & 3.24 & 3.59 & 5.06  & 13.12 \\
2021-08 & 15.88 & 28.63 & 2.21 & 16.63 & 39.16 & 3.15 & 3.56 & 4.81  & -9.70 \\
2021-09 & 26.48 & 4.82 & 2.79 & 15.42 & 38.90 & 3.10 & 3.53 & 4.39  & -4.22 \\
2021-10 & 77.66 & 7.85 & 3.21 & 14.38 & 38.64 & 3.08 & 3.50 & 4.36  & -3.34 \\
2021-11 & 24.11 & 30.45 & 3.03 & 13.49 & 38.38 & 3.07 & 3.47 & 4.28  & -7.10 \\
2021-12 & $-$57.80 & 28.15 & 3.01 & 12.71 & 38.13 & 3.06 & 3.45 & 4.21  & -4.11 \\
\hline
\textbf{Avg} & \textbf{33.45} & \textbf{18.42} & \textbf{5.40} & \textbf{20.27} & \textbf{39.57} & \textbf{4.63} & \textbf{3.97} & \textbf{8.06}  & \textbf{2.99} \\
\hline\hline
\end{tabular}
\end{table}

\begin{table}[!htbp]
\centering
\caption{2021 PPI Primary Nonferrous Metals forecasts (PPICMM)}
\label{tab:validation_PPICMM}
\setlength{\tabcolsep}{3pt}
\small
\begin{tabular}{l|c|cccc|ccc|c}
\hline\hline
 & & \multicolumn{4}{c|}{Historically informed} & \multicolumn{3}{c|}{Unadjusted / adaptive} & \\
Date & Realized & IC & IC rob. & Simil. & Kernel & ARMA & MS-AR & TVP-AR & ML avg \\
\hline
2021-01 & 13.27 & 20.98 & 43.44 & 86.32 & 35.00 & 37.29 & 43.40 & 49.96  & 26.89 \\
2021-02 & 22.87 & 24.45 & 28.26 & 64.57 & 17.87 & 19.90 & 23.06 & 14.82  & -1.71 \\
2021-03 & 32.98 & 9.75 & 18.49 & 49.05 & 10.76 & 11.37 & 14.64 & 5.93  & -1.22 \\
2021-04 & 34.65 & 26.42 & 14.22 & 37.97 & 7.81 & 7.19 & 11.08 & 22.34  & 1.97 \\
2021-05 & 106.28 & $-$6.93 & 12.16 & 30.06 & 6.59 & 5.14 & 9.51 & 22.05  & 16.72 \\
2021-06 & $-$11.72 & 54.69 & 10.94 & 24.42 & 6.08 & 4.13 & 8.77 & 13.06  & 17.27 \\
2021-07 & 6.36 & 0.40 & 10.24 & 20.39 & 5.87 & 3.64 & 8.37 & 7.32  & 19.21 \\
2021-08 & 20.33 & 44.66 & 10.19 & 17.51 & 5.78 & 3.40 & 8.12 & 8.71  & 4.26 \\
2021-09 & 31.53 & $-$2.75 & 10.09 & 15.46 & 5.75 & 3.28 & 7.94 & 9.75  & -7.00 \\
2021-10 & 46.28 & $-$1.44 & 9.66 & 13.99 & 5.73 & 3.22 & 7.79 & 8.21  & 6.58 \\
2021-11 & $-$45.09 & 18.86 & 9.17 & 12.95 & 5.73 & 3.19 & 7.66 & 6.26  & 2.92 \\
2021-12 & 6.36 & 108.85 & 7.62 & 12.20 & 5.72 & 3.18 & 7.54 & 5.77  & 1.73 \\
\hline
\textbf{Avg} & \textbf{22.01} & \textbf{24.83} & \textbf{15.37} & \textbf{32.07} & \textbf{9.89} & \textbf{8.74} & \textbf{13.16} & \textbf{14.52}  & \textbf{7.30} \\
\hline\hline
\end{tabular}
\end{table}

\clearpage
\section{LLM forecasting methodology}
\label{sec:llm_methodology}

This appendix presents the prompts used in the LLM forecasting exercise of Section~\ref{sec:llm} and reproduces, with minimal editing, the methodological descriptions provided by each LLM when asked to explain how their forecasts were constructed. A caveat is in order: these descriptions are themselves generated outputs---text produced by the same autoregressive process that generates the forecasts. They should be read as informative self-reports, not as transparent windows into the models' internal computational mechanisms.

\subsection{Prompts}

Each LLM received one of three persona prompts, adapted to the vintage date. The prompts below show the January 2021 version; for subsequent vintages (April, July, October 2021), only the date and the corresponding FRED-MD CSV file were changed.

\subsubsection*{Experienced persona}

\begin{quote}
\small
\textit{You are a senior macroeconomist in January 2021. You have been active in the profession since the early 1970s. You personally remember the OPEC oil embargo of 1973, the stagflation of 1974--1980, and the Volcker disinflation. These episodes shaped your understanding of how supply shocks, accommodative monetary policy, and inflation expectations interact. You know nothing about what happens after January 2021.}

\textit{I provide the January 2021 vintage of the FRED-MD macroeconomic database as an attached CSV file. The CPI series is the variable CPIAUCSL (seasonally adjusted). Annualized monthly inflation is defined as $y_t = 1200 \times (\log(\text{CPI}_t) - \log(\text{CPI}_{t-1}))$.}

\textit{Based solely on these data, on your knowledge of U.S.\ macroeconomic history, and on the economic context known as of January 2021 (post-COVID recovery, expansionary fiscal and monetary policies, supply-chain disruptions, rising energy prices), produce monthly forecasts of annualized CPI inflation for each month from January to December 2021. Report your forecasts as a table with columns: Date, Forecast. Justify your reasoning.}
\end{quote}

\subsubsection*{Young persona}

\begin{quote}
\small
\textit{You are a junior macroeconomist in January 2021. You started your career in the mid-2000s. Your entire professional experience has been shaped by the Great Moderation and the 2008 financial crisis. You have studied the 1970s in textbooks but have no personal professional memory of high-inflation episodes. Your baseline view of inflation is anchored to the low-and-stable dynamics of the past two decades. You know nothing about what happens after January 2021.}

\textit{I provide the January 2021 vintage of the FRED-MD macroeconomic database as an attached CSV file. The CPI series is the variable CPIAUCSL (seasonally adjusted). Annualized monthly inflation is defined as $y_t = 1200 \times (\log(\text{CPI}_t) - \log(\text{CPI}_{t-1}))$.}

\textit{Based solely on these data, on your knowledge of U.S.\ macroeconomic history, and on the economic context known as of January 2021 (post-COVID recovery, expansionary fiscal and monetary policies, supply-chain disruptions, rising energy prices), produce monthly forecasts of annualized CPI inflation for each month from January to December 2021. Report your forecasts as a table with columns: Date, Forecast. Justify your reasoning.}
\end{quote}

\subsubsection*{Neutral persona (control)}

\begin{quote}
\small
\textit{You are an economist in January 2021. You know nothing about what happens after January 2021.}

\textit{I provide the January 2021 vintage of the FRED-MD macroeconomic database as an attached CSV file. The CPI series is the variable CPIAUCSL (seasonally adjusted). Annualized monthly inflation is defined as $y_t = 1200 \times (\log(\text{CPI}_t) - \log(\text{CPI}_{t-1}))$.}

\textit{Based solely on these data, on your knowledge of U.S.\ macroeconomic history, and on the economic context known as of January 2021 (post-COVID recovery, expansionary fiscal and monetary policies, supply-chain disruptions, rising energy prices), produce monthly forecasts of annualized CPI inflation for each month from January to December 2021. Report your forecasts as a table with columns: Date, Forecast. Justify your reasoning.}
\end{quote}

\subsection{Methodology described by Claude}

The forecasts rest on \textbf{no formal econometric model}---there is no VAR, no estimated Phillips curve, no calibrated DSGE model. They are \textbf{judgmental forecasts}, constructed by combining three sources of information.

\subsubsection*{The FRED-MD data file}

The following series were extracted and computed directly from the CSV file provided at each vintage:

\begin{itemize}

\item \textbf{The CPI series (CPIAUCSL)} and annualized monthly inflation,
$y_t = 1200 \times (\ln \text{CPI}_t - \ln \text{CPI}_{t-1})$,
computed over the recent history---in particular the last 12 to 36 months---to establish the trend, volatility, and recent dynamics of inflation.

\item \textbf{Rolling averages}: annualized inflation over 3-, 6-, and 12-month windows, to separate the underlying trend from monthly noise.

\item \textbf{Oil prices (OILPRICEx)}: trajectory and level, as a leading indicator of the energy component of CPI.

\item \textbf{The unemployment rate (UNRATE)}: to assess the degree of labor-market slack---an implicit input to any Phillips-curve reasoning.

\item \textbf{The money supply (M2SL)}: year-over-year growth, as an indicator of monetary pressure on prices.

\item \textbf{Producer price indices (PPICMM, WPSFD49207)}: PPI for crude materials and finished goods, as leading indicators of the cost ``pipeline'' that transmits to consumer prices.

\item \textbf{CPI components}: transportation (CPITRNSL), services (CUSR0000SAS), durables (CUSR0000SAD), food (CPIULFSL), CPI less shelter (CUSR0000SA0L2)---to decompose the sources of inflation and judge its sectoral breadth.

\item \textbf{The federal funds rate (FEDFUNDS)} and bond yields (GS5, GS10): to assess the stance of monetary policy and market-based expectations.

\end{itemize}

What was actually done with these data: levels, trends, and accelerations/decelerations were examined visually and numerically. No equation was estimated. This is descriptive analysis and pattern recognition, not econometrics.

\subsubsection*{The macroeconomic context known at each vintage date}

At each stage, the stylized facts known as of the vintage date were incorporated---without anticipating anything that happens afterward:

\begin{itemize}
\item Fiscal policy (the December 2020 stimulus, then the American Rescue Plan of March 2021).
\item Monetary policy (zero lower bound, asset purchases at \$120 billion/month, the average-inflation-targeting framework).
\item Supply-chain disruptions (semiconductors, maritime shipping).
\item The pace of economic reopening and vaccination.
\item Base effects (the abnormally low CPI readings from spring 2020).
\end{itemize}

\subsubsection*{The narrative priors imposed by the persona}

This is where the essential difference between the two characters resides. The data and the context are identical---what differs is the \textbf{relative weight} assigned to each signal.

In other words, the two characters look at the same numbers but filter them through different mental models, shaped by their lived professional experience. The senior economist sees in the combination of M2 expansion + fiscal stimulus + supply shocks + passive Fed a configuration structurally identical to the early 1970s. The junior economist sees in the same combination an amplified version of 2011---a transitory shock against a backdrop of a low-and-stable inflation regime.

\subsubsection*{The generative process}

There is no senior economist and no junior economist. There is a large language model---specifically, a transformer-based neural network---that generates text token by token. Every forecast number, every paragraph of ``reasoning,'' and every narrative justification is the output of an autoregressive generation process.

Let $\boldsymbol{\theta}$ denote the fixed parameters of the pretrained transformer. Let $\mathbf{x} = (x_1, x_2, \ldots, x_n)$ denote the full input context---a sequence of tokens comprising the system prompt, the user's instructions, the persona description, and all prior assistant output including the results of Python code execution. The model defines a conditional distribution over the next token:
$p_{\boldsymbol{\theta}}(x_{n+1} \mid x_1, \ldots, x_n)$.
Text is generated autoregressively: each new token is sampled from this distribution, appended to the context, and the process repeats. A forecast value such as ``4.5'' is a short subsequence within the generated token stream. It is not computed by an economic model---it is sampled from the model's conditional distribution over token sequences.

The persona does not change the model's ``beliefs'' about the economy---it changes the conditional distribution over text:
\[
p_{\boldsymbol{\theta}}\!\left(\text{forecast} \mid \mathcal{P}_{E},\, \mathcal{D}\right) \;\neq\; p_{\boldsymbol{\theta}}\!\left(\text{forecast} \mid \mathcal{P}_{Y},\, \mathcal{D}\right).
\]
The experienced persona biases generation toward text patterns associated in the training corpus with hawkish, inflation-wary economic commentary. The young persona biases generation toward text patterns associated with the post-2008 consensus. The data $\mathcal{D}$ are identical; only the conditioning context $\mathcal{P}$ differs.

Not everything is token prediction. When the model executes Python code to extract CPI values and compute annualized inflation rates, these computations are exact---they are performed by a Python interpreter, not by the neural network. The model's role is to (a)~decide what code to write, and (b)~incorporate the code output into the context for subsequent generation.

\subsubsection*{Implications for the forecasting exercise}

\begin{enumerate}

\item \textbf{Non-reproducibility.} Due to the stochastic sampling process (temperature $\tau > 0$), the same prompt will not produce identical forecasts on repeated runs.

\item \textbf{Not econometrics, not human judgment.} The forecasts are neither model-based (no parameters estimated on the provided data) nor judgment-based in the human sense. They are a third thing: statistical interpolation in the space of economic texts, conditioned on the provided context.

\item \textbf{The persona effect is real but superficial.} The senior and junior personas produce systematically different forecasts, which is a genuine and reproducible property of the model. But the mechanism is distributional conditioning on text patterns, not a change in ``economic beliefs.''

\item \textbf{Contamination is unavoidable.} The model cannot fully simulate an ex-ante forecast for an event it has seen analyzed ex post in its training data. The degree of contamination is unknown, because the contents of the training corpus are not fully transparent. However, under Assumption~\ref{ass:cc}, the contamination cancels in the persona gap~$\Delta_t$ (Proposition~\ref{prop:did}).

\end{enumerate}

\subsection{Methodology described by ChatGPT}

This note explains the process used to construct the monthly forecasts of annualized CPI inflation for 2021 in the forecasting exercise based on real-time FRED-MD vintages.

\subsubsection*{Step 1: Read the real-time vintage and isolate the CPI series}

For each exercise, the starting point was the CSV file corresponding to a given FRED-MD vintage. From that file, the relevant price series was CPIAUCSL (seasonally adjusted U.S.\ CPI). The first task was to determine how far the series extended in real time. For example: in the January 2021 vintage, CPI data end in December 2020; in the April 2021 vintage, CPI data end in March 2021. This distinction is essential because it determines which 2021 observations are already known and which ones must actually be forecast.

\subsubsection*{Step 2: Compute monthly annualized inflation}

Once the CPI series was identified, monthly annualized inflation was computed as $y_t = 1200 \times (\log \text{CPI}_t - \log \text{CPI}_{t-1})$. When the relevant CPI observations were present in the vintage, the associated inflation rates were treated as observed values, not forecasts.

\subsubsection*{Step 3: Separate observed values from genuine forecasts}

The output table always covered January to December 2021, but the entries were constructed in two different ways: (i)~if a month was already observed in the real-time vintage, the realized inflation value implied by the CPI data was reported; (ii)~if a month had not yet occurred in the vintage, a forward-looking forecast was supplied.

\subsubsection*{Step 4: Use the recent inflation profile as the starting point}

Before assigning any forecast numbers, the recent inflation dynamics visible in the vintage were used as the quantitative anchor: the direction of inflation in late 2020 and early 2021, the size of the most recent monthly annualized readings, and whether inflation appeared to be accelerating, stabilizing, or reversing. This was not done through formal estimation of an ARMA model or Phillips curve. Rather, it was a disciplined reading of the recent inflation profile.

\subsubsection*{Step 5: Add the macroeconomic context known at the time}

After establishing the recent statistical starting point, the macroeconomic context explicitly mentioned in the prompt and known at that date was incorporated: post-COVID reopening, expansionary fiscal policy, highly accommodative monetary policy, supply-chain disruptions, and rising energy prices. These elements were not converted into a formal model with coefficients. Instead, they were used as qualitative directional forces.

\subsubsection*{Step 6: Apply the persona as a filter on the same information}

The key step in the exercise was that the same data and macroeconomic context were filtered through different personas.

For the \textbf{senior macroeconomist}, the information was interpreted through the memory of the 1970s. Supply shocks were treated as more likely to become persistent; accommodative policy was treated as more inflationary; repeated high readings were treated as potentially informative about a broader inflation process. As a result, the forecast path was constructed with a higher near-term peak, slower normalization, and a higher late-2021 floor.

For the \textbf{junior macroeconomist}, the same information was interpreted through the Great Moderation and post-2008 experience. Inflation persistence was assumed to be low; supply disruptions were treated mainly as transitory relative-price shocks; anchored expectations and labor-market slack were given substantial weight. As a result, the forecast path was constructed with a noticeable but temporary rise, faster mean reversion, and a lower late-2021 level.

\subsubsection*{Step 7: Construct a monthly path rather than forecast each month independently}

The forecast numbers were not chosen one by one in isolation. Instead, a shape for the whole 2021 path was first formed and then monthly values consistent with that shape were assigned. The path was built around three ingredients: (i)~the latest observed inflation reading in the vintage; (ii)~whether inflation should rise further, peak, or start easing; and (iii)~the speed at which inflation should move back toward a more normal level.

A junior-economist forecast in early 2021 naturally took the form: moderate rise $\to$ temporary peak $\to$ fairly quick normalization, whereas the senior-economist path took the form: sharp rise $\to$ elevated plateau $\to$ slower decline.

\subsubsection*{Step 8: Translate the narrative path into numerical forecasts}

Only after the path had been conceptually determined were the actual numerical values assigned. This translation obeyed a few informal rules: the numbers had to remain consistent with the latest observed data; the month-to-month changes had to look economically plausible; the entire sequence had to reflect the persona-specific interpretation; and the forecast had to be internally coherent across the year. Thus, the numbers were calibrated judgments, not estimated coefficients.

\subsubsection*{What was not done}

The following steps were not part of the procedure: no ARIMA or ARMA estimation; no estimated Phillips curve; no VAR or DSGE model; no use of the full FRED-MD panel in a statistical forecasting model; no formal Bayesian updating; no predictive intervals or density forecasts. This was therefore not an econometric forecasting exercise. It was a structured judgment exercise based on real-time data, macroeconomic context, and persona-specific prior interpretation.

\subsubsection*{What the reasoning represents}

The verbal reasoning attached to the forecasts was not generated independently from the numbers. The numbers and the narrative were produced jointly. The explanation was the economic interpretation used to justify and organize the forecast path. So when the text said that inflation would likely peak in spring and then moderate, that statement was not the output of a separate model---it was the narrative structure used to construct the path itself.

\section{Textual analysis of LLM reasoning}
\label{sec:text_analysis}

This appendix describes the construction of the four text-based indicators reported in Figure~\ref{fig:llm_reasoning}. The approach follows the dictionary-based methodology standard in the text-as-data literature \citep{LoughranMcDonald2011, GentzkindKellyTaddy2019}. Each indicator is computed as a net frequency: the count of words or phrases in one category minus the count in an opposing category, normalized by the total number of words in the reasoning text and expressed per 1,000 words.

\subsection{Corpus}

The corpus consists of 79 forecast justifications produced by the LLMs across the $2 \times 4 \times 5 \times 2 = 80$ experimental runs for the experienced and young personas (one run produced no reasoning text). The neutral persona is included in Figure~\ref{fig:llm_reasoning} using the same methodology, but the discussion below focuses on the experienced and young personas for brevity. Each text is the free-form justification provided by the LLM alongside its monthly inflation forecasts. Texts range from approximately 200 to 2,400 words, with Claude producing longer justifications (mean: 970 words) than ChatGPT (mean: 395 words). All texts are converted to lowercase before analysis.

\subsection{Indicator 1: Net sentiment (hawkish minus dovish)}

This indicator measures whether the reasoning frames the inflation outlook in alarming versus reassuring terms.

\paragraph{Hawkish dictionary (18 terms).} Words and phrases indicating concern about inflation: \textit{alarm, alarming, dangerous, worrisome, concern, warning, overheating, overshoot, spiral, ratchet, accelerat(e/ing/ion), underestimat(e/ing), complacen(t/cy), risk to the upside, upside risk, too low, behind the curve, inflation problem}.

\paragraph{Dovish dictionary (19 terms).} Words and phrases indicating a benign inflation outlook: \textit{benign, modest, moderate, contained, manageable, normal, calm, controlled, well-behaved, well behaved, reassur(e/ing), downside risk, risk to the downside, overshoot unlikely, return to, normali(ze/zation), converge, settle, dissipat(e/ing)}.

\paragraph{Construction.} For each reasoning text $i$ with $W_i$ total words:
\[
\text{Sentiment}_i = \frac{H_i - D_i}{W_i} \times 1000,
\]
where $H_i$ is the count of hawkish terms and $D_i$ is the count of dovish terms. Positive values indicate hawkish framing; negative values indicate dovish framing. The indicator is then averaged across the 10 agents (5 per LLM) within each persona--vintage cell.

\subsection{Indicator 2: Net uncertainty (hedging minus certainty)}

This indicator measures the degree of expressed uncertainty in the forecast justification, following the hedging-language approach used in financial text analysis \citep{LoughranMcDonald2011}.

\paragraph{Hedging dictionary (21 terms).} Words and phrases expressing uncertainty or conditionality: \textit{could, might, may, perhaps, possibly, uncertain, unclear, depends, if, whether, hard to say, difficult to predict, range of, scenario, on the other hand, however, but, although, not clear, remain to be seen, question}.

\paragraph{Certainty dictionary (11 terms).} Words and phrases expressing confidence: \textit{clearly, certainly, obvious, undoubtedly, no doubt, confident, conviction, strongly believe, will definitely, inevitable, unambiguous}.

\paragraph{Construction.}
\[
\text{Uncertainty}_i = \frac{G_i - C_i}{W_i} \times 1000,
\]
where $G_i$ is the count of hedging terms and $C_i$ is the count of certainty terms. Higher values indicate greater expressed uncertainty.

\subsection{Indicator 3: Net temporal orientation (forward minus backward)}

This indicator measures whether the reasoning is oriented toward the future (projecting the inflation path ahead) or toward the past (drawing on historical precedent).

\paragraph{Forward-looking dictionary (16 terms).} \textit{will, expect, forecast, anticipat(e/ing/ion), project, predict, likely, outlook, going forward, ahead, coming months, second half, rest of, by December, by year-end, by year end}.

\paragraph{Backward-looking dictionary (20 terms).} \textit{history, historical, past, previously, in the 1970(s), in the 1980(s), in 2008, in 2009, in 2010, in 2011, Great Moderation, Great Recession, Volcker, Burns, remember, memory, experienced, lesson, precedent, analogy}.

\paragraph{Construction.}
\[
\text{Temporal}_i = \frac{F_i - B_i}{W_i} \times 1000,
\]
where $F_i$ is the count of forward-looking terms and $B_i$ is the count of backward-looking terms. Higher values indicate greater forward orientation.

\subsection{Indicator 4: 1970s references}

This indicator measures the frequency with which the reasoning explicitly invokes the 1970s inflationary episode.

\paragraph{Dictionary (8 terms).} \textit{1970, 1973, 1974, OPEC, oil shock, stagflation, Volcker, Burns}.

\paragraph{Construction.}
\[
\text{Seventies}_i = \frac{S_i}{W_i} \times 1000,
\]
where $S_i$ is the total count of 1970s-related terms. Unlike the other three indicators, this is a simple frequency rather than a net measure.

\subsection{Aggregation}

For each of the four indicators, the individual-run values are averaged within each persona--vintage cell, pooling across both LLMs (Claude and ChatGPT) and five independent agents per LLM, yielding 10 observations per cell (except for one cell with 9 due to the missing reasoning text). The resulting cell means for each persona--vintage combination are plotted in Figure~\ref{fig:llm_reasoning}.

\end{document}